\documentclass{aa}  
\usepackage{graphicx}
\usepackage{txfonts}

\usepackage[utf8]{inputenc} 
\usepackage[T1]{fontenc}    
\usepackage[colorlinks = true,
            linkcolor = blue,
            urlcolor  = blue,
            citecolor = blue,
            anchorcolor = blue]{hyperref}
\usepackage{url}            
\usepackage{booktabs}       
\usepackage{amsfonts}       
\usepackage{nicefrac}       
\usepackage{microtype}      
\usepackage{lipsum}
\usepackage{multicol}       
\usepackage{hyperref}       
\hypersetup{colorlinks=true,linkcolor=blue, linktocpage}
\usepackage{textcomp}
\usepackage{gensymb}        
\usepackage{graphicx}       
\usepackage{amsmath}        
\usepackage{amssymb}        
\usepackage{float}          
\usepackage{subcaption}
\usepackage{stfloats}       
\usepackage[export]{adjustbox} 
\usepackage[notquote]{hanging}        
\usepackage[labelfont=bf]{caption}  
\usepackage[font=scriptsize]{caption}   

\usepackage{pdflscape}
\usepackage{afterpage}
\usepackage{capt-of}

\usepackage{times}
\usepackage{tikz}
\usepackage{siunitx} 
\usetikzlibrary{shapes.geometric, arrows}

\begin{document} 

    \title{Characterisation of the X-ray point source variability in the eROSITA south ecliptic pole field}

\author{D. Bogensberger \inst{1} \inst{2}, K. Nandra \inst{1}, M. Salvato \inst{1}, T. Liu \inst{1}, J. Wolf \inst{1}\inst{3}\inst{4}, S. Croom \inst{5}, H. Starck \inst{1}, J. Buchner \inst{1}, G. Ponti \inst{6}\inst{1}, J. Ider Chitham \inst{1}, C. Maitra \inst{1}, J. Robrade \inst{7}, A. Merloni \inst{1}, M. Krumpe \inst{8}}
\institute{Max Planck Institute f{\"u}r Extraterrestrische Physik, Gie{\ss}enbachstra{\ss}e 1, 85748 Garching bei München, Germany 
    \and
        Department of Astronomy, The University of Michigan, 1085 South University Avenue, Ann Arbor, Michigan, 48103, USA
    \and
        Excellenzcluster ORIGINS, Boltzmannstr. 2, 85748 Garching, Germany
    \and 
        Max-Planck Institut für Astronomie, Königstuhl 17, 69177 Heidelberg, Germany
    \and
        School of Physics, The University of Sydney, Camperdown NSW 2006, Australia
    \and
        INAF - Osservatorio Astronomico di Brera, via E. Bianchi 46, 23807 Merate, Italy   
    \and
        Hamburger Sternwarte, Universit{\"a}t Hamburg, Gojenbergsweg 112, 21029 Hamburg, Germany 
    \and
        Leibniz-Institut f{\"u}r Astrophysik Potsdam, An der Sternwarte 16, 14482 Potsdam\\
    \email{\href{mailto:dbogen@umich.edu}{dbogen@umich.edu}}
}
\date{Received <date> / Accepted <date>}

\abstract{
        \textit{\textbf{Aims}}: During the Spectrum Roentgen Gamma (\emph{SRG})/\emph{eROSITA} all-sky surveys, X-ray sources close to the South Ecliptic Pole (SEP) are observed almost every 4 hours. We aim to identify the sources exhibiting the most significant long-term X-ray variability within $3\degree$ of the SEP in the first three surveys, and investigate their properties. 
        
		\textit{\textbf{Methods}}: We determined the variability significance of all sources observed by \emph{eROSITA} within $3\degree$ of the SEP by using thresholds on the Bayesian excess variance (SCATT\_LO) and the maximum amplitude deviation (AMPL\_SIG). Sources exhibiting a variability significance above $3\sigma$ were subdivided into likely Galactic and extragalactic sources, by using spectral and photometric information of their optical counterparts. We quantified the X-ray normalised excess variances of all variable sources, and also calculated the periodograms of the brightest ones.

		\textit{\textbf{Results}}: Out of more than $10^4$ X-ray sources detected by \emph{eROSITA} within $3\degree$ of the SEP, we identified 453 that exhibit significant X-ray variability. SCATT\_LO is significantly more sensitive to detecting variable sources in this field, but AMPL\_SIG helps provide a more complete variability sample. Of those variable sources, 168 were classified as likely extragalactic, and 235 as likely Galactic. The periodograms of most bright and variable extragalactic sources are approximately described by an aliased power law ($P\propto\nu^{-\alpha}$) with an index of $\alpha\approx 1$. We identified a potential tidal disruption event, and long-term transient sources. The stellar X-ray variability was predominantly caused by bright X-ray flares from coronally active stars. } 


\titlerunning{\emph{eROSITA} variable X-ray sources in the SEP}
\authorrunning{David Bogensberger}

\maketitle 

\section{Introduction}

Studies of the X-ray variability on a range of timescales can help unveil the nature of the physical processes at work in different types of astronomical sources \citep{1996ApJ...469L...9S, 2001AdSpR..28..295N, 2002ApJ...572..392B, 2005MNRAS.359..345U, 2009A&A...504...61I, 2016A&A...593A..55V, 2019AN....340..334S}. Most of the brightest strongly variable X-ray point sources are active galactic nuclei (AGNs), stars, or X-ray binaries (XRBs) \citep{1993Sci...260.1769T, 2003A&A...403..247F}. 




AGNs have been observed to feature significant variability across the electromagnetic spectrum, with different energy bands being sensitive to different aspects of the torus, accretion disc, corona, and jet of AGNs \citep{2017A&ARv..25....2P}. X-rays are generated in the corona, located close to the supermassive black hole, through Compton up-scattering of lower energy photons from the accretion disc \citep{1991ApJ...380L..51H}. These also interact with the disc to produce reflection and fluorescence features in the X-ray spectrum \citep{1991MNRAS.249..352G}. In this extreme gravitational field, the interaction of in-falling matter and radiation produces a variety of unique effects, many of which have yet to be properly understood. Due to the location of the X-ray emitting region, the X-ray light curves of AGNs change on the shortest timescales, and can already vary considerably over a timescale of less than 1 hour \citep{2003A&AT...22..661G, 2003MNRAS.345.1271V, 2009A&A...504...61I, 2012A&A...542A..83P, 2013MNRAS.430L..49M}. X-ray variability is essential to investigating the extreme physics close to the event horizon of supermassive black holes \citep{2006Natur.444..730M}.


Several previous studies of AGN X-ray variability have examined continuous light curves, mapping out the high frequency AGN variability domain \citep[e.g.][]{2004MNRAS.348..783M, 2012A&A...542A..83P, 2013MNRAS.430L..49M, 2013MNRAS.431.2441D, 2017MNRAS.466.3309R}. The low frequency AGN variability domain has also been explored, which often involves combining sets of observations obtained with an irregular sampling. Long-term variability studies with irregular sampling often rely on the structure function, which estimates the average difference in the brightness of a source at two instances, as a function of their temporal separation \citep{2017A&A...599A..82M}. In X-rays, several groups \citep[e.g.][]{2001A&A...367..470G, 2017ApJ...849..127Z, 2017MNRAS.471.4398P} have combined data on sources obtained from the same instrument over several years. \citet{2017A&A...599A..82M} combined data from different instruments to extend the time domain studied even further. It has even been possible to study the long-term optical variability of AGNs over a timescale of $>10^4$ years, by investigating the quasar light echo \citep{2009MNRAS.399..129L}. \citet{2018MNRAS.476L..34S} have combined multiple methodologies to investigate the optical AGN structure function on timescales of several days to tens of thousands of years. 

A different powerful tool for describing variable sources is the Power Spectral Density (PSD), which describes how large the contribution at individual frequencies is to the observed overall variability. For a detailed overview, see \citet{1989ASIC..262...27V}. The PSD is a useful quantity to estimate, to understand where the variability originates, and to investigate the structure of the accretion disc. The PSD of a variable source is estimated by the periodogram, which is the squared absolute value of the Fourier transform of the light curve. 


A few studies have also investigated long-term AGN X-ray variability with regular sampling. For instance, \citet{2001ApJ...547..684M} investigated AGN variability from RXTE light curves sampled regularly every 5 days, for a duration of 300 days. However, this was only possible for a total of 9 sources.



Periodograms of AGNs are commonly observed to have a power law relationship between the power $P$, and the frequency $\nu$;  $P(\nu) \propto \nu^{-\alpha}$. The power law index, $\alpha$, is found to often be close to 2 at high frequencies, and close to 1 at low frequencies \citep[][]{1999ApJ...514..682E, 2002A&A...382L...1P, 2004MNRAS.348..207P, 2004MNRAS.348..783M}. The frequency at which the break occurs, has been found to be anywhere in the range of $10^{-6.4} - 10^{-3.3} ~\mathrm{Hz}$ \citep{2012A&A...544A..80G}. The low frequency $\alpha\approx1$ power law is expected to become flatter at even lower frequencies, to avoid a divergence of the integral. This second break in the periodogram power law has been observed by \citet{1990A&A...227L..33B} for the XRB Cygnus X-1. It is expected that AGNs might also feature a similar low frequency break to $\alpha\approx 0$ at very low frequencies. However, this has not yet been observed. 



The variability of regularly, or irregularly sampled data can also be quantified with the normalised excess variance \citep[NEV;][]{1990ApJ...359...86E, 1997ApJ...476...70N}, which is equal to the integral of the periodogram \citep{1989ASIC..262...27V}. The NEV has been found to anti-correlate with the black hole mass and luminosity \citep{1997ApJ...476...70N, 2001MNRAS.324..653L, 2005MNRAS.358.1405O, 2012A&A...542A..83P, 2016ApJ...831..145Y, 2017ApJ...849..127Z, 2017MNRAS.471.4398P}. This means that a measurement of the NEV could be used to constrain the black hole mass.

XRBs are another class of bright, and variable X-ray sources. Many of the spectral and variability properties of XRBs are very similar to those of AGNs, when accounting for the differences in mass \citep{2006Natur.444..730M}. However, the temperature of the accretion disc in XRBs is higher than in AGNs, so the soft X-ray spectrum also features a variable black body component \citep{1988ApJ...324..363W}. There are further spectral differences between systems involving a black hole, or a neutron star \citep{2005ApJ...626..298T}. The X-ray emission of XRBs varies on timescales of milliseconds, as the size of the X-ray emitting region is merely a few tens of km \citep{2012MNRAS.426.1701B}. XRBs also often feature quasi-periodic oscillations \citep{2019NewAR..8501524I}, which are much more rarely detected in AGNs \citep{2005MNRAS.362..235V}.

Stars in the Milky Way are the other main type of significantly variable X-ray sources observed by \emph{eROSITA}. Stellar X-rays are produced in the stellar corona through emission from hot plasma at temperatures above $1~\mathrm{MK}$ by magnetic activity \citep{2004A&ARv..12...71G}. Stellar coronae feature variability on timescales of minutes to days \citep{2012A&A...548A..85F}. Typical stellar X-ray variability features infrequent, asymmetric flares that exceed the non-flaring continuum flux level by a factor of up to $\approx 100$ \citep{2005ApJS..160..469F, 2010PNAS..107.7158T}. 



This paper is structured as follows: Section \ref{eRobs} summarises the observations that were used in this work. Section \ref{SourcedetLCe} describes the \emph{eROSITA} source detection and light curve generation procedure. Section \ref{Sec:varmethod} introduces the variability methods we used, and the challenges faced by an \emph{eROSITA} variability analysis. The detection of significantly variable sources throughout this data set is discussed in Section \ref{VarsampleeR123}. In Section \ref{MatchbeteR}, we describe the matching of variable sources across the three eRASSs. The identification of the counterparts to the X-ray sources is described in Section \ref{VarsampleMultWav}, which also discusses how variable sources were subsequently classified as either likely Galactic or likely extragalactic sources. In Section \ref{sec:propvarsample}, the different distributions of X-ray variability properties for Galactic and extragalactic sources are discussed. Section \ref{AGNPSDs} describes the periodograms of the brightest variable likely extragalactic sources. We highlight some of the most interesting variable sources we identified, in Section \ref{Intvarsrc}. Finally, we summarise our results in Section \ref{Conclusions}. 



\section{\emph{eROSITA} observations}\label{eRobs}

The extended ROentgen Survey with an Imaging Telescope Array \citep[\emph{eROSITA};][]{2021A&A...647A...1P} is a soft X-ray telescope on the Spectrum-Roentgen-Gamma \citep[\emph{SRG};][]{2021A&A...656A.132S} spacecraft, that is sensitive to the energy range $0.2-8.0 ~\mathrm{keV}$. 

The principle aim of the \emph{eROSITA} mission is to map out the entire X-ray sky at an unprecedented sensitivity. Following an initial calibration and performance verification phase after its launch to the $\mathrm{L}_2$ Lagrange Point in July 2019, it commenced its first of eight \emph{eROSITA} All Sky Surveys (eRASSs) in December 2019. For the purpose of achieving a high sensitivity to detecting X-ray sources across the sky, \emph{eROSITA} has a large field of view with a diameter of $1.03\degree$, and a large on-axis effective area of $\approx 1.5\times10^3 ~\mathrm{cm}^2$ \citep{2021A&A...647A...1P}.

During the eRASSs, \emph{SRG} completes one revolution about itself (which is called an eroday) every 4 hours. Combined with the large field of view, this means that sources can be observed for up to $41.2~\mathrm{s}$ per eroday. The direction of the angular momentum vector of \emph{SRG} shifts by an average of $10'$ per eroday along the ecliptic plane, to complete an eRASS in 6 months. 




The consequence of this selected scanning pattern is that X-ray sources lying close to the ecliptic equator are observed on approximately 6 consecutive erodays per eRASS. These observations take place within 24 hours, after which they will not be observed again by \emph{eROSITA} for the next 6 months. 


At greater altitude angles from the ecliptic, the advance of the field of view per revolution is reduced. So sources at a high altitude are observed on a greater number of erodays per eRASS. Of particular interest are the poles of the survey, which are the two points that \emph{eROSITA} scans through during every rotation. The eRASS survey poles approximately correspond to the ecliptic poles, but shift slightly over time. Due to \emph{eROSITA}'s large field of view, all sources within $\approx 0.5\degree$ can therefore, in theory, be observed every 4 hours for the duration of the eRASSs. 

A quick estimate of the average number of erodays ($N_{ed}$) that sources located at an angle $\theta$ (expressed in degrees) relative to either of the ecliptic poles can be observed for per eRASS, can be found from the equation: 



\begin{equation} \label{Nedtheory}
N_{ed} = \begin{cases}
  \frac{2160}{\pi}\sin^{-1}\left[\frac{\sin(1.03\degree)}{2\sin(\theta)}\right] & \textrm{if } 0.5\degree \leq \theta \leq 90\degree \\
  1080 & \textrm{if } \theta < 0.5\degree,
\end{cases}
\end{equation}

\noindent
This equation assumes that the survey poles do not vary over the course of the survey, and that each survey consists of 1080 erodays. It is derived by finding when the length of the chord to the circle of constant angle $\theta$ to a survey pole on the unit sphere is equal to the size of the \emph{eROSITA} field of view, $\sin(1.03\degree)$. The number of erodays are then found by determining how many multiples of the average $10'$ advance of the field of view per eroday are required to obtain the opening angle of the chord. The number of erodays of observations per eRASS no longer increases for $\theta < 0.5$, as all sources in this region are observed during every eroday.

The non-uniform advance of the scanning axis per eroday means that this number is merely an average for a particular angle to either of the two ecliptic poles. The number of erodays of observation can still vary a lot between different sources at the same angle to an ecliptic pole. Other instrumental issues can also affect this number, which will be described in more detail in Appendix \ref{SecNeD}. 

The solid angle of the region that lies within an angle of $\theta$ relative to either of the two survey poles, is defined by:

\begin{equation} \label{Omegangle}
\Omega = 2\pi \left(1-\cos\theta\right).
\end{equation}

\noindent
For small angles, the solid angle contained between $\theta$ and $\theta+d\theta$ is approximately: $d\Omega \approx 2\pi\theta d\theta$. So the number of sources detectable above a fixed lower flux limit decreases linearly with decreasing $\theta$, assuming an isotropic distribution.  








The consequence of Eqs. \ref{Nedtheory} and \ref{Omegangle} is that half of the sky is only be observed on 7 erodays per eRASS, or less. Three-quarters of the sky is only be observed on 9 erodays per eRASS, or less. In contrast, the field within about $0.5\degree$ of the poles is expected to be observed on up to 1080 erodays per eRASS. However, this region is small and only contains $\approx 300$ \emph{eROSITA} detected X-ray sources (see Appendix \ref{Prop3dSEP}). 

The region close to the ecliptic poles is particularly interesting for studying long-term variability. For our variability analysis, we chose to expand the selection of sources to the region within $3\degree$ of the SEP. This selection provides a balance between increasing the number of sources investigated, while ensuring that each source is still observed sufficiently often to enable a long-term variability study. This region contains almost all sources observed with more than 100 erodays per eRASS. From now on, we will refer to this region as the SEP field. 

The German \emph{eROSITA} consortium holds data rights to $|l|>180\degree$, which includes the SEP, the focus of this work. It does not include the NEP, which we do not consider further.

Using Eq. \ref{Nedtheory}, we find that the sources in the SEP field should be observed on 115 to 1080 erodays per eRASS, assuming a constant scanning axis angular velocity, and no instrumental downtime. The SEP field only makes up 0.137\% of the solid angle of a hemisphere, but \emph{eROSITA} spends 3.33\% of its time observing it. 


In this paper, we investigated \emph{eROSITA} observations of the eRASS:3 (the colon indicates that this includes all eRASSs up to that number) data set. These observations spanned from December 13, 2019, when eRASS1 started, until June 16, 2021, when eRASS3 was completed. The light curves we analysed are up to 552 days long.

\section{Source detection and light curve extraction}\label{SourcedetLCe}

The data were processed using the \emph{eROSITA} Science Analysis Software System \citep[eSASS;][]{2022A&A...661A...1B} version 946. Within eSASS, the sky observed by \emph{eROSITA} is subdivided into 4700 rectangular sky tiles, each of which has a size $3.6\degree \times 3.6\degree$, and slightly overlaps with the adjacent sky tiles on all sides \citep{2021A&A...647A...1P}. The SEP field is completely covered by the seven \emph{eROSITA} sky tiles: 082159, 085153, 087156, 090159, 092153, 093156, and 098159, with some overlap. 

X-ray sources were detected in each sky tile using the three-band detection procedure described in \citet{2022A&A...661A...1B}. This includes both point-like and extended sources found with a detection likelihood ($\texttt{DET\_LIKE}$) of greater than 6. Sources were labeled as extended, if they had an extension likelihood ($\texttt{EXT\_LIKE}$) larger than 14 \citep{2022A&A...661A..27L}. All other sources were considered to be point-like. The eSASS version and the detected catalogue used in this work are preliminary. They will differ slightly from the final release of eRASS catalogues that are still in preparation (Merloni et al. in prep.). However, these minor differences are expected to mainly affect the faintest sources near the detection limit, which are of little interest to this work. For the faintest sources to be detected as variable, they must feature a large degree of variability. 

In this preliminary source detection, two sky tiles have small corner regions at the boundary of the SEP field masked out in eRASS2 and 3 because of the large exposure gradients. Sources in these regions were omitted in this work. 



We merged the catalogues detected in the seven sky tiles into one. The sources located in the overlapping regions of the sky tiles are contained in both catalogues. To avoid analysing the same sources multiple times, we identified the matching sources in the overlapping sky tiles within $15 ''$ and removed the duplicates. The choice of this separation is based on the angular resolution of \emph{eROSITA} \citep{2021A&A...647A...1P}. We used independent catalogues for the three eRASSs we investigated, and initially performed independent variability analyses on each eRASS data set individually. We describe the merging of the variable source catalogues in more detail in Section  \ref{MatchbeteR}. A combined eRASS:3 data set was not yet available at the time of writing. 

In each eRASS scan, we extracted source light curves using the eSASS task \texttt{srctool}. We used a bin size of $40~\mathrm{s}$, and the full energy band of $0.2-5.0~\mathrm{keV}$. Energies above $5~\mathrm{keV}$ are less useful for variability study because of the low effective area and strong background \citep{2021A&A...647A...1P}. In the light curve extraction, we adopted the \texttt{srctool} option \texttt{lctype="REGULAR-"}, which ensures regular spaced bins, and removes all empty bins. This option also reduces the output file sizes.

The light curves generated in this way sometimes contain two bins for the same eroday, as the exposure is split between two $40 ~ \mathrm{s}$ bins. To ensure that there is precisely one bin per eroday of observations, we merged such bins into one. 




A key parameter for investigating \emph{eROSITA} light curves is the fractional exposure, $\epsilon$, which denotes what fraction of the chosen bin duration is fully exposed as if it had been observed on-axis. The fractional exposure is calculated to be the product of the fractional time, $\epsilon_t$, and the fractional area, $\epsilon_a$. The fractional time denotes the fraction of the bin duration during which the source was contained inside the \emph{eROSITA} field of view. The fractional area corrects for vignetting effects at different off-axis angles, corresponding to the path of the source through the field of view. Both parameters can vary significantly from eroday to eroday, so the fractional exposure has a wide distribution of values.

In this analysis, we discarded all bins with a fractional exposure of less than 0.1, as they add little to no information, and can complicate the variability analysis. The vast majority of sources we observed have a low count rate in the range of $0.001 - 1.0 ~\mathrm{cts/s}$. Therefore, single eroday observations were not split further to look into $<40~\mathrm{s}$ variability, as there is not enough information available for that. In the following analysis, unless otherwise stated, we denote the number erodays of observation with $\epsilon > 0.1$ as the number of bins of the light curve, $N_{\mathrm b}$.

The standard calculation for the count rate is performed as follows. Within a $\Delta t=40~\mathrm{s}$ bin at time $t_i$ (corresponding to the i\textsuperscript{th} bin), $C(t_i)$ counts were measured in the source extraction region, and $B(t_i)$ counts were measured in the background extraction region, which covers an area $1/A(t_i)$ times as large as the source extraction region. The effective exposure time is obtained by multiplying the fractional exposure ($\epsilon (t_i)$) by $\Delta t$. All of these parameters can then be used to estimate the source count rate in this bin, $R_{\mathrm S}(t_i)$, with: 

\begin{equation}\label{countratedef}
R_{\mathrm S}(t_i) \approx \frac{C(t_i)-A(t_i)B(t_i)}{\epsilon(t_i) \Delta t}.
\end{equation}

\noindent
Due to varying fractional exposures, a simple measure of the difference in the source and background counts divided by the bin time is an inaccurate estimator of the source flux. Instead, this equation calculates the exposure corrected count rate, which is a reasonable estimator of the source flux. However, this equation only works well if the measured values for $C-AB$, are large, and if the background count rate is uniform and equivalent within both the source and background extraction regions. However, most sources in our data set have a relatively low count rate. Even bright sources have bins with small $\epsilon$, in which $C-AB$ is small. Therefore, the varying fractional exposure creates challenges for accurately estimating the source flux in each bin. This effect needs to be taken into account when identifying variable sources. 

Instead, we calculated exposure corrected count rates from the peak of its posterior distribution, calculated in a Bayesian framework from the Poisson distributions of the number of source and background counts measured. This approach yields the same values for the count rate as Eq. \ref{countratedef} for bins with a large number of source counts, but is more accurate in the Poisson regime. The uncertainties on the count rate are defined at the $1\sigma$ credible confidence interval \citep{1991ApJ...374..344K}. This method, and its application to \emph{eROSITA} is described in more detail by \citet{2022A&A...661A..18B}, and \citet{2014ApJ...790..106K}. For simplicity, we will refer to the exposure corrected count rate determined using this method as the count rate in the rest of this paper. 


Appendix \ref{Prop3dSEP} discusses the distribution of various properties of the observations of sources in the SEP field. It shows the distribution of the number of eRODays of observation per eRASS, the source and background count rates, exposure times, fractional exposures, background areas, and detection likelihoods. The properties of this data set differ from those of \citet{2022A&A...661A...5L}, as they focus on a different region of the sky, with a different observing pattern. Above $2.3~\mathrm{keV}$, the fractional area drops relative to lower bands, and the particle background increases. Nevertheless, we chose to study the variability within the entire energy band, rather than for a subset of it. 

\section{Variability methods for analysing \emph{eROSITA} light curves}\label{Sec:varmethod}
\subsection{Challenges for \emph{eROSITA} variability analysis}

A variability analysis of eRASS sources needs to take the properties of the survey into account. The varying fractional exposure within a light curve induces an extra degree of variability in the measured count rates. The lower the fractional exposure is, the more likely it is to measure very high or low count rates. As bins with low fractional exposure occur at the start and end of each period of observation, \emph{eROSITA} light curves often have a "U"-like shape, particularly for not rebinned light curves of low count rate sources. This is, however, merely a result of low count rate statistics, and these bins have correspondingly larger errors associated with them. 



Due to the short exposure times per bin, most sources have count rates described by Poisson statistics in at least some bins. This can be a challenge for variability methods that are defined for Gaussian statistics. Rebinning may help, but also risks losing information of the eroday to eroday variability. 

An additional challenge for eRASS variability analysis are the long, but consistent gaps between short bins. During the gaps between consecutive observations, variable sources can undergo significant changes, that are not detected. As a result of this, the count rate measured in each bin might only be weakly correlated with the count rates in the previous or subsequent bins. This increases the degree of variability measured close to the Nyquist frequency via the aliasing effect \citep{1989ASIC..262...27V, 2005PhRvE..71f6110K}.  See Bogensberger et al. 2024B for a more detailed description of the challenges faced by variability analysis of eRASS light curves.


\subsection{Methods}


Bogensberger et al. 2024B defined thresholds for identifying variable sources in the \emph{eROSITA} SEP data set, which exclude false positives at the $1, 2, $ and $3\sigma$ levels. These thresholds were determined as a function of both the average count rate and the number of bins in the light curve. They were computed for the two variability quantifiers; SCATT\_LO, and AMPL\_SIG, following the results of \citet{2022A&A...661A..18B}.

The Bayesian excess variance \footnote{\href{https://github.com/JohannesBuchner/bexvar}{https://github.com/JohannesBuchner/bexvar}} \citep[bexvar;][]{2022A&A...661A..18B} uses the measured source and background counts, the bin size, the fractional exposure, and the background ratio, to calculate the posterior probability distribution of the mean and standard deviation of the logarithmic count rate distribution. This methodology allows bexvar to remain accurate in the Poisson regime, and never estimate a negative degree of variability. 

SCATT\_LO is defined as the value at the lower 10\% quantile of the bexvar posterior distribution of the standard deviation of the count rate. As \citet{2022A&A...661A..18B} showed, this parameter can accurately distinguish between variable and non-variable sources, even for a wide range of different types of variability.   


The maximum amplitude variation significance \citep[AMPL\_SIG;][]{2016A&A...588A.103B} computes the significance of the difference between the maximum and minimum count rates measured within a light curve. Despite only considering two points in the light curve, it was found to be more sensitive to detecting flaring sources than SCATT\_LO \citep{2022A&A...661A..18B}. Bogensberger et al. 2024B modified AMPL\_SIG from the original definition in \citet{2016A&A...588A.103B}, to instead use the bins with the lowest upper bound, and the highest lower bound, for the application to \emph{eROSITA} data. This change reduces the dependence of AMPL\_SIG on the bins with the smallest fractional exposures, as these frequently are the ones in which the highest and lowest count rates are measured. 


The NEV \citep{1997ApJ...476...70N} is usually defined as: 

\begin{equation} \label{NEVeq}
    \mathrm{NEV} = \frac{\sigma_{\mathrm{obs}}^2 - \overline{\sigma_{\mathrm{err}}^2}}{\overline{R_{\mathrm S}}^2}. 
\end{equation}

\noindent
In this equation, $\sigma_{\mathrm{obs}}^2$ is the measured variance of the count rates, and $\overline{\sigma_{\mathrm{err}}^2}$ is the average of the squared measurement uncertainties of the count rates in the light curve. The difference of these two quantities determines how much of the measured variance is likely due to the variability of the source, rather than measurement uncertainty. The average count rate, $\overline{R_{\mathrm S}}$, normalises this excess variance, allowing the degree of variability to be compared between different sources or epochs, independent of the source flux.

The NEV is an estimator of the normalised intrinsic variance (NIV) of the source during the times of observation. In the high count rate limit, when measurement errors, background contamination, and Poisson noise can be disregarded, the NEV approaches the NIV. There are also systematic offsets, which occur when using Eq. \ref{NEVeq} to determine the variability of a source that has a light curve with only a few source counts per bin, and a variable fractional exposure. 



Bogensberger et al. 2024B introduced a new method for estimating the NIV of a light curve, by converting the bexvar estimate of the standard deviation of the logarithmic count rate distribution into an estimate of the NIV. They found that this estimate of the NIV, which they label as $\mathrm{NEV}_{\mathrm b}$, is more accurate than Eq. \ref{NEVeq}, especially for low count rate sources. The $\mathrm{NEV}_{\mathrm b}$ estimate is based on an assumed pink noise PSD, in which the power law index, $\alpha = 1$, but they found that it is also accurate for red noise ($\alpha =2$), and white noise ($\alpha=0$) PSDs.  


The long gaps between short bins of eRASS light curves induce a significant aliasing effect in their periodograms \citep{1989ASIC..262...27V, 2003MNRAS.345.1271V, 2005PhRvE..71f6110K}. This causes power law periodograms to flatten towards the Nyquist frequency. Bogensberger et al. 2024B also determined the excess noise in a periodogram that is caused by varying fractional exposures within a light curve. For estimating the shape of the PSD of a variable source, both the Poisson and varying fractional exposure noise need to be subtracted from the periodogram at all frequencies. We used the Stingray\footnote{\href{https://docs.stingray.science/}{https://docs.stingray.science/}} timing package \citep{2016ascl.soft08001H} to calculate periodograms of \emph{eROSITA} light curves. 

The NIV itself varies over time due to sampling effects \citep{2003MNRAS.345.1271V, 2013ApJ...771....9A}. Assuming a constant source PSD, the NIV varies around the stationary quantity $\mathrm{NIV}_{\infty}$. Finally, the band-limited power is the integral of the source PSD in a selected frequency range, and is constant, as long as the PSD is stationary. The $\mathrm{NIV}_{\infty}$ is offset from the band-limited power due to aliasing, and the red noise leak \citep{1978ComAp...7..103P}. Bogensberger et al. 2024B described the sampling error of estimating the band-limited power or the $\mathrm{NIV}_{\infty}$ of sources exhibiting pink noise variability by using a measurement of the $\mathrm{NEV}_{\mathrm b}$ or NEV. This is caused by the intrinsic scatter in the NIV.

\section{Variable sample within $3\degree$ of the SEP, in eRASS1, 2, and 3}\label{VarsampleeR123}

We sought to apply the variability methodology developed by Bogensberger et al. 2024B, to study the variability properties of X-ray sources in the SEP field, as observed by \emph{eROSITA} in its first three eRASSs. As a first step, we intended to identify likely variable sources. 



The SCATT\_LO and AMPL\_SIG values were computed for all sources in the SEP fields in eRASS1, 2, and 3. Utilising the variability thresholds determined by Bogensberger et al. 2024B, we classified the significance of the variability into the four variability classes: $< 1 \sigma$, $\geq 1\sigma$, $\geq 2\sigma$, and $\geq 3\sigma$. These thresholds depend on the count rate and the number of bins of the light curve, so they are different for every source. The thresholds determined by Bogensberger et al. 2024B were found through simulations for specific values on a logarithmic grid of the number of bins, $\{50, 135, 370, 1000\}$,  and the average source count rates, $\{0.001, 0.003, 0.01, 0.03, 0.1, 0.3, 1.0, 3.0, 10, 30\}$ $\mathrm{cts/s}$. We estimated the value of the threshold for each individual source by interpolating between the values at the neighbouring grid points. For this purpose, we assumed the threshold between each group of four grid points could be described by the surface $T=(a_1+a_2N_{\mathrm b})(b_1+b_2\overline{R_{\mathrm S}})$. In this equation, $T$ represents the threshold (for either SCATT\_LO or AMPL\_SIG), and $a_1, a_2, b_1, b_2$ are defined by its value at the four corners. This assumed linear relation may however over-, or underestimate the value of the thresholds.

In our analysis, the \emph{eROSITA} observations were split into different eRASSs for the source detection and light curve generation. Due to slight boresight inaccuracies in the preliminary data of each eRASS, the three catalogues have positional offsets of a few arcseconds. This could lead to sources being wrongly matched across the three eRASSs. Wrongly matched sources are much more likely to be identified as variable. To avoid having a significant fraction of the variable sample consist of wrongly matched sources, we analysed each eRASS data set individually. However, this also means that this analysis is less sensitive to long-term variability.

A small number of sources had count rates or number of bins slightly outside of the interval for which the thresholds were determined from simulations. For those sources, we assumed that the thresholds remained at the value of the closest point on the grid.

\begin{figure}[h]
\resizebox{\hsize}{!}{\includegraphics{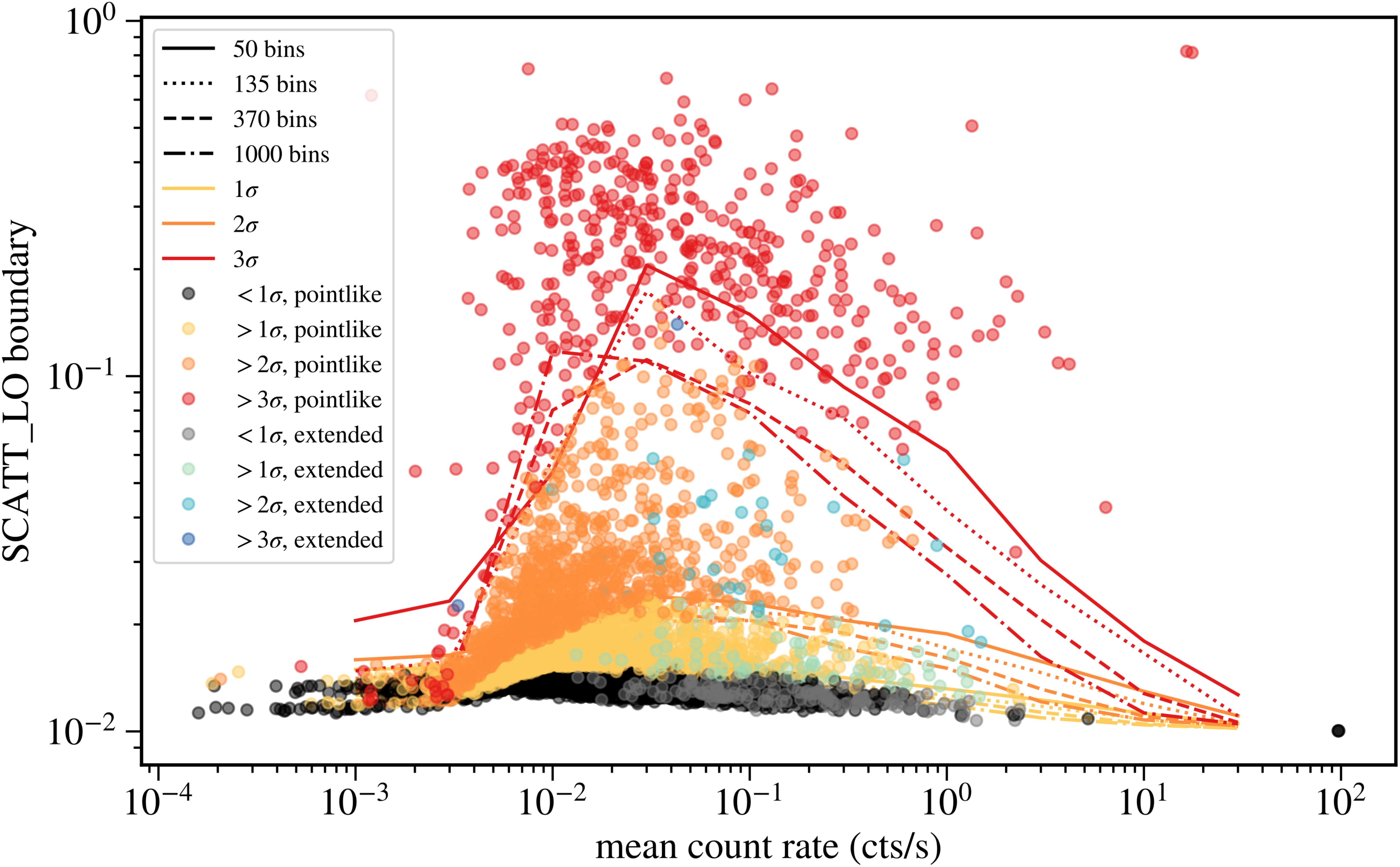}}
\caption{The measured SCATT\_LO parameter of the light curves of all sources detected in eRASS1, 2 and 3, relative to the SCATT\_LO variability thresholds. The colours denote what variability class each source was assigned. This figure also distinguishes between point-like and extended sources.
 \label{BV3DeR1}}
\end{figure}

\begin{figure}[h]
\resizebox{\hsize}{!}{\includegraphics{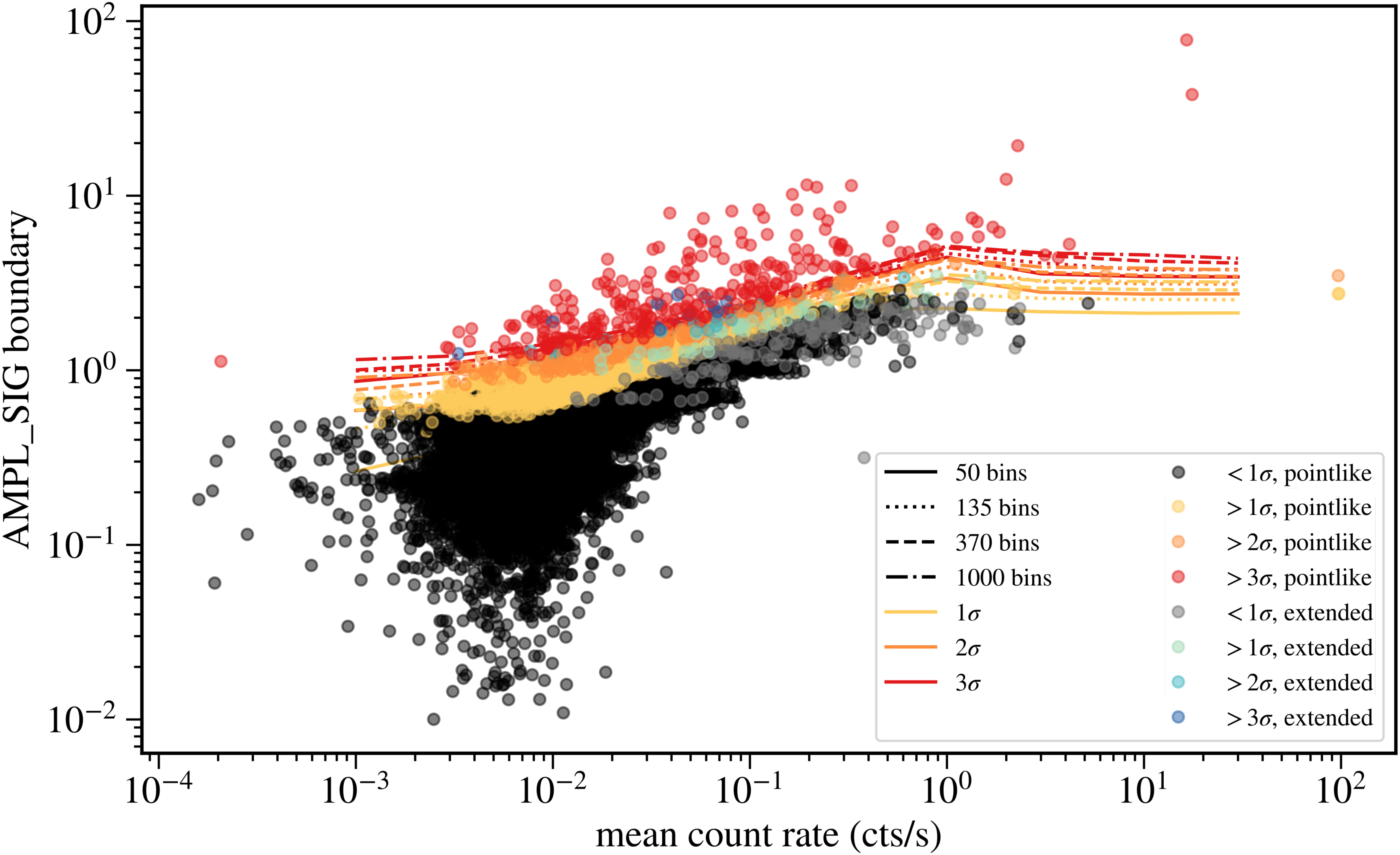}}
\caption{The measured AMPL\_SIG parameter of the light curves of all sources observed in eRASS1, 2, and 3. The colours denote the variability class we assigned each source, based on the thresholds of AMPL\_SIG. This figure also distinguishes between point-like and extended sources.
 \label{MAD3DeR1}}
\end{figure}

Figs. \ref{BV3DeR1} and \ref{MAD3DeR1} depict the distribution of the SCATT\_LO and AMPL\_SIG parameters for sources in eRASS1, 2, and 3, relative to the 1, 2, and 3$\sigma$ variability thresholds. On this logarithmic scale, sources that only vary by a negligible amount have very similar values of SCATT\_LO, but a wide range of values of AMPL\_SIG. Some sources were also observed to have negative AMPL\_SIG values, which are not shown in Fig. \ref{MAD3DeR1}. The thresholds in SCATT\_LO cover a large fraction of the range of values observed for this parameter. In contrast, the difference between the $1$ and $3\sigma$ thresholds in AMPL\_SIG is only a small fraction of the range of values observed for it. 


Many sources have a different AMPL\_SIG and SCATT\_LO variability significance. This is to be expected, as the two methods are sensitive to different types of variability, and have different sensitivities as a function of both the count rate, and the number of bins. In the following analysis, we assigned sources to the higher of the two variability classes. 


The use of both methods increased the total number of sources classified as variable. However, it also affected the false positive rate. For one method alone, the thresholds were defined to keep the likelihood of a false positive detection at 15.9\%, 2.28\%, and 0.135\%, for the 1, 2, and 3$\sigma$ levels, respectively. If SCATT\_LO and AMPL\_SIG were two independent variables, then the false positive rate would almost double, to 29.2\%, 4.50\%, and 0.270\%, for the 1, 2, and $3\sigma$ thresholds, respectively. If SCATT\_LO and AMPL\_SIG were to always yield the same variability significance for all sources, the false positive rate would be unaffected. In reality, the rate of false positives lies somewhere between those two extremes, but is likely closer to being unaffected. Most sources that AMPL\_SIG identified as lying above $3\sigma$ were also detected at this significance in SCATT\_LO.


Table \ref{Tabvarsrcprop} lists the number of sources in each of the three eRASSs that were found above the 1, 2, and $3\sigma$ thresholds for either the SCATT\_LO or the AMPL\_SIG parameter. Of the 8728, 7984, and 7770 sources detected in eRASS1, 2, and 3, we identified 4900 (56.1\%), 4588 (57.5\%), and 4512 (58.1\%) sources to lie below the $1\sigma$ thresholds for both variability quantifiers. These fractions are all noticeably less than the $70.8-84.1\%$ we expected to find, if all sources were intrinsically non-variable. This indicates that at least 19\% of all sources observed are inconsistent with being non-variable in the frequency range we investigated.  

\begin{table}[h]
\centering
\setlength{\tabcolsep}{4pt}
\def\arraystretch{1.1}
\begin{tabular}{l|lll}
    \textbf{Number of variable sources} & \textbf{eRASS1} & \textbf{eRASS2} & \textbf{eRASS3} \\ \hline \hline
    \textbf{Above $\boldsymbol{1\sigma}$} & 3828 & 3396 & 3258 \\
    Of which: \textit{only in SCATT\_LO} & \textit{2375} & \textit{2138} & \textit{2056} \\
    \hspace{1.5cm}\textit{only in AMPL\_SIG} & \textit{711} & \textit{596} & \textit{554} \\
    \hspace{1.5cm}\textit{variable in both} & \textit{742} & \textit{662} & \textit{648} \\ 
    \hspace{1.5cm}\textit{expected FP} & \textit{1385} & \textit{1267} & \textit{1233} \\
    \hspace{1.5cm}\textit{extended} & \textit{96} & \textit{77} & \textit{67} \\ \hline
    \textbf{Above $\boldsymbol{2\sigma}$} &  914 & 822 & 743 \\
    Of which: \textit{only in SCATT\_LO} & \textit{555} & \textit{518} & \textit{475} \\
    \hspace{1.5cm}\textit{only in AMPL\_SIG} & \textit{183} & \textit{153} & \textit{106} \\
    \hspace{1.5cm}\textit{variable in both} & \textit{176} & \textit{151} & \textit{162} \\ 
    \hspace{1.5cm}\textit{expected FP} & \textit{199} & \textit{182} & \textit{177} \\
    \hspace{1.5cm}\textit{extended} & \textit{33} & \textit{27} & \textit{14} \\ \hline
    \textbf{Above $\boldsymbol{3\sigma}$} & 226 & 176 & 188\\
    Of which: \textit{only in SCATT\_LO} & \textit{83} & \textit{67} & \textit{82} \\
    \hspace{1.5cm}\textit{only in AMPL\_SIG} & \textit{46} & \textit{37} & \textit{35} \\
    \hspace{1.5cm}\textit{variable in both} & \textit{97} & \textit{72} & \textit{71} \\ 
    \hspace{1.5cm}\textit{expected FP} & \textit{12} & \textit{11} & \textit{10} \\
    \hspace{1.5cm}\textit{extended} & \textit{6} & \textit{6} & \textit{3} \\ 

\end{tabular}
\caption{Table summarising the results of the variability detection analysis applied to the \emph{eROSITA} eRASS1, 2, and 3 observations of the SEP field. FP is short for false positives. It indicates the number of sources that can be expected to be found above the particular threshold, if all sources were intrinsically non-variable, for either method individually. All sources identified at each level, that are not characterised as being extended, are point-like.
\label{Tabvarsrcprop}}
\end{table}


The purity of the variable sample increased significantly with increasing variability significance. At least 37.1\% of all sources detected above $1\sigma$ are expected to be false positives. However, the fraction of false positives drops to an estimated 22.5\% above $2\sigma$, and down to 5.59\% above the $3\sigma$ thresholds. We decided to use the $3\sigma$ thresholds to identify variable sources for further analysis, and will henceforth refer to this as the set of variable sources.

Out of the 8728, 7982, and 7770 sources detected in eRASS1, 2, and 3, we found 184, 160, and 166 extended sources, respectively. These were not excluded from the data set, because they can act as an independent check of the variability detection technique we used. Extended sources should only feature significant variability, if they were wrongly classified as extended, or if they contain a very bright point source. For instance, a bright AGN in a cluster of galaxies could be identified as a variable extended source. Nevertheless, the fraction of extended sources detected above the variability thresholds should approximately correspond to the expected fraction of false positives.

The fraction of extended sources detected above either the AMPL\_SIG or SCATT\_LO 1, 2, and 3$\sigma$ thresholds is $47.1\%$, $14.5\%$, and $2.94\%$, respectively. This noticeably exceeds even the false positive rate calculated under the assumption of complete independence between SCATT\_LO and AMPL\_SIG. This excess of extended sources identified as variable is predominantly due to AMPL\_SIG, as it identified 15 extended sources to be variable. In contrast, SCATT\_LO only found 2 extended sources to show significant variability, and both of them were also classified as variable by AMPL\_SIG. This occurred even though SCATT\_LO detected 32\% more sources above its $3\sigma$ threshold, than AMPL\_SIG. 

We investigated these 15 variable extended sources, to understand the cause of their variability. Two of these sources were found to be extended in one eRASS data set, but point-like in the others, indicating that they were potentially wrongly identified as extended. Eleven sources were not detected at all in one or two eRASSs, which also suggests they could have been wrongly identified as extended. 

Nevertheless, we cannot demonstrate that all of these sources were wrongly identified as extended. Some of the apparent variability may instead have been caused by the extraction of the light curves, which assumed all sources to be point-like. For instance, several of the variable extended sources had consistent count rates throughout the observations, except for a single eroday in which a significantly greater number of source counts was observed, than was expected. This explains why these sources were found much more commonly by AMPL\_SIG than by SCATT\_LO. However, even when taking all of these effects into account, the likelihood that a non-variable source is falsely identified above the $3\sigma$ thresholds may still be larger than $0.1350\%$, especially for AMPL\_SIG. 



\begin{figure}[h]
\resizebox{\hsize}{!}{\includegraphics{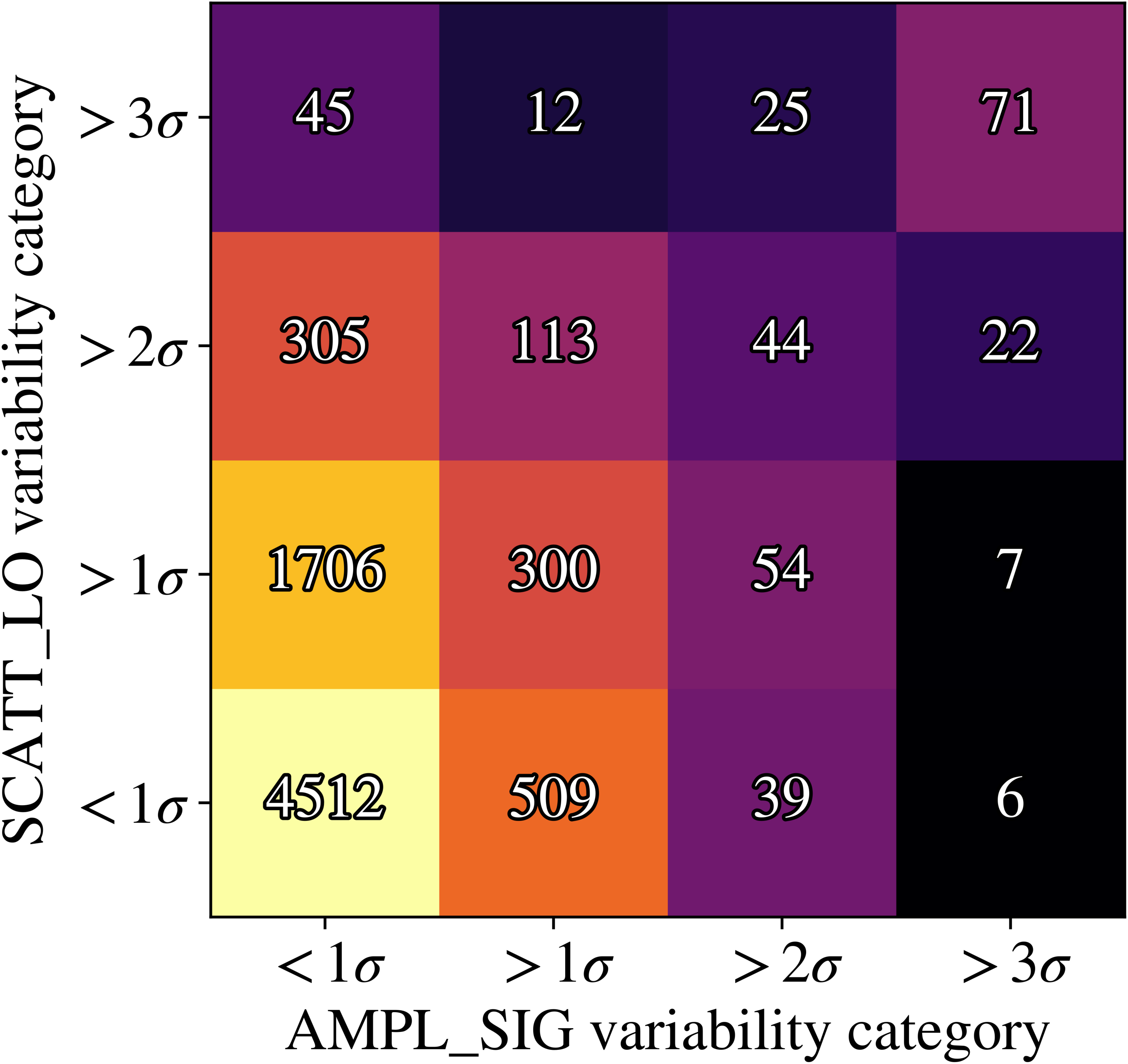}}
\caption{This figure depicts how many sources were identified in each of the four variability categories, for the two variability identifiers SCATT\_LO and AMPL\_SIG. We added the numbers of sources in each of the 16 categories from the three separate analyses of the three individual eRASSs. This means that many sources will appear up to three times in this figure, possibly in different regions. All four categories exclude sources at a higher variability significance. For example, sources that are shown to lie in the bin $>1\sigma$ do not include the sources observed with a variability significance of $>2\sigma$. \label{compBVMAD}}
\end{figure}

Fig. \ref{compBVMAD} compares the number of sources that were placed into each of the four variability categories for the two variability detection methods. Even though both SCATT\_LO and AMPL\_SIG have thresholds defined by excluding the same fraction of intrinsically non-variable sources, SCATT\_LO always identified more sources above a particular variability threshold. Of the 590 instances of a source being detected as variable in one of the three eRASSs using either the SCATT\_LO or the AMPL\_SIG methodology, 80.0\% were identified using SCATT\_LO only, compared to 60.7\% using only AMPL\_SIG. 40.7\% of all variable sources were identified above $3\sigma$ by both methods. Almost equally many were identified to be variable by SCATT\_LO but not AMPL\_SIG. It is worth noting that the second most common of the seven categories of sources identified as variable, as shown in Fig. \ref{compBVMAD}, amounting to 21.2\% of all variable sources, are ones that were detected above the $3\sigma$ threshold in SCATT\_LO, but below the $1\sigma$ threshold in AMPL\_SIG. 


Of these two methods, SCATT\_LO is more sensitive to identifying variable sources in \emph{eROSITA} observations. It also appears to be more reliable, as it identified 87\% fewer extended sources above 3$\sigma$ than AMPL\_SIG did. Nevertheless, AMPL\_SIG identified several significantly variable sources that SCATT\_LO did not find. Therefore, we recommend using both methods for variability detection. 



Figs. \ref{nbhist}, and \ref{TEFEhist} - \ref{DLhist} in Appendix \ref{Prop3dSEP} show the distribution of various observation parameters for the variable sources identified here, alongside the total distribution of all sources in the SEP field. The distributions of most parameters for these two classes of sources are mostly similar. They also show that the methods of detecting variability used here are sensitive to variability throughout the parameter space spanned by the sources in our sample. 

There are slight differences between the distributions of the number of bins (Fig. \ref{nbhist}), and the effective exposure time (Fig. \ref{TEFEhist}), for variable, and non-variable sources. This is because there is a higher likelihood of a source being identified as variable if it is observed for a longer time. There are more significant differences in the distribution of the average source count rates (Fig. \ref{arhist}) and the total source counts (Fig. \ref{cthist}), between variable and non-variable sources. The likelihood of a source being identified as variable increases significantly with an increasing count rate, and both parameters span a larger range of values.

\section{Matching variable sources between eRASSs}\label{MatchbeteR}

After having identified variable sources in each of the three eRASSs individually, we next sought to investigate how many of them were also found to be variable in the other two eRASSs. This allows us to create a list of unique sources that featured significant variability in at least one of the three eRASSs, and investigate their variability over a span of 1.5 years.

We matched the variable sources identified in each of the three eRASSs to the entire catalogue of sources in the other two eRASSs. We used a boresight-corrected eRASS:2 SEP catalogue (T. Liu, priv. comm.), to estimate the average boresight correction needed to determine slightly more accurate X-ray positions for the sources in each of the c946 catalogues of individual eRASSs. We did this by determining the distribution of the shift in RA and Dec between the closest matching sources in the catalogues eRASSi (with i being 1, 2, or 3), and eRASS:2. We subtracted the median shift in RA and Dec between the two catalogues from the eRASSi catalogue positions. Next, this procedure was repeated using the updated positions of sources in the eRASSi catalogues. The previous steps were repeated at least two more times, until the change in the median shift relative to the previous iteration was 0. We performed this estimate individually for each sky tile. In this way, $93\%$ of all sources detected in the SEP field in eRASS1 or 2, could be matched to a source in the boresight-corrected eRASS:2 catalogue within $30''$. Even $87\%$ of eRASS3 sources were matched to the eRASS:2 catalogue within $30''$.


Using these boresight-corrected positions, we matched the sources in each of the three eRASSs to the boresight-corrected eRASS:2 SEP catalogue. The requirements for this matching were that each source in the eRASS:2 catalogue could only be matched to one source in each of the individual eRASS catalogues, and vice versa. Additionally, source $A_i$ in catalogue A, and source $B_j$ in catalogue B were only matched, if $A_i$ was the closest source to source $B_j$ in catalogue A, and if source $B_j$ was the closest source to source $A_i$ in catalogue B. We also set an upper limit on their separation of at most $30''$. This separation was chosen to be about twice as large as the angular resolution of \emph{eROSITA} \citep[see][]{2021A&A...647A...1P}. Sources in different eRASSs, that were matched to the same eRASS:2 source using this procedure were subsequently considered to be the same source, and their light curves and the information about them were combined. 


Two sources separated by merely $38.6''$ were found to show significant, and nearly identical variability in all three eRASSs. This was caused by the overlapping of their point spread functions, and likely originated from just one of them. Therefore, we identified the source exhibiting a larger degree of variability, and discarded the other source from the list of variable sources. This reduced the total number of instances of a source being identified as variable within a single eRASS, from 590 (as shown in Table \ref{Tabvarsrcprop}), to 587. 

By matching all sources with the eRASS:2 catalogue, the 587 instances of a source exhibiting significant variability within one of the three eRASSs, were identified to be 453 unique sources. Of those, 39 sources were independently identified to be variable in all three eRASSs. A further 56 sources were found to be variable in two of the three eRASSs. However, the remaining 358 sources, the vast majority of sources identified as variable, were only found to be variable in one of the three eRASSs. Of these, 148 sources were found to only be variable in eRASS1, 98 in eRASS2, and 112 in eRASS3. 


There are several reasons why sources might only have been detected to have significant variability in one eRASS. In many instances, the variable sources showed a smaller degree of variability in other eRASSs, and were subsequently not found above the $3\sigma$ thresholds. Some flaring sources only exhibited flares in one of the three eRASSs, and maintained a near-constant luminosity during the others. There were also instances of sources becoming so dim that they were not even detected in other eRASSs. Finally, there might also be some issues with the matching of sources. Some variable sources could have been wrongly matched to a different source in a different eRASS. 


By visually inspecting the light curves of matched sources, we identified four instances in which a variable source was almost certainly matched to a wrong source in a different eRASS. This check is only possible for sources that were observed during the transition from one eRASS to the next. In these four instances, we detected a sudden sharp rise or fall in the count rate during the transition from one eRASS, in a manner inconsistent with the variability observed at all other times. These effects were almost certainly caused by an incorrect matching of sources. Therefore, we disassociated the matched sources in these four instances. This has already been incorporated in the previously stated number of matched variable sources. 

It is likely that other variable sources were also wrongly matched between eRASSs. Several sources feature significantly different average count rates in different eRASSs. However, it is to be expected that variable sources could have significantly different average count rates on either side of a gap spanning several months. Therefore, unless the transition from one eRASS to the next is observed, a change in the average count rate could also have been caused by the variable nature of the source, rather than an incorrect matching.







\section{Multiwavelength source classification of the variable sample}\label{VarsampleMultWav}

\subsection{Methods}\label{classmethodsec}

After having identified 453 unique sources exhibiting significant variability, we next sought to investigate what these sources are, to identify the mechanism causing the observed variability.
For this purpose, we determined the most likely counterparts of the variable sources in catalogues at other energy bands. We subsequently used spectral, photometric, and parallax information to distinguish between Galactic and extragalactic sources.


We first used NWAY\footnote{\href{https://github.com/JohannesBuchner/nway}{https://github.com/JohannesBuchner/nway}} \citep{2018MNRAS.473.4937S} to find the most likely counterparts to the X-ray source positions of the eRASS:2 boresight-corrected catalogue of the SEP field (described in Section \ref{MatchbeteR}), in the CatWISE2020 \citep{2021ApJS..253....8M} catalogue. An eRASS:3 boresight-corrected catalogue was not available at the time of writing. We used NWAY to select only the most likely counterpart to each eRASS:2 X-ray source. The parameter \texttt{match\_flag} indicates whether multiple matching solutions with comparable likelihoods are found. We only account for primary solutions by requiring \texttt{match\_flag}$=1$. Therefore, for the following analysis, we only selected sources with $\texttt{match\_flag}=1$. NWAY is a robust Bayesian tool for counterpart identification across different wavelength bands. The CatWISE2020 positions of matched \emph{eROSITA} variable sources in the SEP field were subsequently also cross-matched with the NSC data release 2 \citep{2021AJ....161..192N}, the VHS data release 5 \citep{2021yCat.2367....0M}, and the GAIA early data release 3 \citep{2021A&A...649A...1G} catalogues, all to within $2''$. 


\citet{2022A&A...661A...3S} introduced two methods for classifying \emph{eROSITA} sources. In particular, they distinguished between Galactic and extragalactic sources. The first relation they found utilises the IR band W1 magnitude, at a wavelength of $3.4~\mu\mathrm{m}$, and the optical g, r, and z band magnitudes. They identified that most sources with $z-W1-0.8(g-r)+1.2 > 0$ are extragalactic. To use this relationship, we converted the W1 magnitudes obtained from the CatWISE2020 counterparts into AB magnitudes. The second method introduced by \citet{2022A&A...661A...3S} uses the X-ray flux in the $0.5-2~\mathrm{keV}$ interval ($F_{0.5-2}$, expressed in units of $\mathrm{ergs}~\mathrm{cm}^{-2}~\mathrm{s}^{-1}$), and the W1 magnitude. Most sources detected with $W1+1.625\log(F_{0.5-2})+6.101 > 0$ are extragalactic, whereas most sources at negative values are Galactic. We used both of these linear relations whenever all the required magnitudes in the various energy band were known.



We also classified sources utilising the method described by \citet{2015MNRAS.448.1305K}. They found that almost all sources whose W1 and J band magnitudes satisfy the inequality $W1-J<-1.7$ are extragalactic. However, this boundary might be too strict, as many extragalactic sources are excluded from this selection. To use this method, we converted the J magnitudes obtained from VHS to match those used by 2MASS. 

A third way to distinguish between Galactic and extragalactic sources is to use the measured parallax significance from GAIA. For $p/\sigma_p>5$ (where $p$ is the parallax, and $\sigma_p$ is its uncertainty), there is a high degree of confidence that the source is Galactic \citep{2018A&A...616A...9L}. Sources with a less significant parallax can still be either Galactic or extragalactic, and need to be investigated further using the other methods we described. We also investigated the possibility of using the CatWISE2020 proper motions for distinguishing between likely Galactic and extragalactic sources. However, we found that those were not ideal for this task, as we found a number of extragalactic sources with measured proper motions several times larger than the associated errors. There was also an insufficiently large difference in the distribution of the proper motion significance between Galactic and extragalactic sources, to use it to distinguish between them. 


\begin{figure*}[h]
\resizebox{\hsize}{!}{\includegraphics{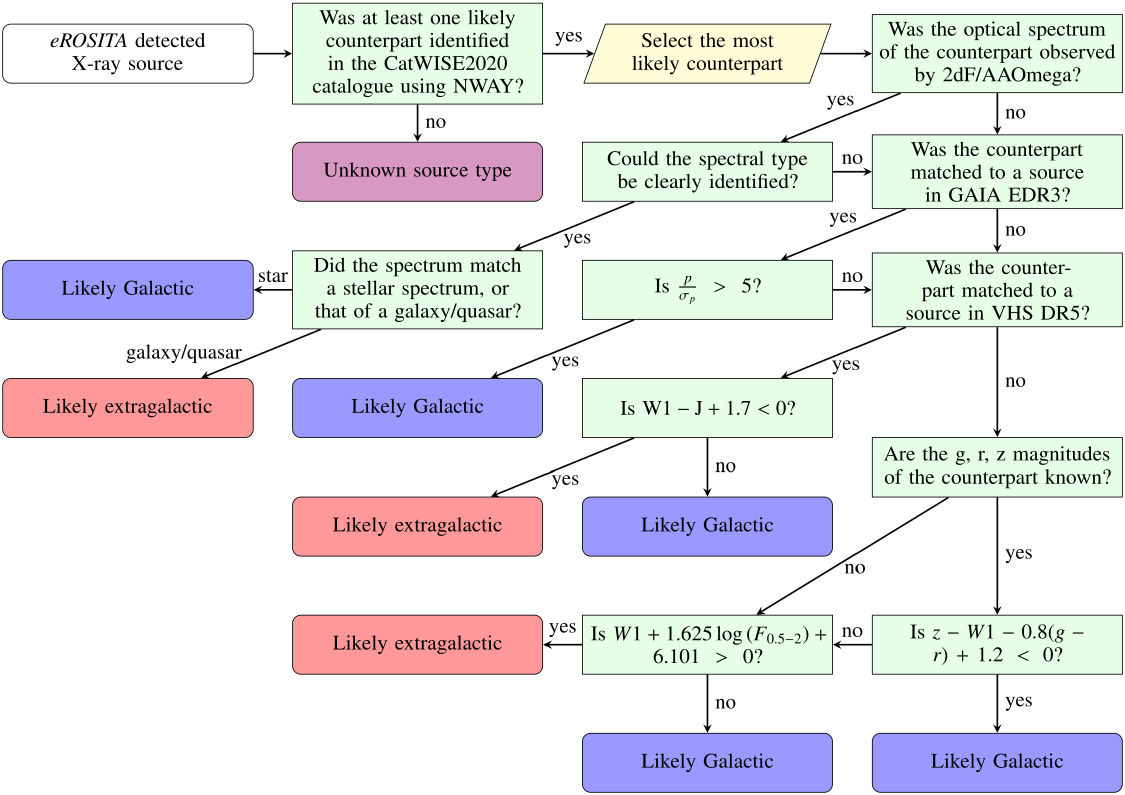}}
\caption{Flowchart showcasing the methodology used to separate the selected variable X-ray sources into the three groups of likely Galactic, likely extragalactic sources, and sources of unknown type. We used the X-ray flux, and information that was collected on the most likely counterparts to the X-ray detections by CatWISE2020, GAIA, VHS, and the optical spectra observed by the \emph{AAT}. \label{FlowchartGalExtragal}}
\end{figure*}

We followed up the observations of a subset of \emph{eROSITA}-detected sources in the SEP by obtaining optical spectroscopy of them with the 2 degree Field fibre positioner \citep[2dF;][]{2002MNRAS.333..279L} and the AAOmega \citep{2004SPIE.5492..389S, 2006SPIE.6269E..0GS} dual beam spectrograph, on the Anglo-Australian Telescope \citep[\emph{AAT};][]{2012opsa.book...27H}. The observations with good visibility took place on February 8 - 11, September 5, November 6, 2021, and January 5, 2022. Using the four methods described above, we selected likely extragalactic sources, and the positions of their respective optical counterparts, for observation. This selection was carried out over the entire SEP field, for all sources, not just for the variable sample. However, we did assign a higher priority to observe variable sources. To maximise the quality of the optical spectra, and the number of sources for which we could obtain accurate spectra within these observations, we limited the r-band magnitudes of the optical counterparts we selected, to between 17.0 and 22.5. We targeted a total of 2644 likely extragalactic X-ray sources in these observations.

These observations are part of an ongoing survey of the optical spectroscopic properties of \emph{eROSITA} detected, likely extragalactic sources, near the SEP. A detailed description of the selection of fields and targets for the observations, the data analysis, the results, and a catalogue of the X-ray variability, and the optical spectroscopic properties of the observed sources in the SEP field will be described in future work. However, we will mention some results of the spectroscopic analysis relating to the variable sources in the following sections. The observed optical spectra can clearly distinguish between Galactic and extragalactic sources, if they are sufficiently well determined. 


We applied these five methods to separate the selected variable sources into the three groups of: likely Galactic, likely extragalactic, and unknown sources. Fig. \ref{FlowchartGalExtragal} depicts the flowchart that we used to categorise all sources into one of these three groups. For this procedure, we had to prioritise between these four methods, so that sources could be categorised even if the methods disagreed about the source type. We decided that the reliability of the Galactic-extragalactic distinction based on \emph{AAT} spectra with unambiguously fitted spectral lines was the greatest. Extragalactic sources should have a non-detectable parallax, so sources that GAIA measured to have $p/\sigma_p>5$ are very likely to be Galactic. Next we used the $W1-J<-1.7$ methodology to distinguish between different types of sources. 


Following \citep{2022A&A...661A...3S}, if the relevant information for all of the previous methods was not available, we classified sources with $z-W1-0.8(g-r)+1.2 < 0$ as likely Galactic. Sources with a positive value were instead classified using $W1+1.625\log(F_{0.5-2})+6.101$. If they had a positive value for this parameter, they were classified as likely extragalactic, otherwise they were described as likely Galactic.

We associate a Galactic-extragalactic distinction parameter, $D$, to all sources, to indicate which of the three source types it was identified as, in each of the four methods we describe here. To unambiguously associate each value of $D$ with exactly one combination of classifications by all methods, we define it as: 

\begin{align}
    \begin{split}
    D & = \sum_{j=0}^3 3^j d_j \\
    d_j & = \begin{cases}
        1 & \textrm{if likely extragalactic}  \\
        0 & \textrm{if unknown type} \\
        -1 & \textrm{if likely Galactic},
    \end{cases}
    \end{split}
\end{align}

\noindent
In this equation, $j=0$ refers to the $W1+1.625\log(F_{0.5-2})+6.101 > 0$, and $z-W1-0.8(g-r)+1.2 > 0$ methodologies. $j=1$ corresponds to the $W1-J<-1.7$ distinction. $j=2$ describes the $p/\sigma_p>5$ method, and $j=3$ corresponds to the results of the \emph{AAT} spectral analysis. In evaluating this parameter, we label all sources with $p/\sigma_p\leq5$ as `likely extragalactic sources', even though many of them are likely to be Galactic sources. 


In general, sources with positive values of the distinction parameter were identified as likely extragalactic sources, sources with negative values as likely Galactic sources, and sources with 0 as sources of unknown type. However, because sources with $p/\sigma_p\leq5$ could be both Galactic and extragalactic sources, all sources with $5\leq D\leq8$ were still identified as likely Galactic sources. 

\subsection{Identification and classification of variable sources}

There are several issues to be aware of regarding the matching of eRASS3 sources. Variable sources that lie below the detection limit in eRASS1 and 2, but became significantly brighter in eRASS3 will not be contained in the eRASS:2 catalogue. Those sources will either be matched to wrong counterparts or be labelled as unknown sources. The estimate of the boresight correction that we used to match eRASS3 sources to eRASS:2 might be insufficiently accurate for matching unknown sources to catalogues in other parts of the electromagnetic spectrum. 


Out of the 453 individual variable sources we identified in Sections \ref{VarsampleeR123} and \ref{MatchbeteR}, 403 were matched to a source in the CatWISE2020 catalogue. Therefore, following Fig. \ref{FlowchartGalExtragal}, we classified the 50 remaining variable sources to be of unknown type. Out of those, 23 were only detected in eRASS3. 


During the \emph{AAT} observing campaign, we obtained the optical spectra of 120 variable \emph{eROSITA} sources with a match in the CatWISE2020 catalogue. Of those, 81 spectra were identified to be sufficiently well observed that an unambiguous determination of their redshift, and source classification could be performed. The GAIA parallax significance was determined for 292 sources. A total of 79 sources were matched to the VHS DR5 catalogue. These are the sources for which we could determine a Galactic-extragalactic distinction using $W1-J<-1.7$. The $z-W1-0.8(g-r)+1.2 > 0$ method was used to classify 224 variable sources, for which all the magnitudes were known. Finally, all 403 of the matched variable sources could be classified using the $W1+1.625\log(F_{0.5-2})+6.101 > 0$ method. Using all data available on them, and the flowchart of Fig. \ref{FlowchartGalExtragal}, we first classified the sources with a counterpart in the CatWISE2020 catalogue into 164 likely extragalactic, 239 likely Galactic, and 50 unknown sources. 



In addition to the classification methods described in Section \ref{classmethodsec}, we analysed each source individually, searching through various source catalogues at the detected position of the X-ray source and its matched infrared counterpart, to check the classification. For instance, we found a variable source that was matched to an AGN but was wrongly identified as a likely Galactic source, because it had $p/\sigma_p = 5.14$. This might have been caused by an underestimation of the error in the parallax. We determined the mean value of $p/\sigma_p$ for the extragalactic sources in our sample, and found it to be 1.08. This indicates that the errors might be underestimated for some of these sources, as it should have a value of $\sqrt{2/\pi}\approx 0.798$ for a normal distribution with well-defined errors \citep{10.1093/biomet/27.3-4.310}. We also incorrectly classified the bright star $\mathrm{\eta^2}$ Dor as a source of unknown type. It is so bright that CatWISE2020 masked it out, so no counterpart to the X-ray detection could be identified. Additionally, seven sources were identified to be XRBs in the Large Magellanic Cloud (LMC). A few more updates to the classification of variable sources, similar to these ones were performed. The 453 variable sources were finally classified into 168 likely extragalactic sources, 235 likely Galactic sources, 7 XRBs, and 43 sources of unknown type. All of the XRBs are located in the LMC. Many of the Galactic sources correspond to stars in the LMC. 


Nevertheless, some of the variable sources might still be wrongly classified. Out of the 81 variable sources for which we had obtained unambiguous optical spectra with the \emph{AAT}, 14 sources (17.3\%) were identified to have a different source class than what was found when using the other methods. 





\begin{figure}[h]
\resizebox{\hsize}{!}{\includegraphics{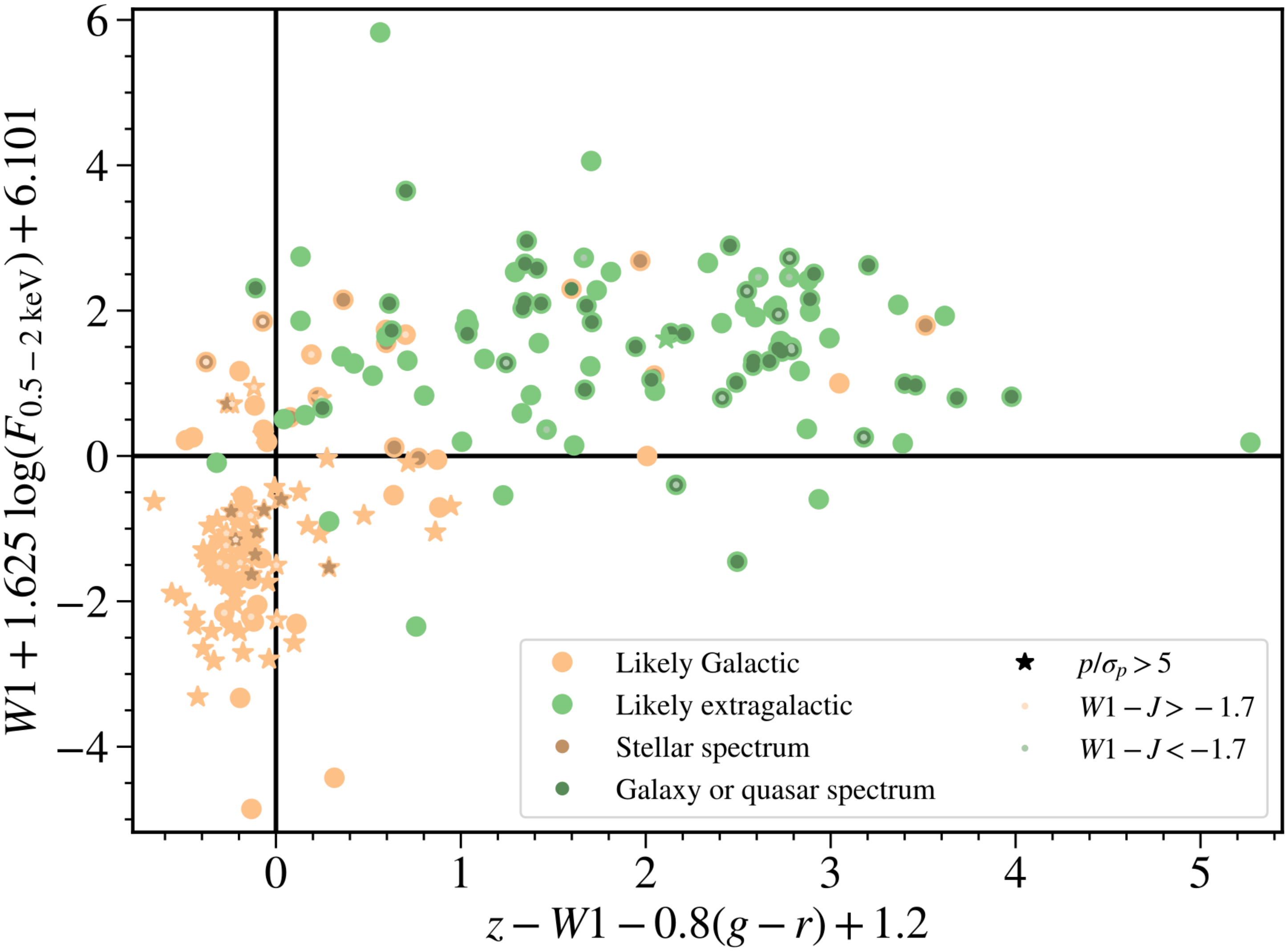}}
\caption{The classification of variable sources, according to the four methods outlined in Section \ref{classmethodsec}, and the flowchart of Fig. \ref{FlowchartGalExtragal}. Sources that were not matched to an optical counterpart, or whose counterparts lacked information on the $W1$, $g$, $r$, or $z$ magnitudes, are not shown. The largest circles indicate the final classification of sources. Smaller circles and stars indicate the results of different methods. 
 \label{SrcClass}}
\end{figure}

Fig. \ref{SrcClass} displays the classification of all variable sources with counterparts and known $W1$, $g$, $r$, and $z$ magnitudes. It shows that all four methods are mostly reliable and consistent with each other, but that there are still some disagreements between them. One source was observed with an extragalactic spectrum, but was finally classified as a Galactic source, as the galaxy is located close to a nearby star, which is the cause of the observed X-ray variability.


\section{Properties of the variable sample}\label{sec:propvarsample} 

For many sources located between $0.5-3\degree$ of the SEP, the set of observations performed within each eRASS consists of two parts; one at the start, and one at the end of the eRASS, with a long break in between. Even the light curves of sources located very close to the SEP feature gaps, when the survey mode operation was interrupted for a few days for calibration observations, orbit corrections, or telescope downtime. For all of these sources, it is more sensible to split the light curves into the segments within which observations occurred every eroday, rather than to split them by eRASS. This is particularly relevant for computing periodograms, to ensure a consistent sampling for as long as possible. It is also relevant for computing $\overline{\mathrm{NEV}_{\mathrm b}}$ values, if these are to accurately describe the degree of variability within a particular frequency interval, especially if their values are compared between different sources. To remain consistent throughout, we performed the following variability analysis on individual segments of the eRASS:3 light curves, regardless of how these are located relative to the start and end times of each eRASS.

To compare the variability strength of different sources over the same frequency range, we split the segments into smaller parts of 20, 50, and 100 consecutive bins. The $\mathrm{NEV}_{\mathrm b}$ was determined for each of these smaller segments individually. These were subsequently used to calculate a single geometric mean $\mathrm{NEV}_{\mathrm b}$ for the entire observed eRASS:3 light curve of each variable source. 



We associated sampling errors to the $\overline{\mathrm{NEV}_{\mathrm b}}$ measurements in accordance with the findings of Bogensberger et al. 2024B. As many of the smaller segments are adjacent, we used the sampling errors of adjoined segments rather than those for randomly spaced segments. However, there are long gaps between segments, so the sampling errors might be overestimated.  





\begin{figure}[pt]
\resizebox{\hsize}{!}{\includegraphics{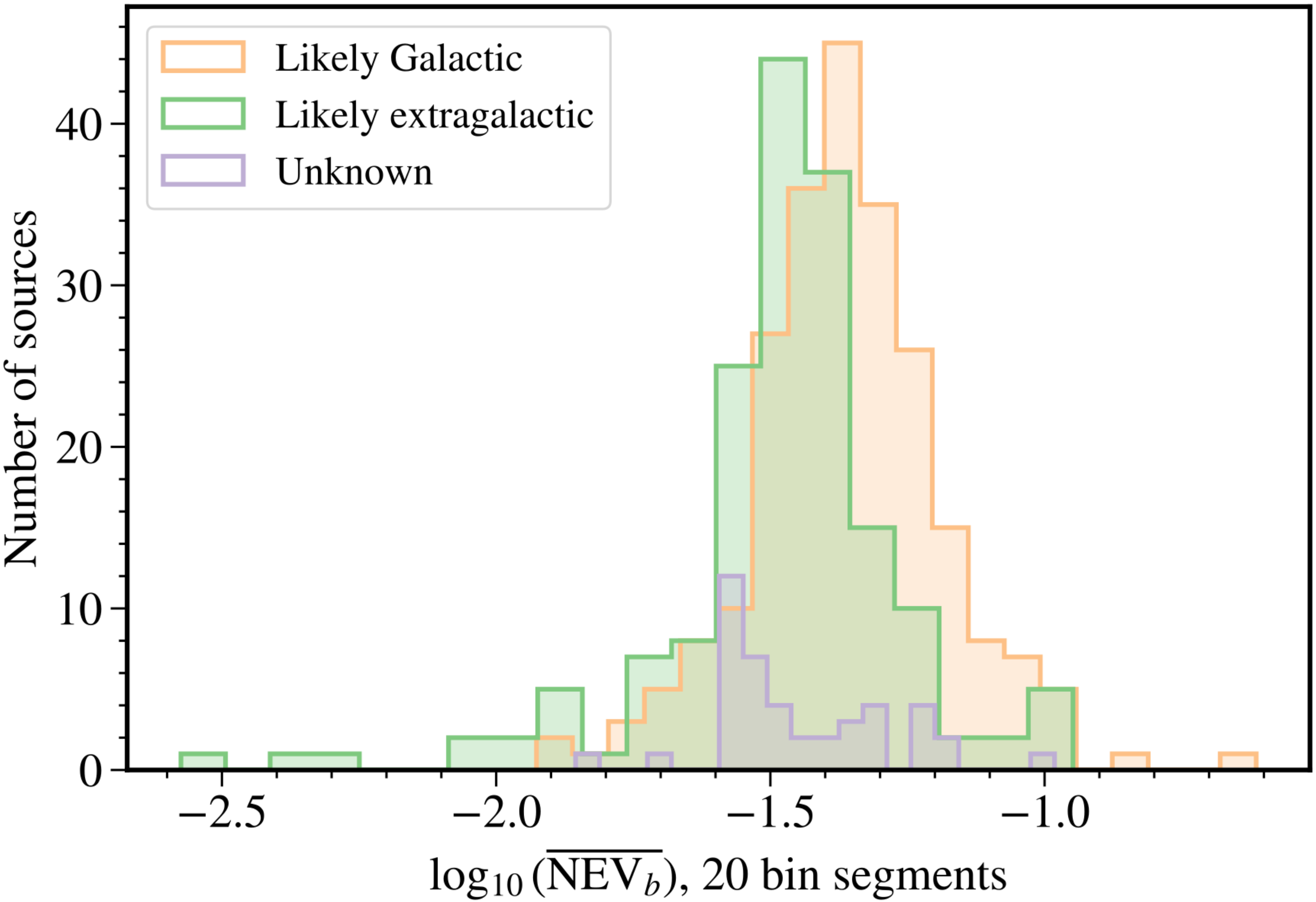}}
\caption{Distribution of the geometric mean $\mathrm{NEV}_{\mathrm b}$ of 20 bin segments of the variable sources identified in the SEP field, distinguished into likely Galactic, likely extragalactic and unknown sources. 
 \label{NEV20dist}}
\end{figure}

Fig. \ref{NEV20dist} depicts the distribution of the geometric mean $\mathrm{NEV}_{\mathrm b}$ over all segments of 20 consecutive and consistently spaced bins in the combined eRASS1, 2, and 3 light curves of all variable sources. This figure distinguishes between the distributions of this parameter for likely Galactic, likely extragalactic, and unknown sources. The $\mathrm{NEV}_{\mathrm b}$ distribution for both the likely Galactic and the likely extragalactic sources is approximately Gaussian. The variable likely Galactic sources were detected within the $\overline{\mathrm{NEV}_{\mathrm b}}$ range of $0.0118-0.243$, with a geometric mean of $0.0477$. The likely extragalactic sources were instead found in the range $0.00266 - 0.113$, with a geometric mean of $0.0385$. We used Welch's t-test to determine whether there is a difference in the mean of the $\overline{\mathrm{NEV}_{\mathrm b}}$ distribution for likely Galactic and likely extragalactic sources. We found $t=5.96$ and an associated p-value of $p=6.94\times10^{-9}$, so these are most likely distinct distributions. 

The difference in the distributions of the $\overline{\mathrm{NEV}_{\mathrm b}}$ predominantly reflects the different variability properties of the variable Galactic and extragalactic sources. Stellar variability is dominated by infrequent flares. The NIV of such light curves is typically larger than for continuously variable sources with power law PSDs, which usually apply to AGNs. The most variable sources have light curves which feature infrequent large flares over a low continuum. 


We did not analyse the properties of the XRBs we detected to show variability. These will be investigated in Kaltenbrunner et al. in prep. These XRBs are some of the brightest, most variable sources that we identified, and are often found towards the upper right corner in Figs. \ref{BV3DeR1} and \ref{MAD3DeR1}.

\begin{figure}[pt]
\resizebox{\hsize}{!}{\includegraphics{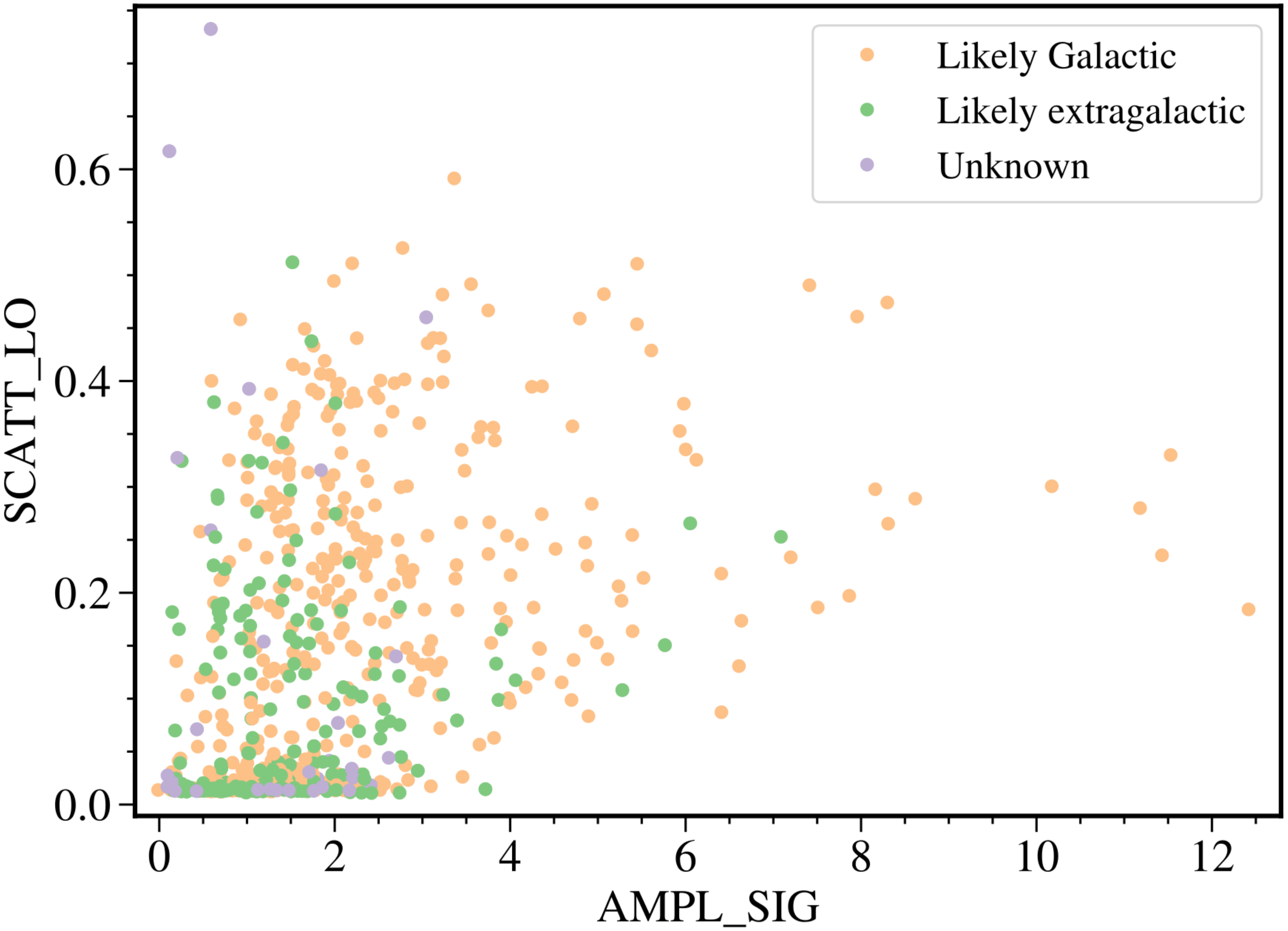}}
\caption{Comparing the SCATT\_LO and AMPL\_SIG values found for different types of variable sources in eRASS1, 2, and 3. Each dot represents a single SCATT\_LO and AMPL\_SIG value of a variable source in one of the three eRASSs. It also shows the values of the variability parameters for eRASSs in which the variable sources were not found above the $3\sigma$ thresholds. A likely Galactic source at an AMPL\_SIG value of 78, and a SCATT\_LO value of 0.82, is omitted from this figure, for display clarity.
 \label{ASvsSL}}
\end{figure}

The different types of variability exhibited by stars and AGNs is also the cause of the difference in the distribution of the SCATT\_LO and AMPL\_SIG parameters for them, which is shown in Fig. \ref{ASvsSL}. Variable sources classified as likely Galactic tend to have larger AMPL\_SIG values than likely extragalactic sources. By comparing the SCATT\_LO and AMPL\_SIG values of variable sources, it could be possible to classify sources of unknown type. However, as Fig. \ref{ASvsSL} shows, the distinction is imperfect and can often lead to erroneous results. The position of a source on the SCATT\_LO and AMPL\_SIG plane can vary a lot from eRASS to eRASS. It can also depend significantly on the number of bins in the light curve, the average count rate of the source, and whether there is a long gap between observations within an eRASS, or not. Nevertheless, $88\%$ of the sources identified as variable in an individual eRASS by exceeding both the SCATT\_LO and AMPL\_SIG $3\sigma$ thresholds, whose likely source type could be determined, were found to be likely Galactic. In contrast, $62\%$ of sources located above the $3\sigma$ SCATT\_LO threshold, but below the $3\sigma$ AMPL\_SIG threshold, that were matched to an optical counterpart, were found to likely be extragalactic. 

We investigated whether the variability, as measured by $\mathrm{NEV}_{\mathrm b}$ differed between different segments. To do so, we investigated the distribution of the parameter $Y=(\mathrm{NEV}_{b,i}-\mathrm{NEV}_{b,j})/\sqrt{\sigma_i^2+\sigma_j^2}$, between segments $i$, and $j$, where $i\neq j$. In this equation, $\sigma_i$ is the error for $\mathrm{NEV}_{b}$ in segment $i$ ($\mathrm{NEV}_{b,i}$). If the variability is stationary, we would expect $Y$ to be distributed with a standard deviation of 1. 

Initially, we used the measurement error of $\mathrm{NEV}_{b,i}$, and calculated $Y$ for all pairs of segments for all variable sources. We found the distribution of $Y$ to have a mean of close to 0, and a standard deviation of 13.7. This large standard deviation was caused by a few sources with very large values of $Y$. We repeated the analysis, but using the total error, that is calculated by combing the measurement error and the intrinsic scatter in the NIV. This time, the distribution again has a mean of close to 0, but a standard deviation of $0.78$. Distinguishing by the total number of counts, we found that $Y$ had a smaller standard deviation for fainter sources. This is caused by the lack of information to detect significant changes in the variability with a small number of counts. However, for the 60 variable sources observed with at least $10^3$ background subtracted source counts in eRASS:3, $Y$ was distributed with a standard deviation of $1.01$. Likely Galactic and likely extragalactic sources had a standard deviation of $1.07$, and $1.00$, respectively.  

The intrinsic scatter in the NIV used to estimate the error for detecting significant changes in the variability in different segments was calculated for the assumption of a pink noise PSD. For sources with significantly different PSDs, these errors might have been over-, or underestimated. This particularly applies to stars. In contrast, AGNs are expected to have approximately pink noise PSDs in the frequency range probed by \emph{eROSITA}, so the intrinsic scatter estimated for them is accurate. This indicates, that the observed AGN variability is consistent with a stationary variability process within eRASS:3. We did not find any instance of a significant change in the variability, of $Y>3$, between any two segments of observation of the same source.

We collated a list of the variable X-ray sources identified in the SEP field in eRASS1, 2, and 3, at the webpage: \href{https://projects.mpe.mpg.de/heg/erosita/SEP_var/}{https://projects.mpe.mpg.de/heg/erosita/SEP\_var/}. It lists some of the most relevant properties that we determined for these sources. It includes the boresight-corrected locations of the \emph{eROSITA} detected X-ray sources, as well as the positions of their most likely optical counterparts. In addition, it lists the most likely redshifts, spectral types, and measures of the spectral quality, for the sources whose optical spectra were observed by the \emph{AAT}. The website contains the Galactic-extragalactic distinction parameter, $D$, and the most likely source classification. It additionally includes the number of bins, the average source count rate in each segment, the SCATT\_LO and AMPL\_SIG variability significances of the eRASS1, 2, and 3 data sets, the $\mathrm{NEV}_{\mathrm b}$ values of all the segments of the light curve, and the geometric mean $\mathrm{NEV}_{\mathrm b}$ of all segments of 20 consecutive bins within the eRASS:3 light curve. Images of the rebinned, and non-rebinned light curves of all variable sources are also included. 

\section{Periodograms of variable likely extragalactic sources} \label{AGNPSDs}

\begin{figure}[h]
\resizebox{\hsize}{!}{\includegraphics{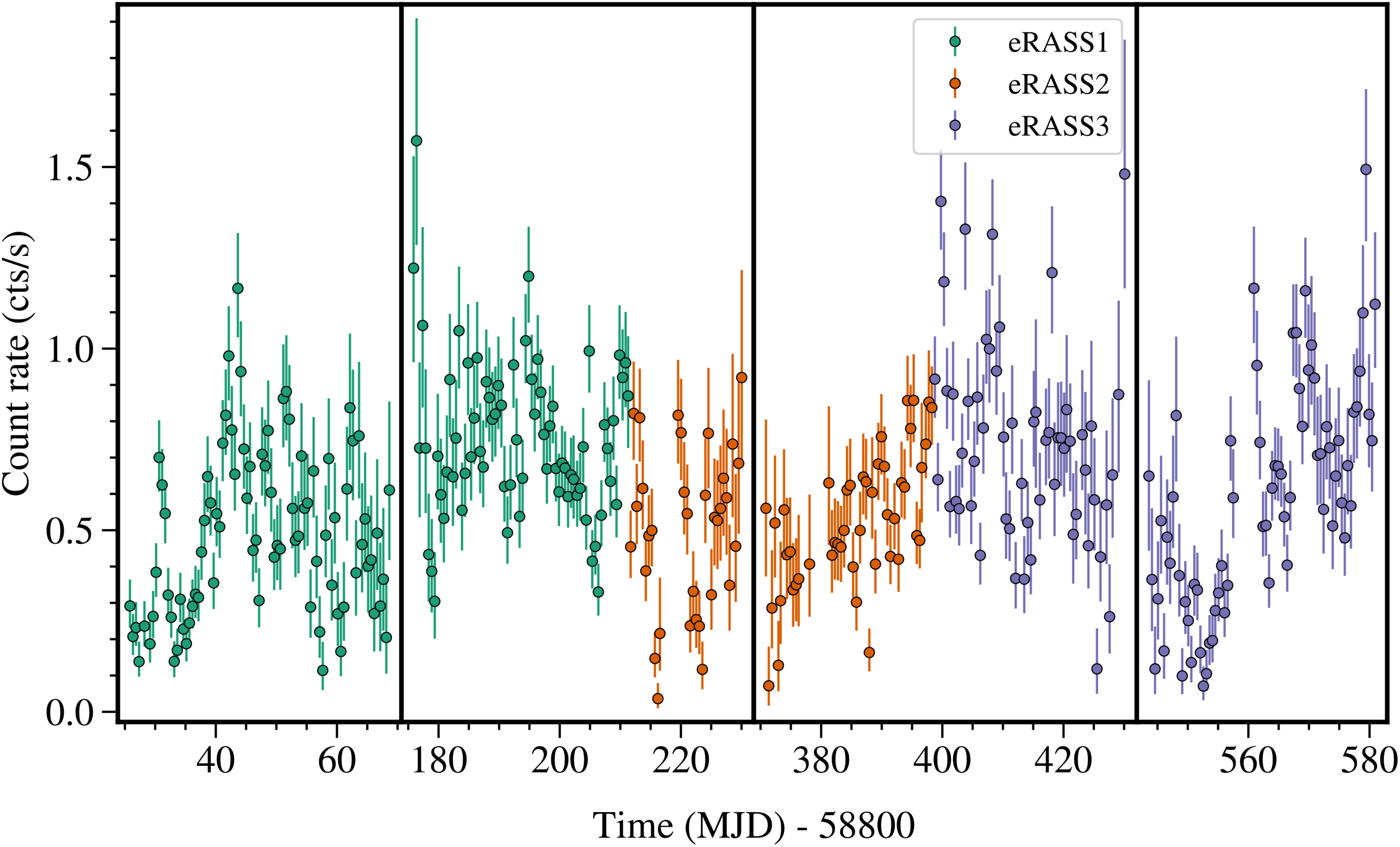}}
\caption{Light curve of the likely extragalactic variable source eRASSt J055033.4-663653, as observed by \emph{eROSITA} in eRASS1, 2, and 3. Black vertical lines distinguish the four different segments of observations, which are separated by several months. The light curve has been rebinned by a factor of 3, for display clarity. 
 \label{PSlc111}}
\end{figure}

\begin{figure}[h]
\resizebox{\hsize}{!}
{\includegraphics{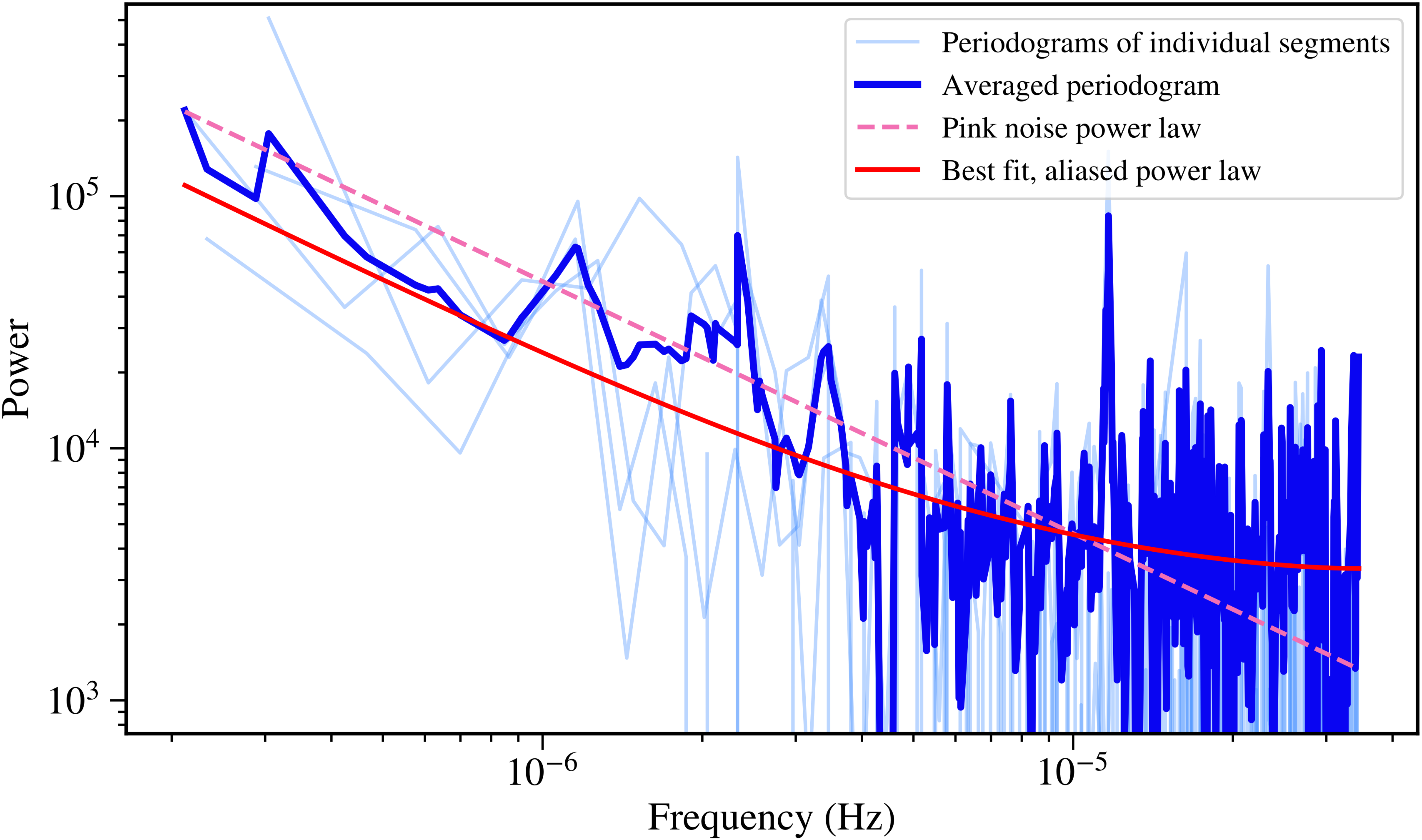}}
\caption{Periodograms of the likely extragalactic source eRASSt J055033.4-663653, whose light curve is shown in Fig. \ref{PSlc111}. The periodograms of individual segments of the light curve are shown in light blue. The averaged periodogram is displayed in dark blue. The red line denotes the best fit to all the periodograms of segments of the light curve, using an aliased single power law with $\alpha = 1.093\pm0.050$. The figure also features a dashed pink line, which depicts a pink noise trend.  
\label{PS111}}
\end{figure}

We investigated the properties of the periodograms of variable sources identified to likely be extragalactic. We only selected the 28 sources with at least $10^3$ background-subtracted source counts in the first three eRASSs for this analysis, as there was insufficient information to constrain the variability power at different frequencies for most sources fainter than that. We generated separate periodograms of each segment of the light curves of these sources. We only selected continuous segments of at least 20 consecutive bins, without any breaks in between. The Poisson and fractional exposure noise were subtracted from the periodograms following the results of \citet{1990A&A...227L..33B} and Bogensberger et al. 2024B. 

We jointly fitted the periodograms of all segments of a variable source using an aliased power law, and aliased broken power law models. In these fits, we set the power law indices to be the same for all periodograms of the same source, but allowed the normalisations to differ. This accounts for the sampling errors and differing red noise contributions in different segments. The fits were performed to minimise the Bayesian likelihood function for a periodogram, that was defined by \citet{2010MNRAS.402..307V}. This likelihood function only works for positive values of the periodogram, so we deleted individual spurious negative powers prior to fitting. It overestimates the normalisation of the periodogram, but we are predominantly interested in measuring the power law slopes. The following figures showing the fitted periodograms have a normalisation adjusted to the periodograms. We used the UltraNest \footnote{\href{https://johannesbuchner.github.io/UltraNest/}{https://johannesbuchner.github.io/UltraNest/}} package \citep{2021JOSS....6.3001B} to minimise the likelihood function, and obtain the best fit parameters and their uncertainties. 

The broken power law models were selected to have a sharp break between the two power laws, at a break frequency, which was allowed to vary freely. In all of the periodograms we investigated, the single power law model was preferred over the broken power law model, using the Bayesian likelihood ratio test. 


Table \ref{TabPSprop} lists the best fitting single power law index, $\alpha$, for each of the 28 brightest variable likely extragalactic sources. It also lists the frequency range over which the periodograms were fitted, the redshifts of the sources, the $\overline{\mathrm{NEV}_{\mathrm b}}$ for 20 bin segments, as well as the number of source counts, and the number of bins in their eRASS:3 lightcurves. 

The break in the power law has been found to depend on the black hole mass, and to a lesser extent also on the bolometric luminosity \citep{2004MNRAS.348..207P, 2006Natur.444..730M, 2011A&A...526A.132G, 2012A&A...544A..80G, 2012A&A...542A..83P, 2017MNRAS.471.4398P}. However, none of the black hole masses of the brightest AGNs in the SEP field are known, so we could not estimate what frequency to expect the breaks to occur at. Based on the dependence of the break on the black hole mass found by \citep{2012A&A...544A..80G}, we estimate that it only falls into the frequency range probed by \emph{eROSITA} for SMBHs with a mass $>1.7\times10^7~\mathrm{M}_{\odot}$, at zero redshift. 

\begin{figure}[h]
\resizebox{\hsize}{!}{\includegraphics{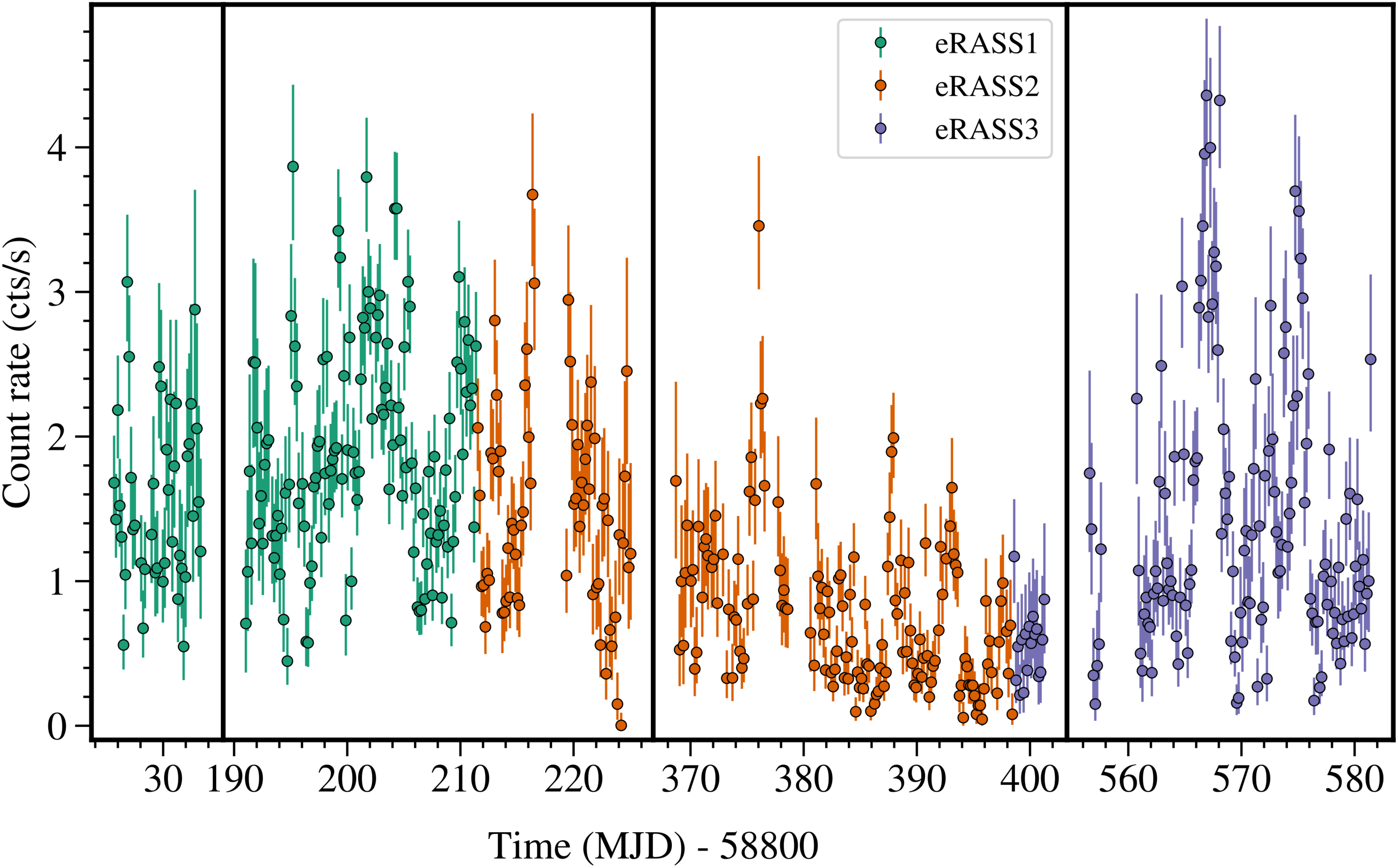}}
\caption{Light curve of the likely extragalactic source eRASSt J061504.1-661717, as observed by \emph{eROSITA} in eRASS1, 2, and 3. The light curve has not been rebinned. 
 \label{PSlc158}}
\end{figure}

\begin{figure}[h]
\resizebox{\hsize}{!}{\includegraphics{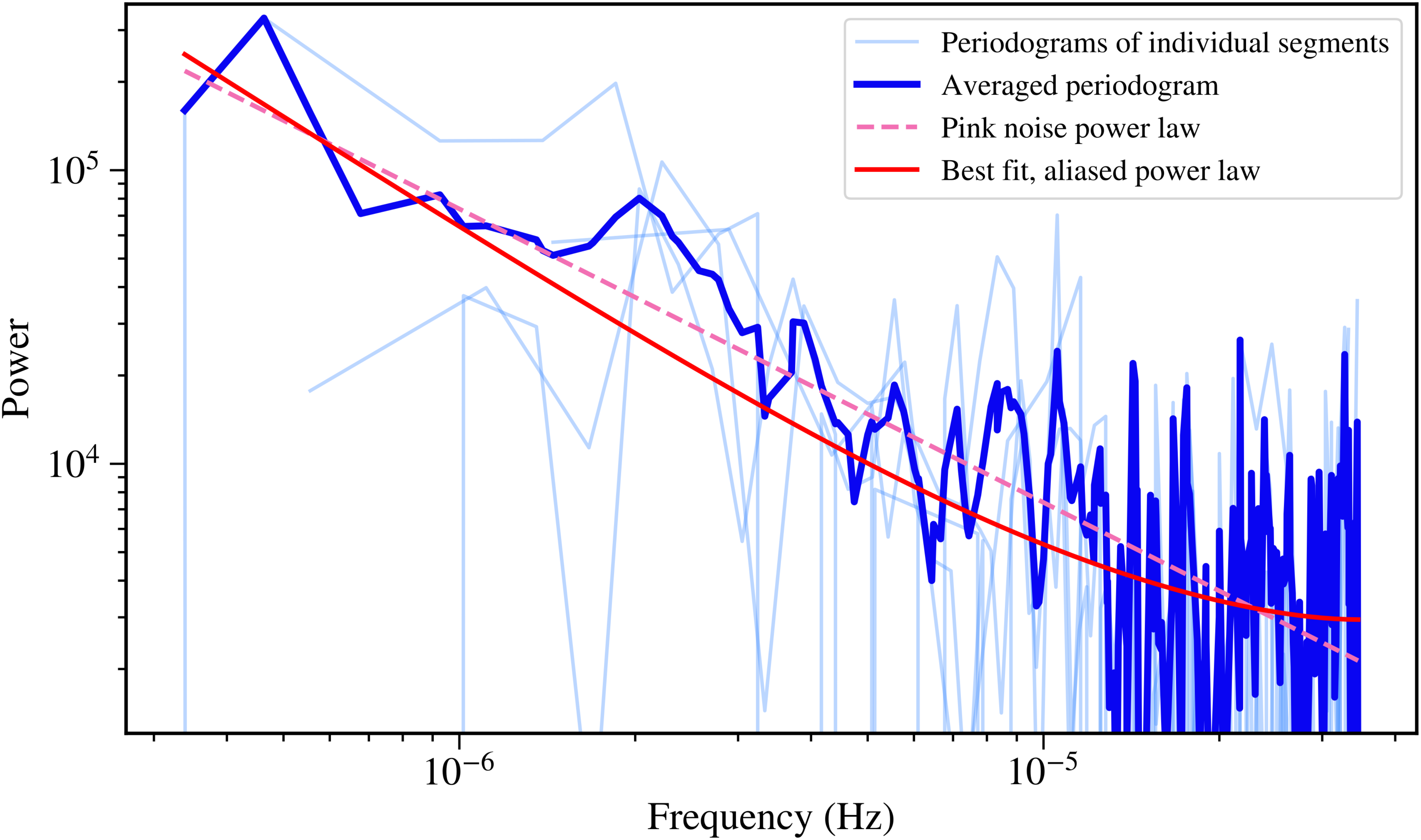}}
\caption{Periodograms of the likely extragalactic source eRASSt J061504.1-661717. The \emph{eROSITA} light curve this is computed for is depicted in Fig. \ref{PSlc158}. The periodograms were best fitted by a single aliased power law model, with $\alpha=1.369\pm0.079$.
 \label{PS158}}
\end{figure}

\begin{table*}[pt]
\centering
\setlength{\tabcolsep}{6pt}
\def\arraystretch{1.3}
\begin{tabular}{c|c|c|c|c|c|c}
    \textbf{Variable source} & $\boldsymbol{z}$ & $\boldsymbol{\sum C_S}$ & $\boldsymbol{N_{\mathrm b}}$ & $\boldsymbol{\overline{\mathrm{NEV}_{\mathrm b}}} ~ (10^{-3})$ & $\boldsymbol{\nu_{min} ~ (10^{-8}~\mathrm{Hz})}$ & $\boldsymbol{\alpha}$ \\ \hline
    eRASSt J061148.4-662435 & $0.230^{a}$ & 45328 & 702 & $2.66^{+0.96}_{-0.56}$ & 24.7 & $1.210 \pm 0.053$ \\
    eRASSt J054641.8-641524 & $0.323^{b}$ & 19343 & 465 & $4.3^{+2.2}_{-1.1}$  & 42.1 & $1.179 \pm 0.056$ \\
    eRASSt J055943.0-660909 & 0.547 & 13419 &  2766 & $13.1^{+6.2}_{-2.3}$  & 6.2 & $0.912 \pm 0.024$ \\
    eRASSt J061504.1-661717 & 0.178 & 10490 &  520 & $80^{+61}_{-33}$  & 34.0 & $1.268 \pm 0.059$ \\
    eRASSt J055357.6-665003 & 0.311 & 10420 &  1628 & $10.6^{+5.3}_{-2.0}$  & 14.8 & $0.963 \pm 0.034$ \\
    eRASSt J055033.4-663653 & $0.0752^{c}$ & 9898 &  1070 & $20^{+11}_{-5}$  & 21.2 & $1.043 \pm 0.039$ \\
    eRASSt J061543.9-653153 & $0.227^{b}$ & 7261 &  361 & $41^{+28}_{-15}$  & 50.7 & $1.142 \pm 0.067$ \\
    eRASSt J060420.2-670234 & 0.678 & 5401 &  1858 & $14.2^{+9.5}_{-2.8}$  & 13.0 & $0.862 \pm 0.038$ \\
    eRASSt J055934.5-653833 & 0.297 & 5191 &  1103 & $12.1^{+8.1}_{-2.8}$  & 20.7 & $0.905 \pm 0.046$ \\
    eRASSt J060534.6-640045 &  & 4836 &  296 & $5.3^{+4.3}_{-1.8}$  & 60.4 & $1.190 \pm 0.094$ \\
    eRASSt J054913.6-642931 & 0.317 & 4514 &  514 & $9.5^{+7.5}_{-2.7}$  & 37.1 & $0.977 \pm 0.073$ \\
    eRASSt J054334.4-642300 &  & 4484 &  447 & $12.9^{+9.9}_{-4.0}$  & 39.7 & $1.166 \pm 0.059$ \\
    eRASSt J060139.2-655756 & 0.603 & 3612 &  1620 & $16^{+13}_{-4}$  & 7.8 & $0.955 \pm 0.027$ \\
    eRASSt J055801.4-665655 & 0.23 & 3305 &  2523 & $22^{+20}_{-5}$  & 9.4 & $0.774 \pm 0.035$ \\
    eRASSt J055745.4-664453 & 0.224 & 3230 &  3117 & $29^{+25}_{-7}$  & 6.2 & $0.619 \pm 0.054$ \\
    eRASSt J055333.8-665751 & 0.121 & 3221 &  1100 & $20^{+39}_{-5}$  & 19.0 & $0.934 \pm 0.022$ \\
    eRASSt J054750.3-672803 & $1.01^d$ & 2895 &  488 & $9.5^{+8.5}_{-2.8}$  & 40.4 & $0.681 \pm 0.093$ \\
    eRASSt J055419.9-663340 & 0.251 & 2829 &  2402 & $26^{+73}_{-6}$  & 6.4 & $0.853 \pm 0.030$ \\
    eRASSt J061932.9-645955 & 0.255 & 2396 &  269 & $11^{+12}_{-4}$  & 66.8 & $0.79 \pm 0.11$ \\
    eRASSt J053423.0-652004 &  & 2047 &  388 & $27^{+28}_{-10}$  & 45.4 & $1.184 \pm 0.082$ \\
    eRASSt J060528.6-652008 & 0.478 & 1812 &  548 & $14^{+14}_{-4}$  & 33.7 & $0.611 \pm 0.082$ \\
    eRASSt J055454.3-653829 &  & 1579 &  1133 & $23^{+29}_{-6}$ & 14.9 & $0.831 \pm 0.043$ \\
    eRASSt J055831.6-663948 & 0.328 & 1380 &  3118 & $34^{+76}_{-8}$ & 6.2 & $0.744 \pm 0.040$ \\
    eRASSt J055927.9-662926 &  & 1374 &  1053 & $26^{+32}_{-7}$  & 19.4 & $0.611 \pm 0.068$ \\
    eRASSt J054831.4-664259 & 0.116 & 1350 &  786 & $21^{+29}_{-6}$ & 28.0 & $0.809 \pm 0.068$ \\
    eRASSt J060221.4-671025 & 0.426 & 1345 &  1720 & $26^{+33}_{-6}$ & 16.3 & $0.576 \pm 0.056$ \\
    eRASSt J060119.3-662918 &  & 1258 &  3117 & $70^{+150}_{-21}$ & 6.2 & $0.948 \pm 0.032$ \\
    eRASSt J061134.3-674708 & 0.376 & 1127 &  721 & $18^{+27}_{-5}$ & 32.6 & $0.896 \pm 0.065$ \\
\end{tabular}
\caption{Properties of the brightest variable likely extragalactic sources with at least $10^3$ background-subtracted source counts observed by \emph{eROSITA} in eRASS1, 2, and 3 in the SEP field. This table lists the redshift, $z$, the total background-subtracted source counts in eRASS:3 ($\sum C_S$), and the number of bins with $\epsilon > 0.1$ in eRASS:3. The sources are sorted by the number of total background-subtracted source counts observed. The table also lists the geometric mean $\mathrm{NEV}_{\mathrm b}$, for 20 bin segments in the light curve. The associated error includes both measurement and sampling errors. Finally, the table also describes the best fit properties of the periodograms of these sources. It lists the minimum frequency for which the power was determined, and the best fitting power law index, $\alpha$, when fitting the periodograms with a single aliased power law model. The maximum frequency probed in all of these periodograms is $6.94\times10^{-5} ~ \mathrm{Hz}$. Most redshifts were determined from \emph{AAT} spectroscopy. We also included the redshifts found from previous analyses of matched optical counterparts, for sources that were not observed by the \emph{AAT}. These redshifts are denoted with letters as superscripts; $a$ is from \citet{2008A&A...482..113M}, $b$ is from \citet{2001AJ....121.2843B}, $c$ is from \citet{1991ApJS...75..297H}, and $d$ is from \citet{2003AJ....125....1G}. The redshift of eRASSt J055357.6-665003 was also measured by \citet{2012ApJ...747..107K}, who measured it to be 0.32. 
\label{TabPSprop}}
\end{table*}

Most periodograms of the variable, bright, likely extragalactic sources fitted with a single aliased power law, were found with a best-fitting index of $\alpha \approx 1$. This agrees with previous findings by other instruments, in which AGN periodograms were observed to mostly follow a pink noise power law within the frequency range probed by \emph{eROSITA} \citep{2002A&A...382L...1P, 2004MNRAS.348..207P, 2012A&A...544A..80G}.

The variable likely extragalactic source eRASSt J055033.4-663653 has a light curve shown in Fig. \ref{PSlc111}. Using SCATT\_LO, it was found to be variable above $3\sigma$ in all three eRASSs. In contrast, it was never found to be variable above $3\sigma$ in any eRASS, when using the AMPL\_SIG thresholds. It has periodograms (Fig. \ref{PS111}) that are best fitted with aliased power laws with $\alpha = 1.043\pm0.039$. 

However, not all periodograms conformed to this standard shape. Most of the best fit power law indices were found to be inconsistent with $\alpha=1$. Nevertheless, $\alpha$ was found to be distributed around $\alpha=1$, between values of $1.268\pm0.059$, and $0.576\pm0.056$. 

One instance of a source with a periodogram best fit with $\alpha > 1$, is eRASSt J061504.1-661717. Its light curve (Fig. \ref{PSlc158}) features large, flaring-like events, as well as an epoch with a significantly lower average count rate in eRASS2. This corresponds to a periodgram that is best described by an aliased power law with $\alpha=1.268\pm0.059$ (Fig. \ref{PS158}). This values lies between the two power law indices commonly observed in AGN periodograms, and is inconsistent with $\alpha=1$. It is also the most variable source out of this sample of the brightest likely extragalactic sources in the SEP field in eRASS:3, with $\overline{\mathrm{NEV}_{\mathrm b}}=0.080^{+0.061}_{-0.033}$ for 20 bin segments in its light curve. 

Seven other sources have periodograms that are best fitted with power laws with $\alpha>1$. The steeper slope in these periodograms might be caused by a break to a steeper slope of $\alpha \approx 2$ at high frequencies, which could not be resolved in the data. 



\begin{figure}[h]
\resizebox{\hsize}{!}{\includegraphics{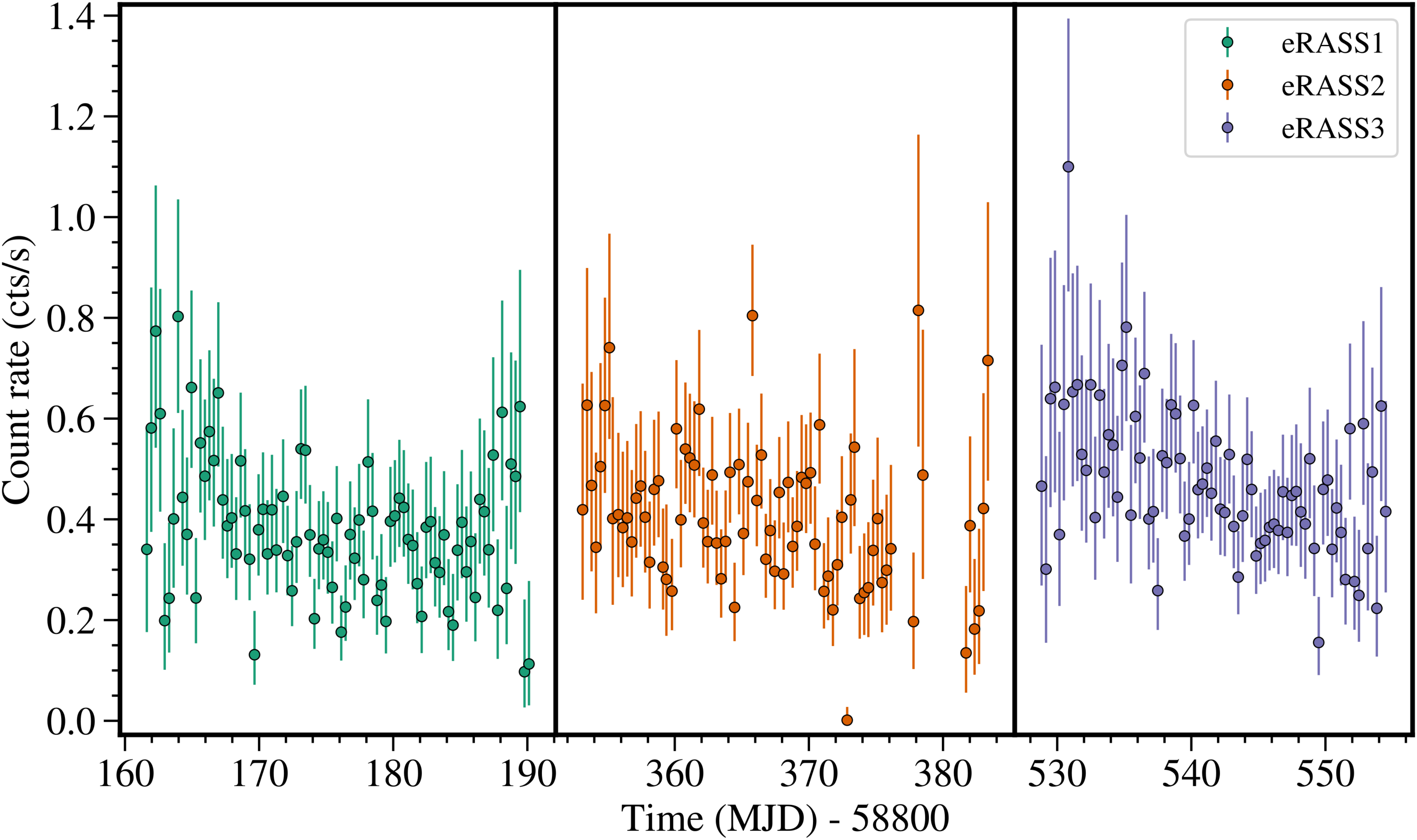}}
\caption{Light curve of the likely extragalactic source eRASSt J054750.3-672803, as observed by \emph{eROSITA} in eRASS1, 2, and 3. The light curve has been rebinned by a factor of 2 for visual clarity. 
 \label{PSlc89}}
\end{figure}

In contrast, the periodograms of eRASSt J054750.3-672803 are best fitted by a power law with $\alpha = 0.681 \pm 0.093$. Comparably low power law indices were found for the periodograms of several other likely extragalactic sources, predominantly for ones with lower total source counts, and count rates. These sources often have a low degree of variability, and have light curves that somewhat resemble white noise, and do not have clearly defined large long-term variability. This can be seen in the light curve of eRASSt J054750.3-672803, which is shown in Fig. \ref{PSlc89}. It was only detected to be variable by AMPL\_SIG in eRASS2, when it was only detected above $1\sigma$ by SCATT\_LO. Both variability quantifiers placed it below $1\sigma$ variability significance in both eRASS1 and 3. 

An aliased power law with $\alpha = 0.5$ is approximately flat for the range of frequencies we are investigating. Some of these sources have periodograms that are dominated by stochastic aliased high-frequency power, which has a negligible dependence on frequency. Additionally, low count rates, and the limitations of the fitting procedure in regards to negative powers, results in a preference for fitting shallower slopes. This systematic error increases with decreasing source counts, which is why we did not investigate sources with fewer than 1000 total source counts. Therefore, the shallow power law slopes are less reliable.



\begin{figure}[h]
\resizebox{\hsize}{!}{\includegraphics{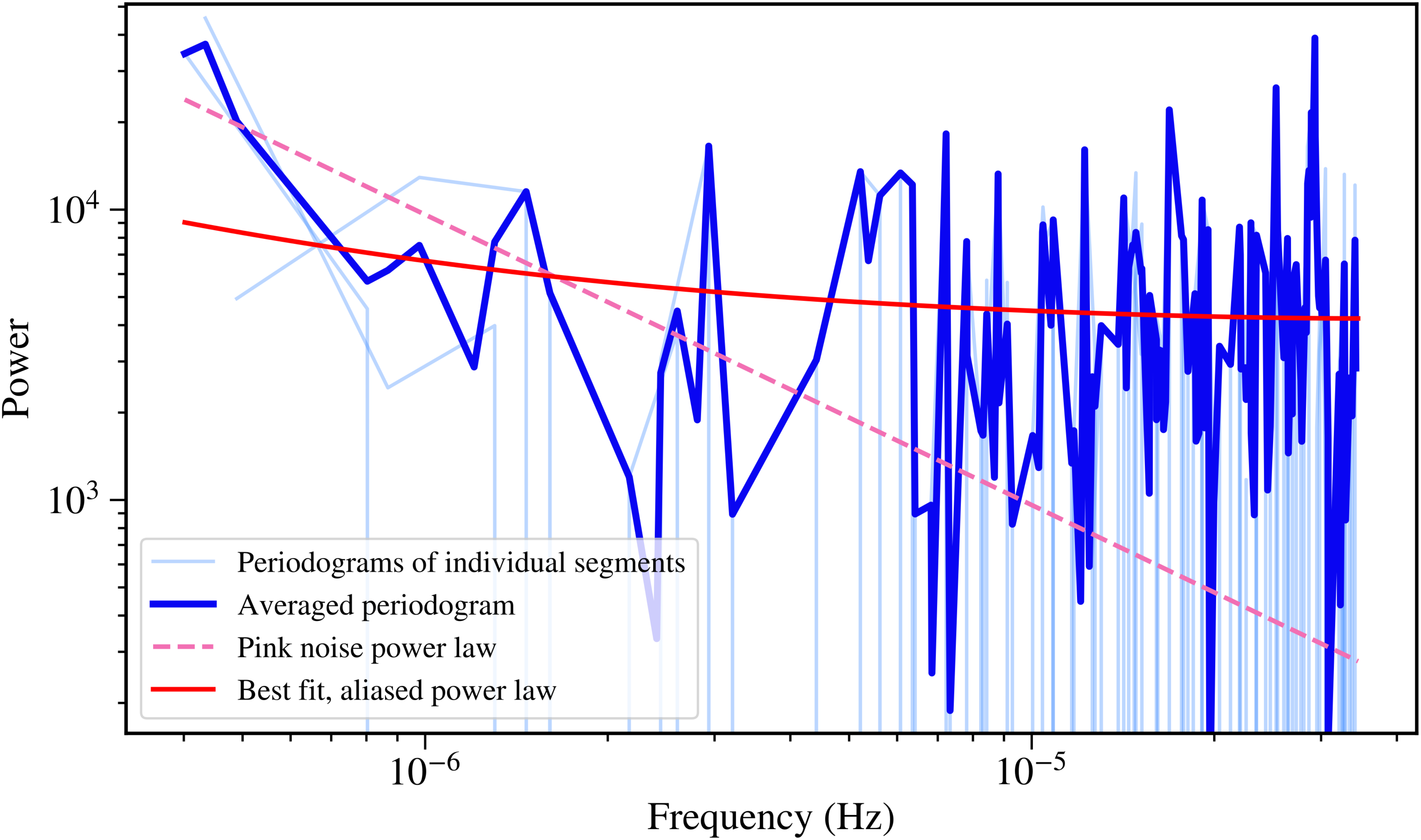}}
\caption{Periodograms of the likely extragalactic source eRASSt J054750.3-672803. The \emph{eROSITA} light curve this is computed for is depicted in Fig. \ref{PSlc89}. The periodograms were best fitted by a single aliased power law model, with $\alpha=0.681 \pm 0.093$.
 \label{PS89}}
\end{figure}


Figs. \ref{PS111}, and \ref{PS158} indicate that a pink noise relationship is a good first-order approximation of the frequency dependence of the variability power of many AGNs observed by \emph{eROSITA}. Many systems have a significantly steeper or shallower power law, but in all cases $\alpha$ was found to be closer to a value of 1, than a value of 2, or 0.

\begin{figure}[h]
\resizebox{\hsize}{!}{\includegraphics{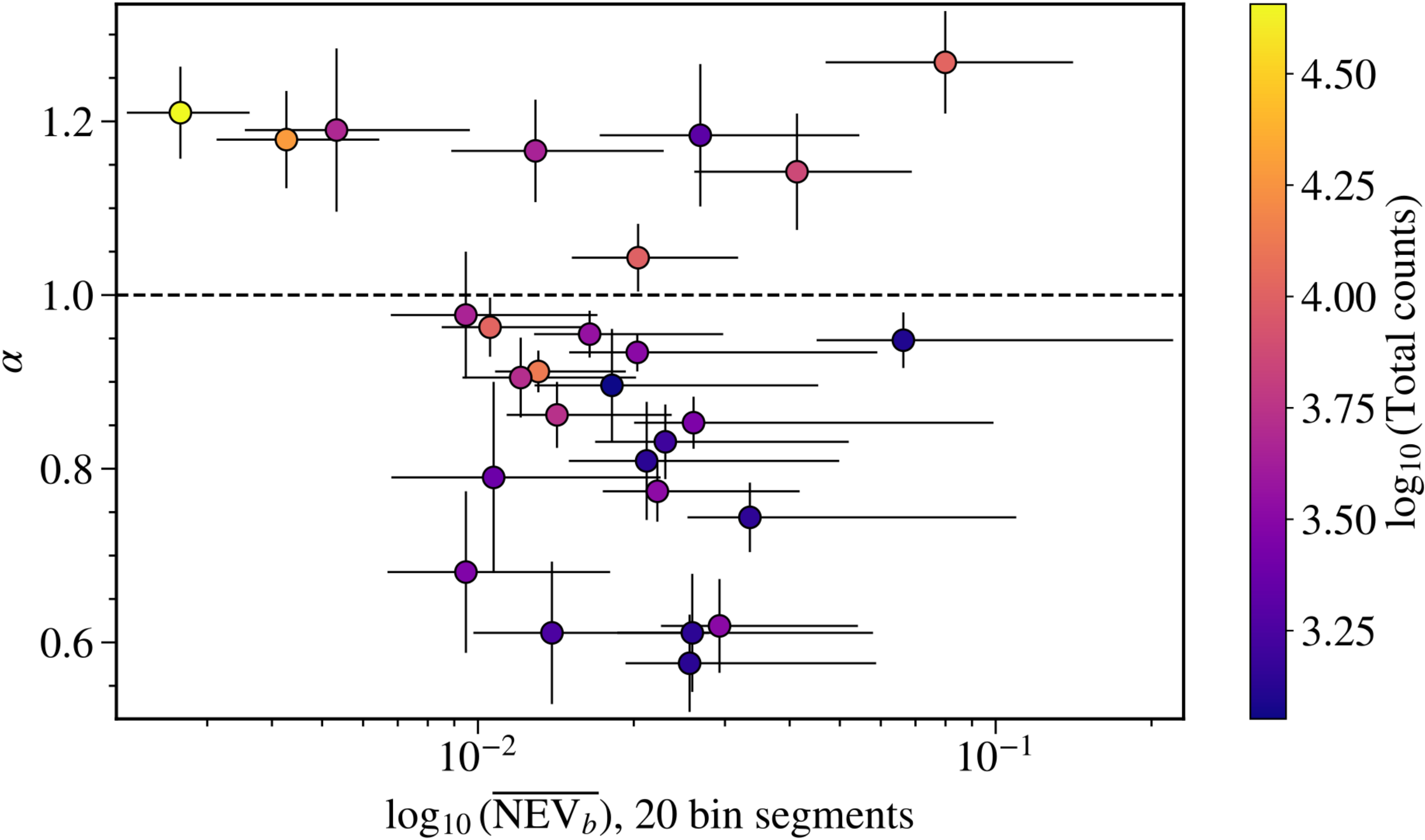}}
\caption{Best fitting aliased power law slopes of the 28 brightest likely extragalactic sources, as a function of the geometric mean $\mathrm{NEV}_{\mathrm b}$ in 20 bin segments, and the total background subtracted source counts. The depicted errors on the $\mathrm{NEV}_{\mathrm b}$ combined the measurement and sampling errors. 
 \label{fig:PSDslopevsNEV}}
\end{figure}

Fig. \ref{fig:PSDslopevsNEV} depicts the best fit power law slopes as a function of the $\overline{\mathrm{NEV}_{\mathrm b}}$, and also showcases the dependence on the total counts. There is no clear correlation between $\alpha$, and $\mathrm{NEV}_{\mathrm b}$. Nevertheless, the variability thresholds are more sensitive to low degrees of variability for sources observed with a large number of source counts. The periodogram fitting process also has an increased sensitivity to steeper power-law slopes for bright sources. Consequently, higher total count sources in this sample have a larger range of $\overline{\mathrm{NEV}_{\mathrm b}}$ values, and are more commonly found with a steeper power-law index. It is possible that the lack of a correlation between $\alpha$ and $\mathrm{NEV}_{\mathrm b}$ relates to the likely differing black hole masses, luminosities, or spins in this sample.

\section{Variable X-ray sources of particular interest}\label{Intvarsrc}

This section discusses the variability properties of some of the most interesting variable X-ray sources in the SEP field. We primarily investigated variable likely extragalactic sources but also inspected likely Galactic sources for unusual variability. 


\subsection{Extragalactic transients}


The \emph{eROSITA} variable source eRASSt J053942.0-653038 was detected barely above the background level, with a mean count rate of $0.0056^{+0.0024}_{-0.0019}~\mathrm{cts/s}$, throughout most of eRASS1. However, on MJD 58869, it suddenly became significantly brighter, reaching an average count rate of $0.294^{+0.033}_{-0.030}~\mathrm{cts/s}$ in the $0.2-5.0~\mathrm{keV}$ energy band in the days that followed it. In this final period of eRASS1 observations, the count rate varied a lot from bin to bin, but appeared to show a downward trend after the peak. At the end of the eRASS1 observations, it was again observed at a very low count rate. However, that could be a fluke detection, resulting from the low fractional exposures in those bins. 



About 150 days later, eRASSt J053942.0-653038 was observed again during eRASS2, this time with a mean count rate of $0.1168^{+0.0069}_{-0.0067}~\mathrm{cts/s}$ in the $0.2-5.0~\mathrm{keV}$ energy band. The mean count rate also appeared to decrease during the next 30 days, in which it was observed every 4 hours by \emph{eROSITA}. Finally, it was observed again in eRASS3, this time with an even lower mean count rate of merely $0.0680^{+0.0071}_{-0.0064}~\mathrm{cts/s}$, which is still larger than the source count rate observed for it before the start of this eruption. 

We determined its most likely optical counterpart to have a GAIA parallax significance of $p/\sigma_p=0.884$. It was also observed to have: $W1+1.625\log(F_{0.5-2})+6.101 = 5.83$, and $z-W1-0.8(g-r)+1.2 = 0.843$. This source was not matched to the VHS DR5 catalogue, so we could not use the $W1-J$ Galactic-extragalactic distinction. We observed the spectrum of the most likely optical counterpart of this source with 2dF/AAOmega. However, by the time we observed it, on January 5, 2022, it had become too faint at optical wavelengths to be identified above the background. Therefore, we were unable to determine its redshift. Combining these results, we classified it as likely extragalactic with a distinction parameter of $D=10$. However, the field in which it is located is very crowded, so associating the X-ray detection with a particular optical source is challenging.  

The light curve bears some resemblances with the evolution of Tidal Disruption Events \citep[TDEs][]{1988Natur.333..523R,2020MNRAS.497.1925G}. It features a sudden large flare that significantly exceeds the pre-flare flux, followed by a gradual decline over hundreds of days. \citet{1989IAUS..136..543P} determined theoretically, that a TDE should evolve with $R_{\mathrm S}\propto T^{-5/3}$, where $T$ is the time since the start of the TDE. However, \citet{2009MNRAS.392..332L} found that the initial drop in luminosity is shallower than $R_{\mathrm S}\propto T^{-5/3}$. \citet{2011MNRAS.410..359L} described that that relationship is only approximately accurate for about one year after the start of the TDE, in the X-ray band. \citet{2011MNRAS.410..359L} found the long-term decay to approximately follow: $R_{\mathrm S}\propto T^{-5/12}$.

\begin{figure}[h]
\resizebox{\hsize}{!}{\includegraphics{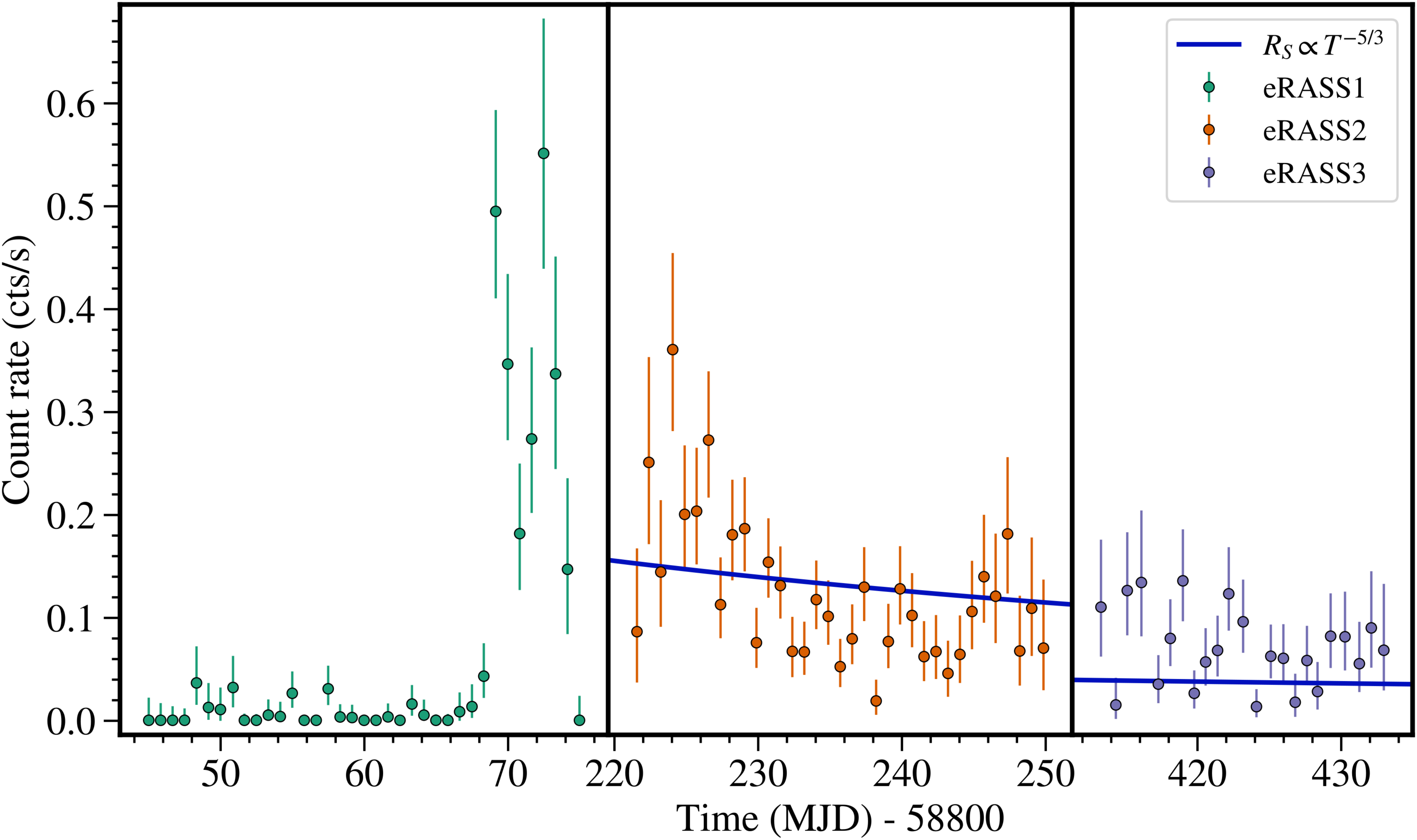}}
\caption{Light curve of eRASSt J053942.0-653038, observed by \emph{eROSITA} in eRASS1, 2, and 3. We also indicate the typical TDE luminosity decay $R_{\mathrm S}\propto t^{-5/3}$, fitted to the eRASS2 and 3 data. The light curve has been rebinned by a factor of 5 for visual clarity.
 \label{TDElc}}
\end{figure}

We investigated whether the decay in the count rate of eRASSt J053942.0-653038 after the start of the flare could be described by $R_{\mathrm S}\propto T^{-5/3}$. The observations immediately after the start of the eruption in eRASS1 were too close to the start to be useful for determining the long-term decay rate. We combined all the observations of segments 2 and 3 (shown in Fig. \ref{TDElc}), and fitted the decaying count rates with the function: $R_{\mathrm S} = A (t-58869) ^{-\gamma}$, where $t$ denotes the time of each measurement, expressed in MJD. These observations were best fitted with $\gamma=0.91\pm0.27$ for the full energy band, which is significantly less than $5/3$, but is also steeper than $5/12$. Fig. \ref{TDElc} shows the best fit using a power law with $\gamma=5/3$. This relation is too steep, and underestimates the mean count rate observed in segment 3. The best fit value of $\gamma$ may correspond to the time after the eruption in which both the $T^{-5/3}$, and the $T^{-5/12}$ power laws are relevant. 


Alternatively, it could be a different type of transient extragalactic source. Without a redshift for it, we are unable to estimate the luminosity at the peak of the eruption. We also lacked follow-up observations of this source during the interval in which it was bright, but noticeably declining. We could not clearly visually identify a galaxy in optical images at the location of this transient. Therefore, the cause of the eruption of eRASSt J053942.0-653038 could not be definitively determined. 

\begin{figure}[h]
\resizebox{\hsize}{!}{\includegraphics{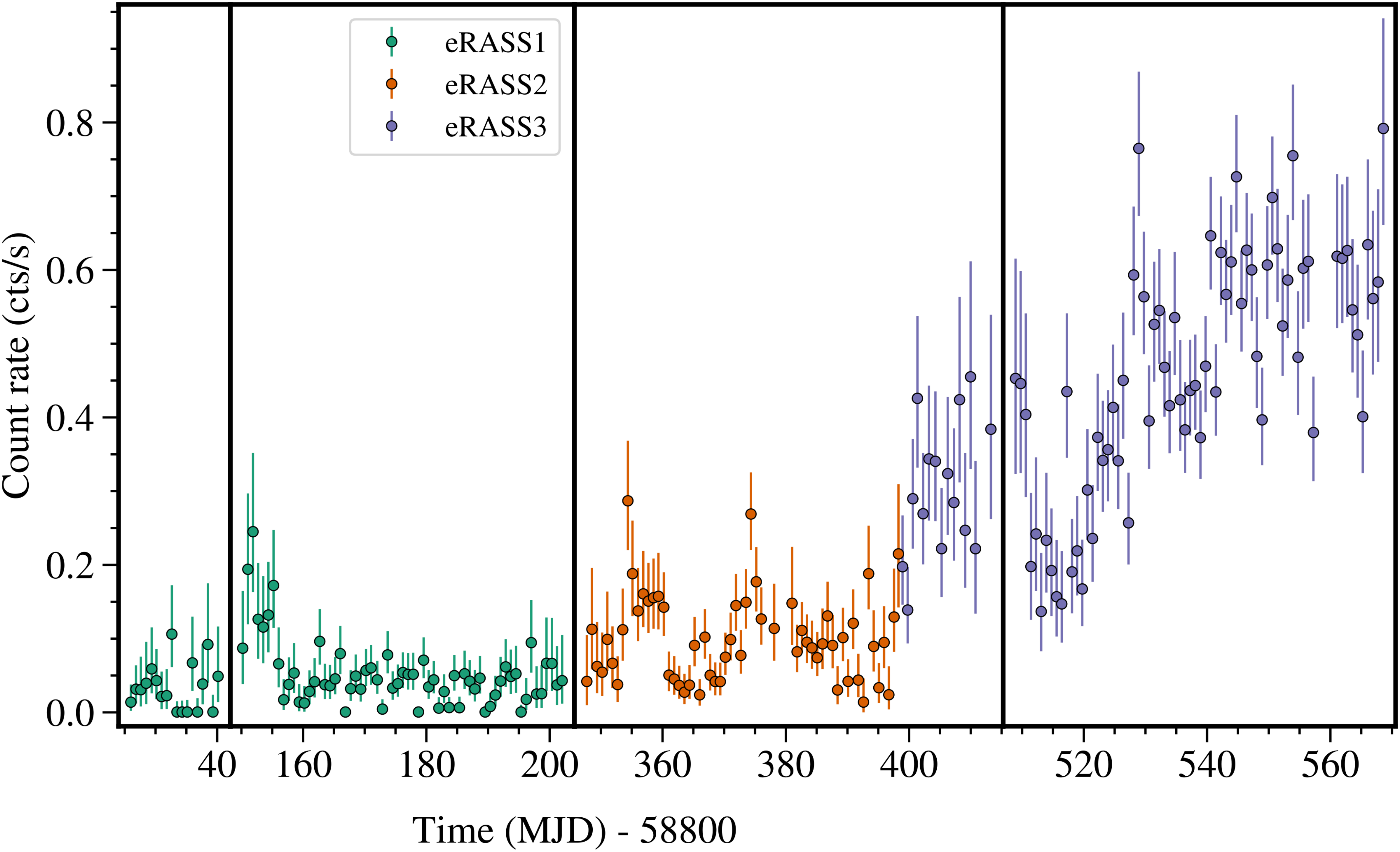}}
\caption{The eRASS1, 2, and 3 light curve of eRASSt J055333.8-665751. The light curved was rebinned by a factor of 5, for display clarity.  
 \label{LC234}}
\end{figure}

We also observed some other likely extragalactic sources, which featured a significant increment or decrement in their observed count rates throughout the observations of them in eRASS:3. Some of these could be AGN ignition and shut-down sources. Fig. \ref{LC234} depicts the light curve of one such source, eRASSt J055333.8-665751. Its count rate increased by a factor of $13\pm4$. In eRASS1, it had a mean count rate of $0.0468\pm0.0031~\mathrm{cts/s}$. In contrast, by the end of eRASS3, after MJD 59345, it had a count rate of $0.599\pm0.17~\mathrm{cts/s}$.

We observed the optical spectrum of the most likely counterpart of this source with the \emph{AAT}. It was well fitted by a quasar model spectrum, at a redshift of $z=0.121$. 

Before the increase in the source flux, we observed this source to feature three flares, each lasting for a few days. These could be episodes of failed ignition of the source. 

In other instances, variable sources were not detected in some eRASSs, which may indicate a possible large change in the source flux to a level below the detection limit. Many more sources in the SEP field might feature such an evolution, but did not satisfy the variability thresholds within any of the three eRASS, and were subsequently not labelled as variable. To detect all of these sources requires a dedicated independent variability detection using data from multiple eRASSs. However, the lack of a counterpart of a source in another eRASSs could also indicate a wrong matching of sources, or an issue with the source detection mechanism. 

Another particularly interesting variable source will be described in Bogensberger et al. in prep. 

\subsection{Variable stars}

More than half of the variable sources we identified were classified as being likely Galactic. In many cases, we found flare stars and active binaries as likely counterparts. Constantly variable emission is seen in the brighter stellar X-ray sources. Many variable stars are instead observed to feature infrequent flares that significantly exceed the non-flaring level. In the following, we present some examples that are typical of the detected sources.

\begin{figure}[h]
\resizebox{\hsize}{!}{\includegraphics{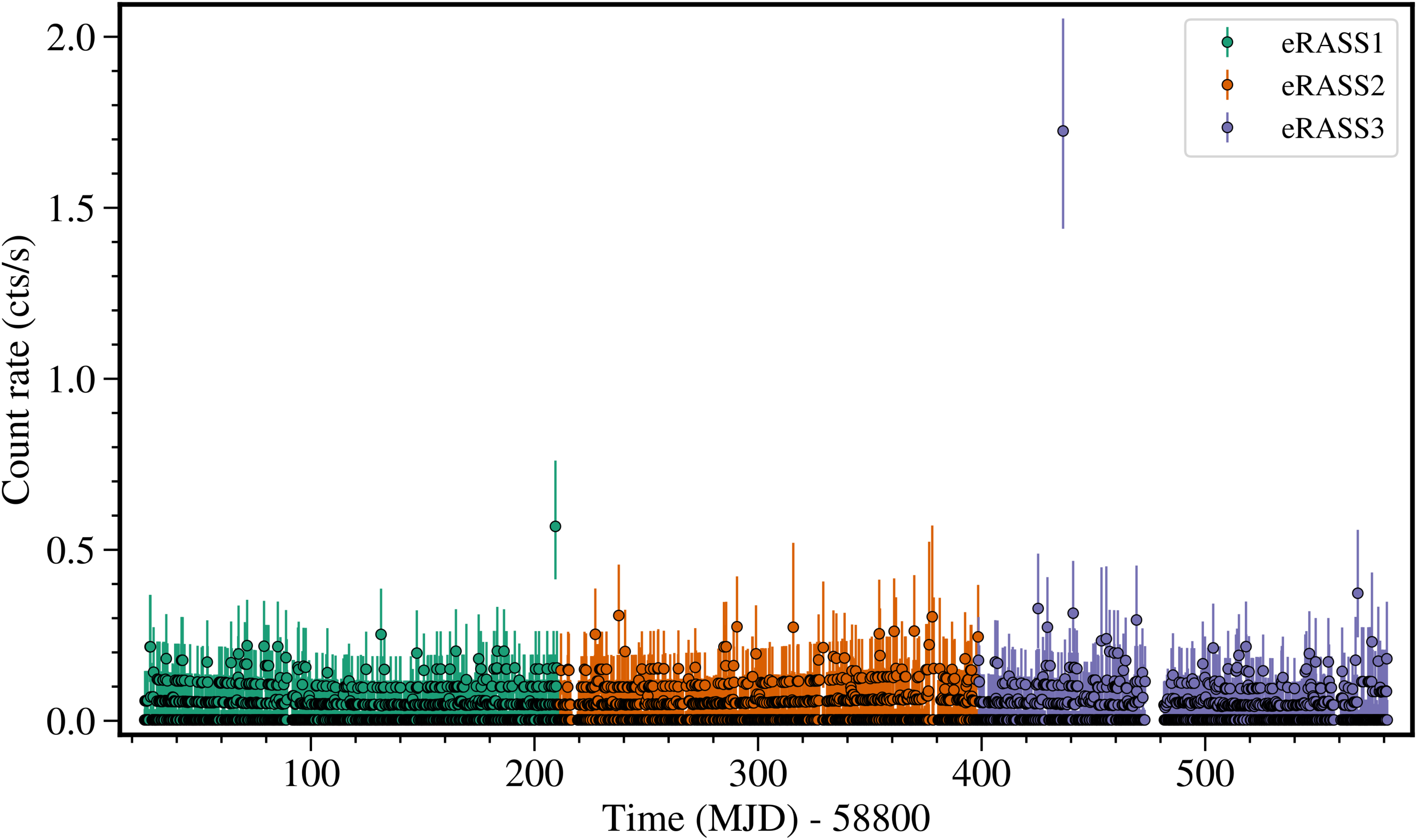}}
\caption{Light curve of the likely Galactic source eRASSt J055930.9-663008, as observed by \emph{eROSITA} in eRASS1, 2, and 3. The light curve has not been rebinned.
 \label{LC223}}
\end{figure}

For instance, the likely Galactic source eRASSt J055930.9-663008 is located merely $4.55'$ away from the SEP. It is matched to the eclipsing binary star OGLE LMC563.08.000004. Except for a few breaks from survey mode observations, such as an 8.5 day interval in eRASS3, this source was observed almost every 4 hours throughout eRASS:3. The light curve of eRASSt J055930.9-663008 is shown in Fig. \ref{LC223}. It had a steady mean count rate of $0.0393^{+0.0012}_{-0.0006} ~ \mathrm{cts/s}$ throughout all 3 eRASSs. However, it was also observed to feature two flares, significantly above its usual, constant flux level. The first, on MJD 59009.38, consisted of a sudden increase to a count rate of $0.57^{+0.19}_{-0.15} ~ \mathrm{cts/s}$. The second flare occurred on MJD 59236.44, when it even became as bright as $1.72^{+0.33}_{-0.29} ~ \mathrm{cts/s}$. In both cases, the source dimmed quickly, and returned to its usual flux by the next eroday. 


\begin{figure}[h]
\resizebox{\hsize}{!}{\includegraphics{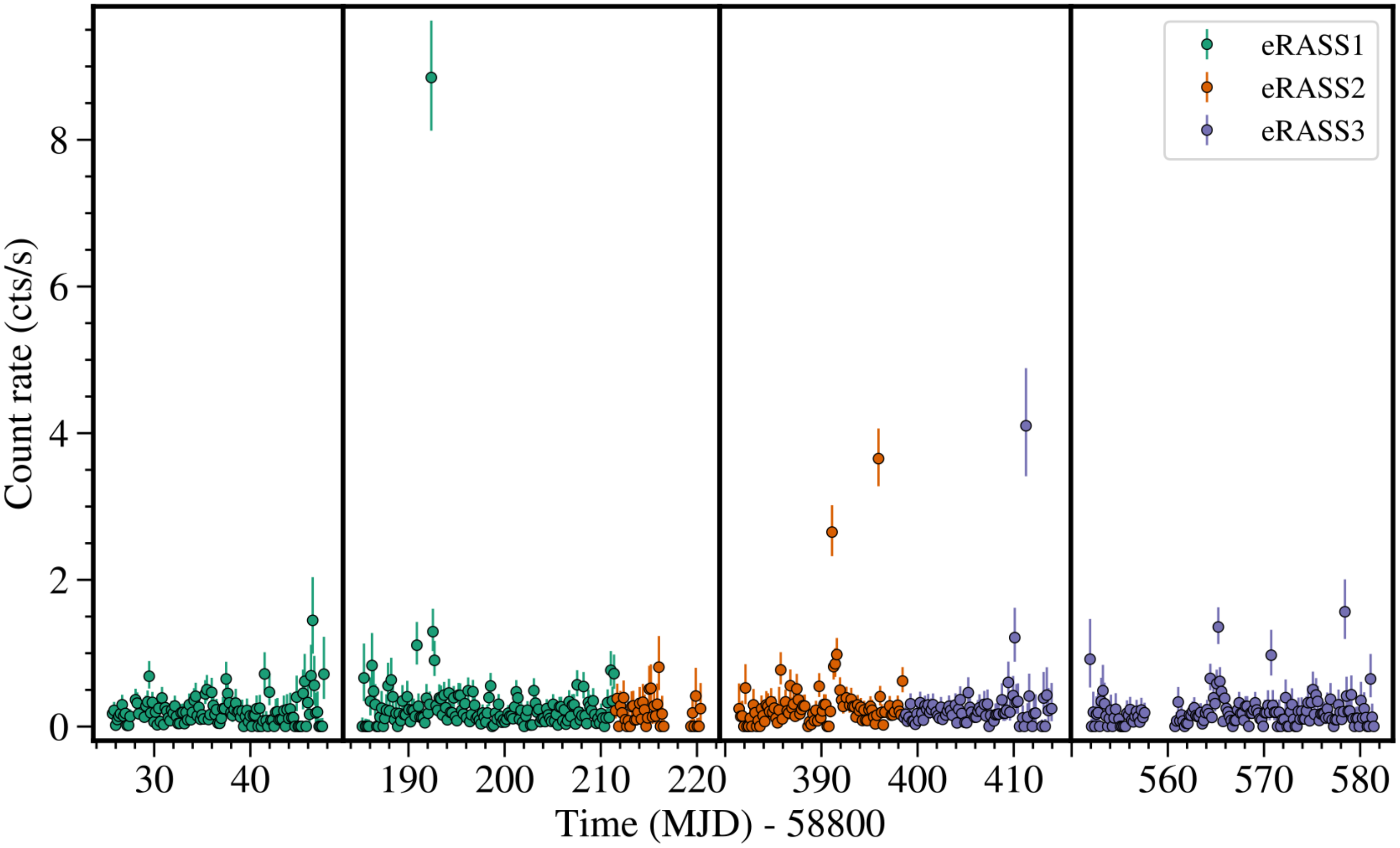}}
\caption{Light curve of the likely Galactic source eRASSt J054557.1-663957, as observed by \emph{eROSITA} in eRASS1, 2, and 3. The light curve has not been rebinned.
 \label{LC74}}
\end{figure}

As the source featured a low mean source count rate and two prominent, but short flares, it had one of the largest $\mathrm{NEV}_{\mathrm b}$ values of all variable sources. For the first segment of consistent observations, from the start of eRASS1 until the start of the extended break in observations during eRASS3, we calculated  $\mathrm{NEV}_{\mathrm b} = 2.5 ^{+3.9}_{-1.7}$. As the segment after the end of the break did not feature any comparable flare, we measured a significantly lower degree of variability for it, of only $\mathrm{NEV}_{\mathrm b} = 0.030 ^{+0.34}_{-0.030}$. The large flares in eRASS1 and 3 caused this source to be identified as significantly variable by both SCATT\_LO and AMPL\_SIG. However, the eRASS2 light curve was classified as not variable by both methods. 

eRASSt J054557.1-663957 (matched to the eruptive variable star WOH G 539) had even larger, and more frequent flares, which are shown in Fig. \ref{LC74}. The largest of these reached a peak count rate of $8.85^{+0.78}_{-0.72}~ \mathrm{cts/s}$, above a consistent non-flaring count rate of $0.2064^{+0.0057}_{-0.0040}~ \mathrm{cts/s}$. Some of the flares only lasted for one eroday, just like the ones seen in eRASSt J055930.9-663008. However, other flares continue for a few erodays, before reaching the non-flaring level again. 

\begin{figure}[h]
\resizebox{\hsize}{!}{\includegraphics{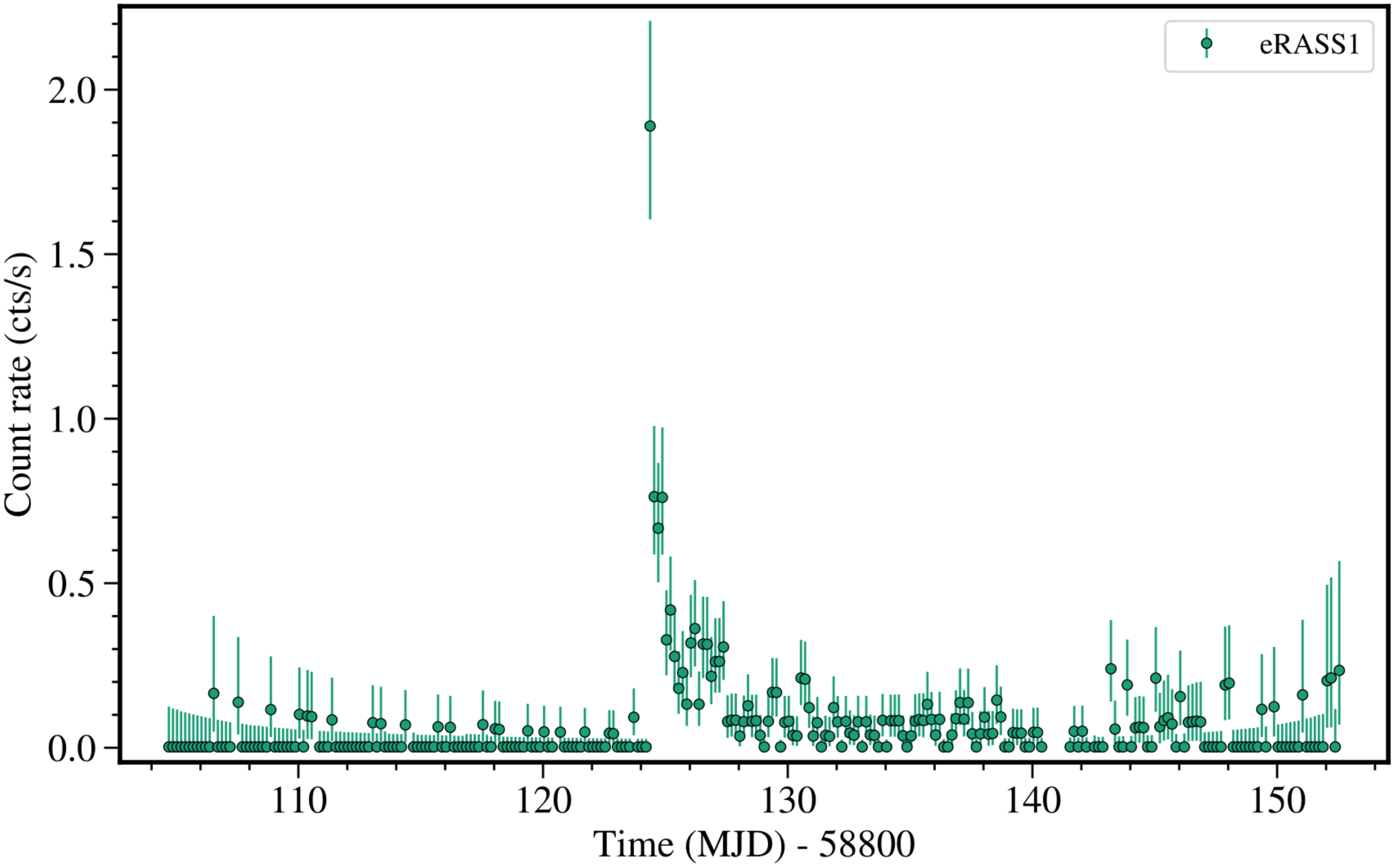}}
\caption{Light curve of the likely Galactic source eRASSt J055718.1-652930, as observed by \emph{eROSITA} in eRASS1. The light curve has not been rebinned.
 \label{LC188}}
\end{figure}

An example of a long-term flare is seen in the light curve of eRASSt J055718.1-652930, in Fig. \ref{LC188}. On MJD 58924.38, the source suddenly brightened by more than two orders of magnitude, from an average of $0.0164^{+0.0056}_{-0.0023}~\mathrm{cts/s}$ prior to the flare, to $1.89^{+0.32}_{-0.28}~\mathrm{cts/s}$ at the peak. The count rate dropped gradually over several days, but the source maintained a higher count rate than the pre-flare level. The average count rate after MJD 58941 was still $0.0511^{+0.0060}_{-0.0032}~\mathrm{cts/s}$. The source became too faint later on, and was not detected in either eRASS2 or 3.

\begin{figure}[h]
\resizebox{\hsize}{!}{\includegraphics{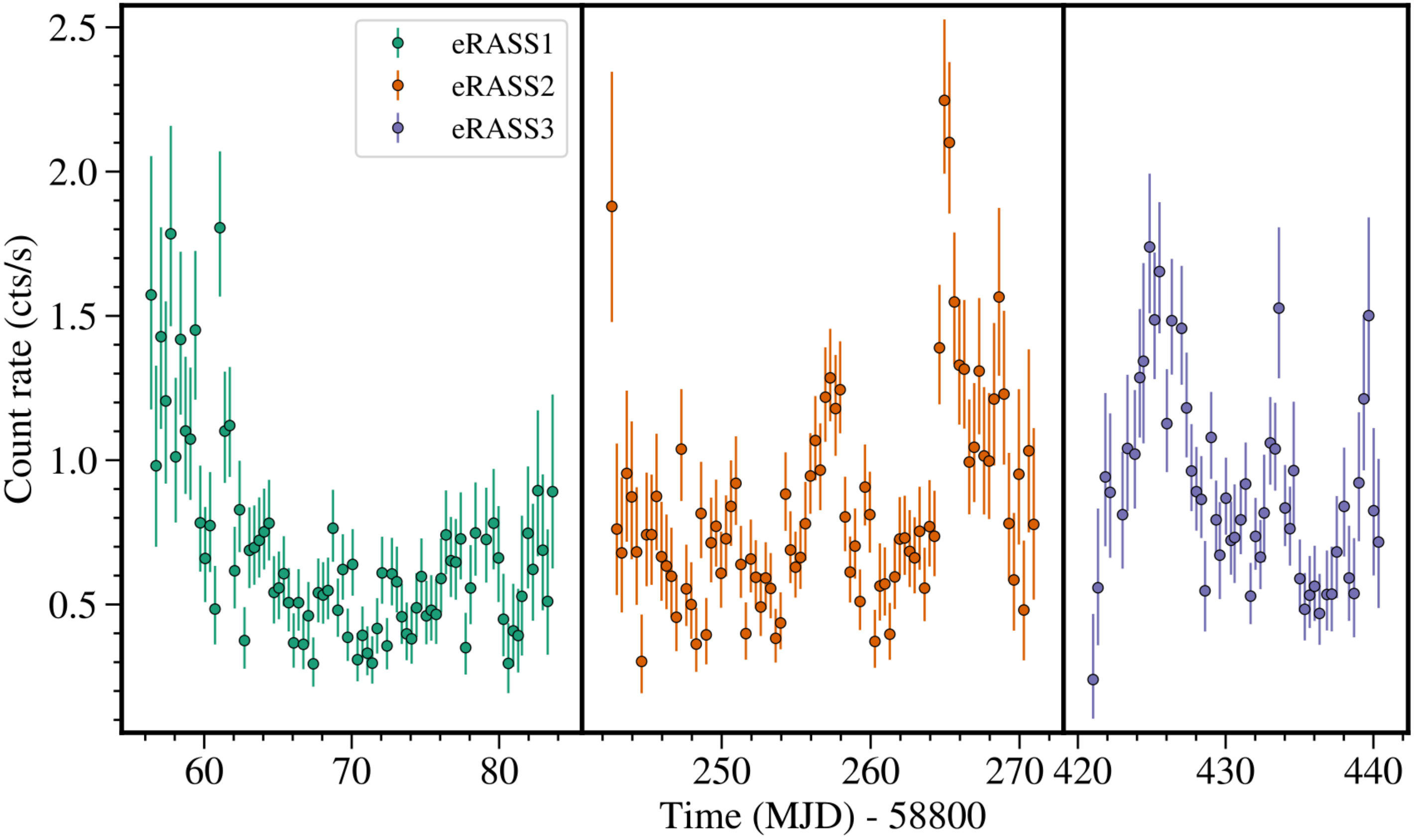}}
\caption{Light curve of the likely Galactic source eRASSt J053620.6-644936, as observed by \emph{eROSITA} in eRASS1, 2, and 3. The light curve has been rebinned by a factor of 2.
 \label{LC29}}
\end{figure}

eRASSt J053620.6-644936 is another variable Galactic source, that is also matched to the star GSC 08887-00400. Its light curve is depicted in Fig. \ref{LC29}. It shows a more continuous variability, rather than the large amplitude flares seen in the X-ray emission of other stars. The light curve looks very similar to the typical variability observed in AGNs. However, we were unable to identify a galaxy that could have caused of the observed X-ray variability. 

The bright star $\mathrm{\eta^2}$ Dor, an M~giant with a J band magnitude of 1.63 \citep{2002yCat.2237....0D}, was wrongly identified as a variable source due to having an apparent triangular flux evolution within each eRASS. However, this effect was caused by optical loading coupled with a variable vignetting strength, and not by variation in its X-ray emission.

\section{Conclusions}\label{Conclusions}

The frequency and consistency of \emph{eROSITA} observations of the SEP field during its eRASSs enables a deep investigation of the mid- to long-term X-ray variability properties of thousands of X-ray sources. However, it also comes with the challenges of varying fractional exposures, and a diverse data set of sources, exposures, and count rates. 

We used thresholds on the SCATT\_LO and AMPL\_SIG variability quantifiers to detect variable sources. Out of the $\approx8200$ X-ray sources observed in each eRASS located within $3\degree$ to the SEP, we identified an average of 197 sources which showed a significant degree of variability, per eRASS. By matching sources between the three eRASSs, we identified 453 distinct variable sources in this field. 


Of the two methods, SCATT\_LO is better suited for detecting variability, as it managed to detect 80.0\% of all variable sources identified by either method. Additionally, 39.3\% of all variable sources were only identified as such by SCATT\_LO. In contrast, AMPL\_SIG is more susceptible to wrongly classify non-variable sources as variable. Nevertheless, AMPL\_SIG detected a number of variable sources that were missed by SCATT\_LO.


Identified variable sources were subsequently distinguished into likely Galactic, likely extragalactic, and unknown sources. We obseved the optical spectra of \emph{eROSITA} selected X-ray sources with the \emph{AAT}, to determine their source type, and redshift. We also used optical photometric catalogues, and known relationships between the magnitudes in various energy bands, to distinguish between likely Galactic and likely extragalactic sources. In this process, the 453 variable sources were classified into 168 likely extragalactic sources, 235 likely Galactic sources, 7 XRBs in the LMC, and 43 sources for which we lacked the information to classify them into one of the other categories.  

We also analysed the periodograms of the 28 brightest significantly variable, likely extragalactic sources. Most of these were best fitted with a single aliased power law with an index of $\alpha \approx 1$. Some sources had steeper power laws, which might be due to an unresolved break at higher frequencies. Several other sources instead had shallower periodograms, whose light curves were dominated by bin to bin variability. 



Finally, we discussed the properties of several especially interesting variable X-ray sources. The light curve of eRASSt J053942.0-653038 featured a sudden flare, followed by a gradually declining flux over the next year. It bears some resemblance to a TDE, but this association could not be verified. Many of the 235 likely Galactic variable sources we investigated featured large, but infrequent and short duration flares.

A table listing the properties of the variable X-ray sources we identified can be found at: \href{https://projects.mpe.mpg.de/heg/erosita/SEP_var/}{https://projects.mpe.mpg.de/heg/erosita/SEP\_var/}.

\section{Acknowledgements}

This work is based on data from \emph{eROSITA}, the soft X-ray instrument aboard \emph{SRG}, a joint Russian-German science mission supported by the Russian Space Agency (Roskosmos), in the interests of the Russian Academy of Sciences represented by its Space Research Institute (IKI), and the Deutsches Zentrum für Luft- und Raumfahrt (DLR). The \emph{SRG} spacecraft was built by Lavochkin Association (NPOL) and its subcontractors, and is operated by NPOL with support from the Max Planck Institute for Extraterrestrial Physics (MPE).

The development and construction of the \emph{eROSITA} X-ray instrument was led by MPE, with contributions from the Dr. Karl Remeis Observatory Bamberg \& ECAP (FAU Erlangen-Nuernberg), the University of Hamburg Observatory, the Leibniz Institute for Astrophysics Potsdam (AIP), and the Institute for Astronomy and Astrophysics of the University of Tübingen, with the support of DLR and the Max Planck Society. The Argelander Institute for Astronomy of the University of Bonn and the Ludwig Maximilians Universität Munich also participated in the science preparation for \emph{eROSITA}.

The \emph{eROSITA} data shown here were processed using the eSASS software system developed by the German \emph{eROSITA} consortium.


This work made use of the software packages astropy \citep[\href{https://www.astropy.org/}{https://www.astropy.org/};][]{2013A&A...558A..33A, 2018AJ....156..123A}, bexvar \citep[\href{https://github.com/JohannesBuchner/bexvar}{https://github.com/JohannesBuchner/bexvar};][]{2022A&A...661A..18B}, numpy \citep[\href{https://www.numpy.org/}{https://www.numpy.org/};][]{harris2020array}, matplotlib \citep[\href{https://matplotlib.org/}{https://matplotlib.org/};][]{Hunter:2007}, scipy \citep[\href{https://scipy.org/}{https://scipy.org/};][]{2020SciPy-NMeth}, Stingray \citep[\href{https://docs.stingray.science/}{https://docs.stingray.science/};][]{2016ascl.soft08001H}, and Ultranest \citep[\href{https://johannesbuchner.github.io/UltraNest/}{https://johannesbuchner.github.io/UltraNest/};][]{2021JOSS....6.3001B}.

GP acknowledges funding from the European Research Council (ERC) under the European Union’s Horizon 2020 research and innovation programme (grant agreement No 865637), support from Bando per il Finanziamento della Ricerca Fondamentale 2022 dell’Istituto Nazionale di Astrofisica (INAF): GO Large program and from the Framework per l’Attrazione e il Rafforzamento delle Eccellenze (FARE) per la ricerca in Italia (R20L5S39T9). MK acknowledges support from DLR
grant FKZ 50 OR 2307.

\bibliographystyle{aa} 
\bibliography{bibliography.bib}      

\appendix

\section{Properties of the SEP field}\label{Prop3dSEP}

The \emph{eROSITA} source detection pipeline \citep{2022A&A...661A...1B} detected 8728 X-ray sources in eRASS1, 7984 in eRASS2, and 7770 in eRASS3 in the SEP field. These amount to more than $10^4$ unique sources. Despite being located within a small region of the sky, the properties of these observations are inhomogeneous, and vary a lot as a function of angle from the SEP. 


\subsection{Number of erodays of observation}\label{SecNeD}

In an idealised and simplified model of the survey, as assumed by Eq. \ref{Nedtheory}, the number of erodays of observations of sources in the SEP field per eRASS ranges from 119 to 1080. In practice, orbit corrections, computer resets, calibration observations, and other events prevented continuous survey mode operation. The non-uniform rotation rate of the scanning plane around the ecliptic equator also caused sources located at the same ecliptic latitude to be observed a different number of times. Removing bins with a fractional exposure of less than $0.1$ further reduced the number of erodays of observation useful for this variability analysis. The slight shift of the survey pole also affected the number of observations of sources located close to the SEP. This reduced the size of the region within which sources were observed on every eroday. 

\begin{figure}[h]
\resizebox{\hsize}{!}{\includegraphics{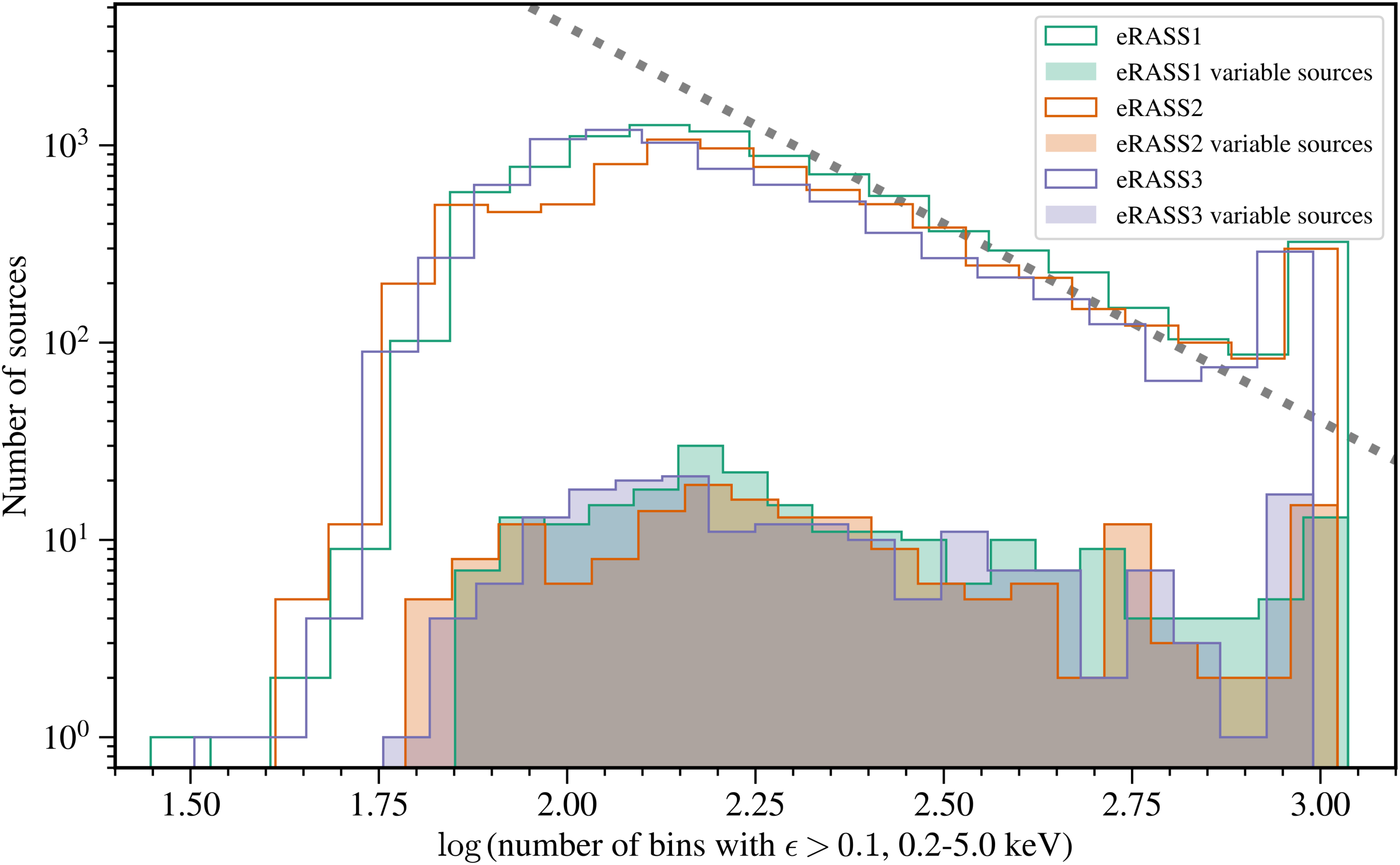}}
\caption{Histogram showing the distribution of the number of erodays, that all sources in the SEP field were observed for, in eRASS1, 2, and 3. The coloured in histogram depicts the distribution of the number of erodays of observation of variable sources. The grey dotted line depicts an inverse square relationship, which approximately matches the decrease in the number of sources observed at a particular logarithmic interval on the number of bins. We only counted bins with a fractional exposure larger than $0.1$.
 \label{nbhist}}
\end{figure}

The combination of these effects is illustrated in Fig. \ref{nbhist}. It displays the distribution of the number of erodays during which each X-ray source detected in the SEP field was observed with a fractional exposure of at least $0.1$, in eRASS1, 2, and 3. Fig. \ref{nbhist}, and the other histograms in this section also show the distribution of the relevant parameters for the variable sources that were identified in Section \ref{VarsampleeR123}. The greatest number of sources per logarithmic interval on the number of bins occurred at $145$, $149$, and $113$ bins, in eRASS1, 2, and 3, respectively. This is close to the expected lower limit, as there is a greater solid angle between $\theta$ and $\theta+\delta\theta$ for larger values of $\theta$ (see Eq. \ref{Omegangle}). eRASS3 had a maximum at a lower value than in the previous two surveys, as it had more days of telescope downtime, which caused the entire distribution to be shifted to lower values.


The distribution of the number of bins per source per eRASS with $\epsilon > 0.1$ extends significantly below the value of 119 in the idealised model of the survey. The minimum number of erodays during which a source was observed in the SEP field was 28, 41, and 32, in eRASS1, 2, and 3, respectively. However, only a few sources were observed with fewer than 50 bins. This tail of the distribution is due to sources at the edge of the field of view. These sources also lost a significant fraction of their observing interval due to a combination of detrimental effects. 


Using Eq. \ref{Omegangle}, the number of sources observed within logarithmically spaced intervals on the number of bin ($\rho(N_{\mathrm b})$), can be estimated to decrease with the inverse square of the number of bins: $\rho(N_{\mathrm b}) \propto N_{\mathrm b}^{-2}$. This relation assumes a constant detected source density per unit solid angle, and only applies to small angles relative to the ecliptic poles. It does not apply to angles smaller than $0.5\degree$, as sources lying closer to the ecliptic poles were observed approximately the same number of times. However, a greater number of bins implies a greater exposure depth and an enhanced ability to detect fainter sources. As a result, the source density is not identical throughout this region, but increased towards the SEP. The interplay between these two effects explains most of the shape of the distribution of Fig. \ref{nbhist}. 


The effect of decreasing solid angles closer to the SEP dominates over the effect of enhanced source detection ability. The resulting distribution is slightly less steep than an inverse square relationship. This is shown in Fig. \ref{nbhist}, in which an inverse square function is displayed as a grey dotted line on top of the distribution. 

The spike at the highest number of bins is caused by sources lying within $0.5\degree$ of the SEP. We observed a maximum number of bins per source, of 1088, 1054, and 978, in eRASS1, 2, and 3, respectively. The sources at this maximum were observed on almost every eroday of operational survey mode during the eRASS in question. The differences between those three numbers is due to the different number of erodays of telescope downtime in these three eRASSs.





Another point of interest is the shift of the survey pole away from the SEP. The SEP has coordinates of $\mathrm{RA}=06\mathrm{h}~ 00\mathrm{m}~ 00.00\mathrm{s}$, $\mathrm{Dec}=-66\degree~ 33'~ 38.55 ''$. In contrast, the average direction of the pole in eRASS1, 2, and 3 varied, and differed slightly from the coordinates of the SEP. We identified the average position of the survey pole to have been located at: $\mathrm{RA}=05\mathrm{h}~ 58\mathrm{m}~ 27\mathrm{s}$, $\mathrm{Dec}=-66\degree~ 33'~ 41 ''$  ($0.1535\degree$ away from the SEP) in eRASS1, $\mathrm{RA}=06\mathrm{h}~ 01\mathrm{m}~ 02\mathrm{s}$, $\mathrm{Dec}=-66\degree~ 32'~ 33 ''$ ($0.1050\degree$ away from the SEP) in eRASS2, and $\mathrm{RA}=05\mathrm{h}~ 58\mathrm{m}~ 29\mathrm{s}$, $\mathrm{Dec}=-66\degree~ 33'~ 54 ''$ ($0.1512\degree$ away from the SEP) in eRASS3. In all three eRASSs, the separation of the survey pole from the SEP was comparable. The average eRASS1 and eRASS3 survey poles only had a small separation of $18 ''$. In contrast, the pole of eRASS2 was located approximately on the other side of the SEP, compared to the poles of eRASS1 and 3. These differences in the direction of the survey pole are small, but they affected the frequency of observation of sources as a function of the angle to the SEP.

\subsection{Fractional exposure}

\begin{figure}[h]
\resizebox{\hsize}{!}{\includegraphics{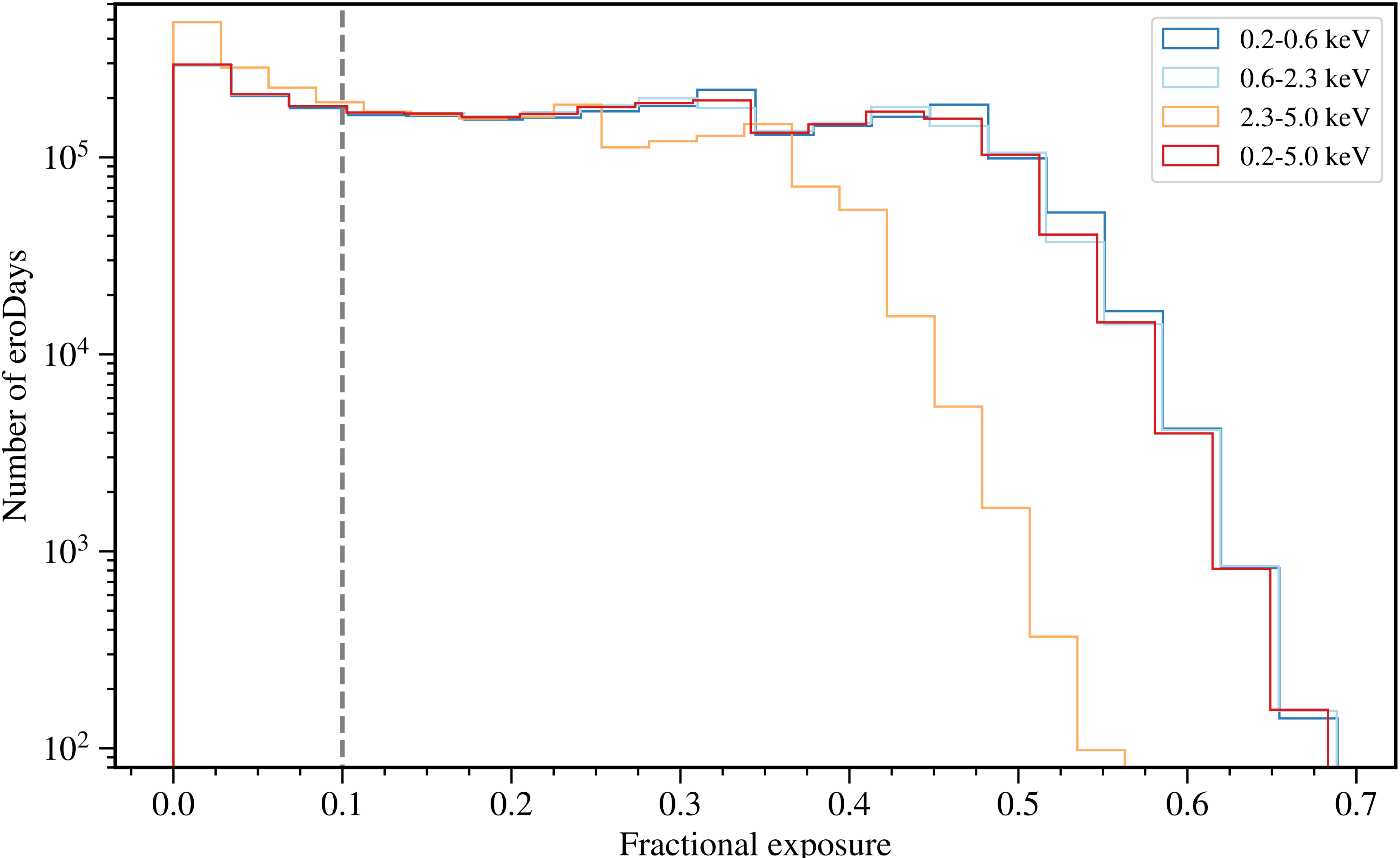}}
\caption{Histogram showing the distribution of the fractional exposure per eroday, for the 4 different energy bands, for all sources in the SEP field in eRASS1. The vertical dashed line at $\epsilon=0.1$ indicates our chosen lower limit. 
 \label{TotFEhist}}
\end{figure}

Fig. \ref{TotFEhist} displays the distribution of the fractional exposure of all sources in the SEP field, for all erodays of observation, in eRASS1. The fractional exposure was calculated for $40 ~ \mathrm{s}$ bins. This figure distinguishes between the four energy bands of $0.2-0.6~\mathrm{keV}$, $0.6-2.3~\mathrm{keV}$, $2.3-5.0~\mathrm{keV}$, and $0.2-5.0~\mathrm{keV}$. The fractional exposure in these different bands was mostly similar. However, the distribution in the $2.3-5.0~\mathrm{keV}$ energy band was shifted to lower fractional exposure values due to the stronger vignetting effect at higher energies. In all four energy bands, the peak of the distribution in the fractional exposure, grouped into linear bins, occurs at a value close to 0. Nevertheless, when grouped in logarithmic intervals, the peak of the distribution occurs at $\epsilon = 0.30$. About $26\%$ of all bins had a fractional exposure below the threshold for the full energy band. 

The distribution of the fractional exposure within this field is reasonably uniform between 0.1 and 0.45, but drops rapidly at larger fractional exposures, with a maximum of just under 0.7. There is a small spike in the fractional exposure at a value of 0.32, probably due to the geometry of the scans close to the SEP. The distribution seen in eRASS2 and 3 is almost identical to that in eRASS1, but shifted to a lower total number of erodays, as those surveys contained fewer sources, and fewer erodays of observation. The mean fractional exposure per source, per eroday for all sources in the SEP field, follows an approximately normal distribution, with a mean of $0.30$, and a standard deviation of $0.042$. The variable sources identified in Section \ref{VarsampleeR123} have a mean fractional exposure distribution slightly shifted to larger values, with a mean of $0.33$, and a standard deviation of $0.054$.

\subsection{Total effective exposure time}

The total effective exposure time denotes how long an observation of a selected source at the center of the field of view would have to be performed, for it to have the same exposure depth as in the actual observations. It describes the product of the fractional exposure with the duration of each observation, summed over all erodays of observation: $\sum_{i=1}^{N_{ed}}40\epsilon$. 



\begin{figure}[h]
\resizebox{\hsize}{!}{\includegraphics{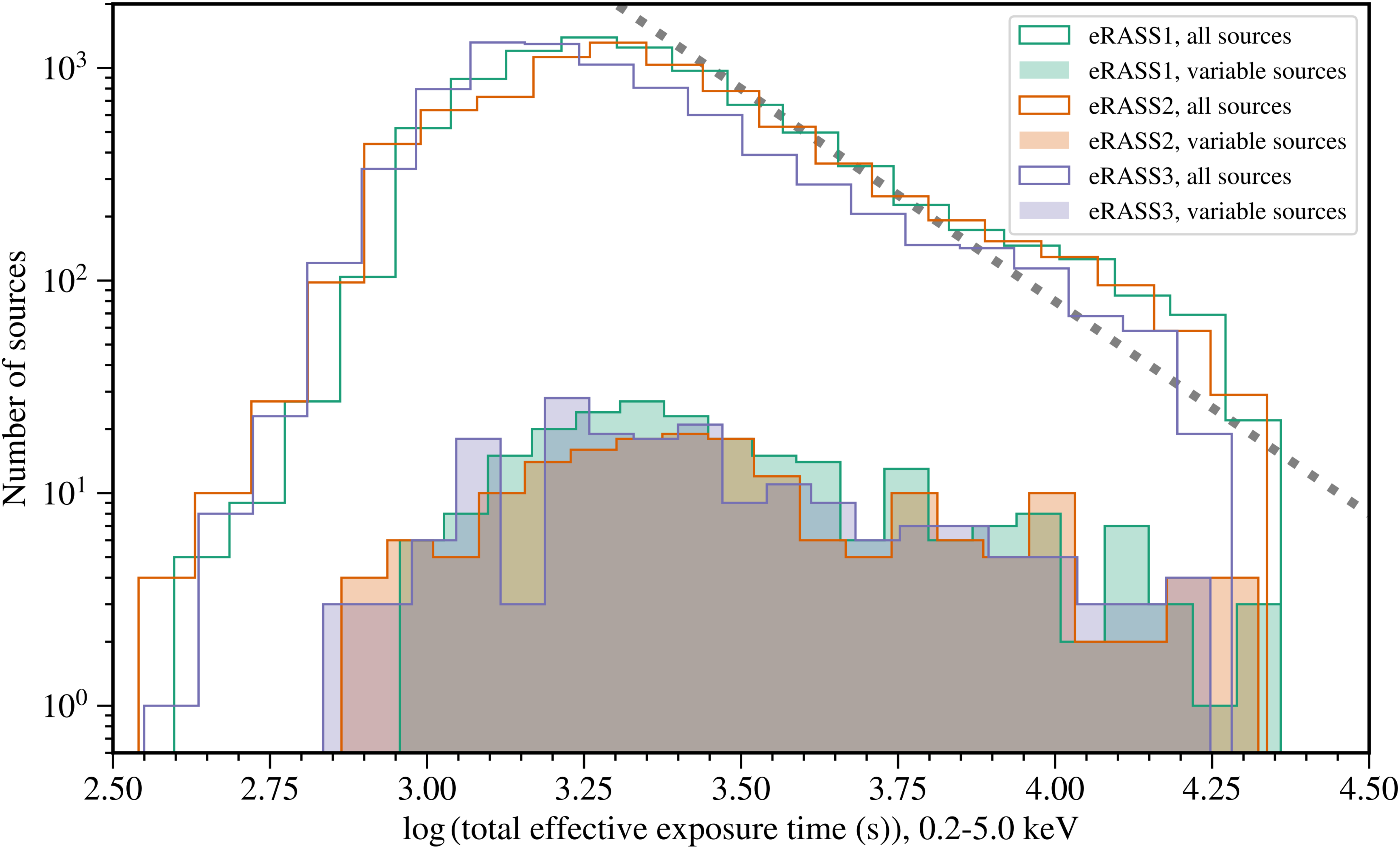}}
\caption{The distribution of the total effective exposure time. The distribution of the effective exposure time of the variable sources identified in each eRASS is also indicated. The grey dotted line depicts an inverse square relationship, showing that the drop in the number of sources within a logarithmic interval on the total effective exposure time is dominated by the decrease in solid angle at a decreasing angle from the SEP. 
 \label{TEFEhist}}
\end{figure}

Fig. \ref{TEFEhist} depicts the distribution of the total effective exposure time for the full energy band, $0.2-5.0~\mathrm{keV}$. The overall shape is still similar to the distribution of the number of bins (Fig. \ref{nbhist}) of sources in the SEP field, but with two notable differences. Firstly, the low exposure tail is more pronounced, as vignetting effects exacerbate the issues causing this tail. Secondly, the high exposure peak corresponding to sources observed almost every eroday is gone in Fig. \ref{TEFEhist}. Even though the number of bins hardly changed at angles $< 0.5\degree$ from the survey pole, the average fractional exposure per bin of these sources kept increasing with decreasing angle from the SEP. Sources lying $0.5\degree$ away from the survey pole were observed with a low fractional exposure in a greater number of bins than sources lying closer to the SEP. 


The effective exposure time of sources in the SEP field varied between $0.40-22.88~\mathrm{ks}$ in eRASS1, $0.35-21.75~\mathrm{ks}$ in eRASS2, and $0.36-19.11~\mathrm{ks}$ in eRASS3. The peak of the distribution of the effective exposure time, per logarithmic interval, occured at $1.866~\mathrm{ks}$ in eRASS1, $1.983~\mathrm{ks}$ in eRASS2, and $1.367~\mathrm{ks}$ in eRASS3. The vignetting effect at large off-axis angles significantly reduced the effective exposure time compared to the total exposure time. 

\subsection{Total exposure time}\label{AppTET}

\begin{figure}[h]
\resizebox{\hsize}{!}{\includegraphics{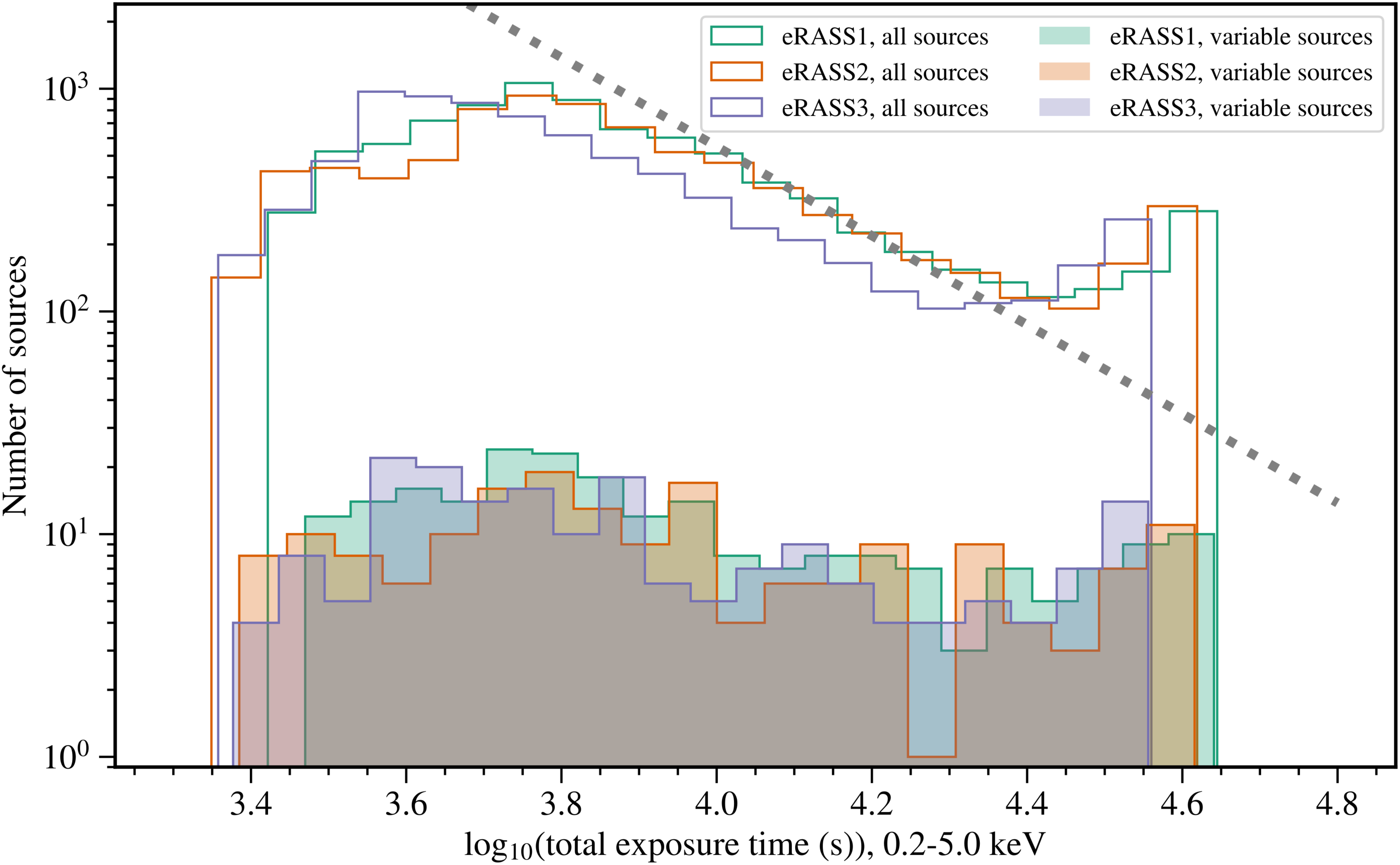}}
\caption{The distribution of the exposure time of sources in the SEP field, for eRASS1, 2, and 3. This quantity describes the total time that a source was inside the field of view, and does not consider vignetting effects. The distribution of the total exposure time for the sources we identify as variables is also shown. The grey dotted line depicts an inverse square relationship, showing that the drop in the number of sources within a logarithmic interval on the total exposure time is dominated by the decrease in the solid angle at a decreasing angle from the SEP. 
 \label{TEFThist}}
\end{figure}

Fig. \ref{TEFThist} depicts the distribution of the total exposure time for sources in the SEP field, in eRASS1, 2, and 3. This parameter describes the total time that each source was inside the field of view during each of the surveys, and is calculated as the product of the fractional time and the maximal duration of an observation, summed over all bins: $\sum_{i=1}^{N_{\mathrm b}}40\epsilon_t$. It does not consider any vignetting effects and is, therefore, closely related to the number of bins. So the distribution of the number of bins per source (Fig. \ref{nbhist}), and the distribution of the total exposure time, are similar. A noticeable difference is that the peak at the highest exposure times is broadened. Even though sources within $0.5\degree$ of the survey pole centre were observed during almost every eroday, sources further away from the survey pole spent less time in the field of view, on average. 


The total exposure time of sources in the SEP field ranged between $2.6 - 44.1~\mathrm{ks}$ in eRASS1, $2.2 - 41.6~\mathrm{ks}$ in eRASS2, and $2.3 - 36.3~\mathrm{ks}$ in eRASS3. The maximum of the distribution of total exposure time per eRASS, in logarithmic intervals, occurs at $5.77~\mathrm{ks}$ in eRASS1, $6.10~\mathrm{ks}$ in eRASS2, and $3.83~\mathrm{ks}$ in eRASS3.

\subsection{Source count rate}


The average source count rate can be determined by replacing all terms of Eq. \ref{countratedef} with their sums over all $N_{\mathrm b}$ bins that exceed the fractional exposure limit: 

\begin{equation}\label{countrateavg}
\overline{R_{\mathrm S}} = \frac{\sum_{i=1}^{N_{\mathrm b}}C(t_i)-A(t_i)B(t_i)}{\sum_{i=1}^{N_{\mathrm b}}\epsilon(t_i) \Delta t}.
\end{equation}

\noindent
This is reasonably accurate for almost all sources in the SEP field, due to the number of bins of the light curves. This equation is also equal to the weighted average of the count rate of each bin, using the fractional exposure as a weight, if the background ratio is constant for all bins.


\begin{figure}[h]
\resizebox{\hsize}{!}{\includegraphics{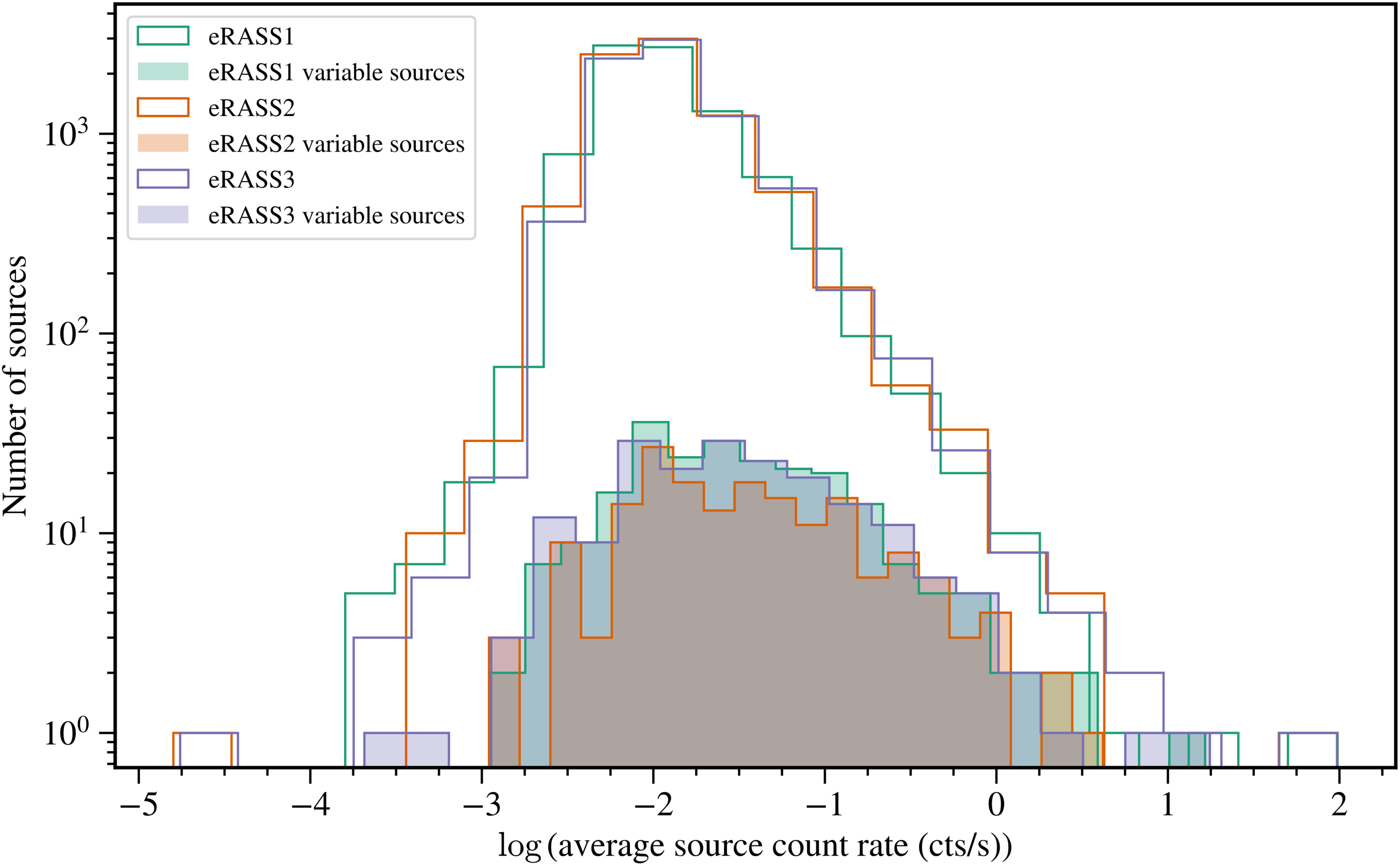}}
\caption{The distribution of the average count rates of sources in the SEP field, in eRASS1, 2, and 3. The average count rates were determined by Eq. \ref{countrateavg}. Due to the high exposures in the SEP field, \emph{eROSITA} could detect sources at much lower count rates than in the rest of the sky. This figure does not show the six sources with a slightly negative average count rate. 
 \label{arhist}}
\end{figure}

Fig. \ref{arhist}, shows the distribution of the average source count rate measured by \emph{eROSITA} for all sources in the SEP field, in the $0.2-5.0~\mathrm{keV}$ band. Almost all detected sources had average source count rates in the range $1.8\times10^{-4} - 4.2~\mathrm{cts/s}$. The most common average source count rate, per logarithmic interval, was $8.9\times10^{-3}~\mathrm{cts/s}$. The distribution increases rapidly from low count rates towards the maximum. Above the peak, the distribution decreases less steeply to higher count rates, in a manner that is consistent with a power law with an index of $\approx -1.2$. Unlike previous plots, the distribution of the average source count rate in eRASS1, 2, and 3 is very similar.



The brightest source in the SEP field is the supernova remnant SNR B0535-66.0, for which we measured an average source count rate of $\approx 97.5 ~\mathrm{cts/s}$. For such a bright source, the measured count rate is inaccurate, due to pileup effects (Merloni et al. 2024, accepted). 

We found six sources with slightly negative average source count rates. These sources were extremely faint. Despite being observed for several ks, only very few source counts were detected for them, slightly below the average background count rate. These lowest count rate estimates are not only unreliable, but are also from potentially spurious detections. 

\begin{figure}[h]
\resizebox{\hsize}{!}{\includegraphics{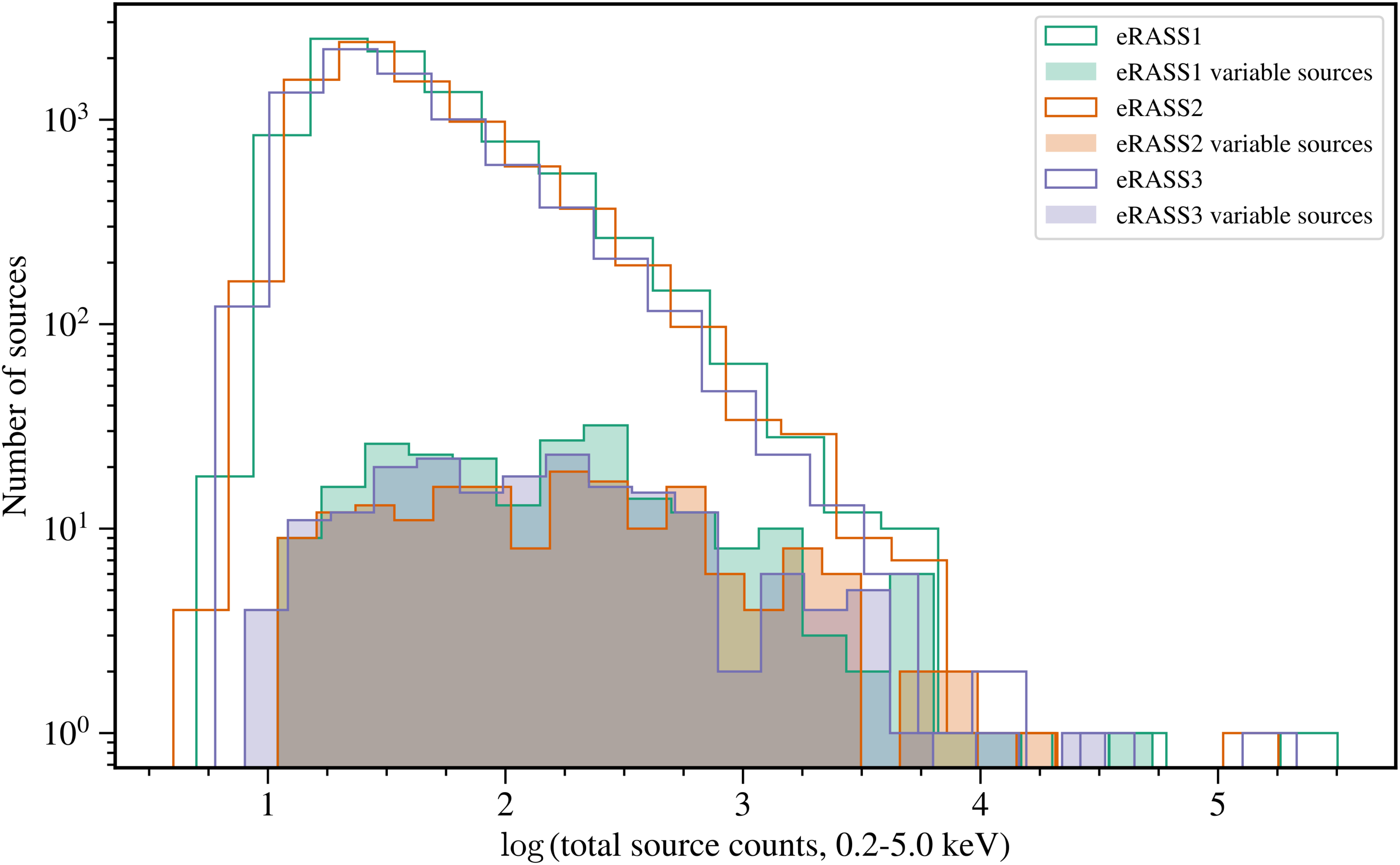}}
\caption{The distribution of the total source counts for all sources in the SEP field, in eRASS1, 2, and 3, for the energy band $0.2-5.0~\mathrm{keV}$. 
 \label{cthist}}
\end{figure}

In Fig. \ref{cthist}, the distribution of the total measured counts in the source extraction region for the $0.2-5.0~\mathrm{keV}$ energy band is displayed. The distribution features a steep rise from the lowest number of counts, of merely 4, to a peak at 25 counts. This is the value at which there were the greatest number of sources per logarithmic interval. Above the peak, the distribution is dominated by a decline that also approximately follows a power law, with an index of  $\approx-1.2$. This power law continues until at least 5600 counts per source, per eRASS. The greatest number of source counts measured for a single source in one eRASS was $3.19\times10^5$ for SNR B0535-66.0 in eRASS1. The strongly positively skewed distribution in Fig. \ref{cthist} results from the positive skew of both the average count rate distribution, and the effective exposure time distribution. 


\subsection{Background count rate}

\begin{figure}[pt]
\resizebox{\hsize}{!}{\includegraphics{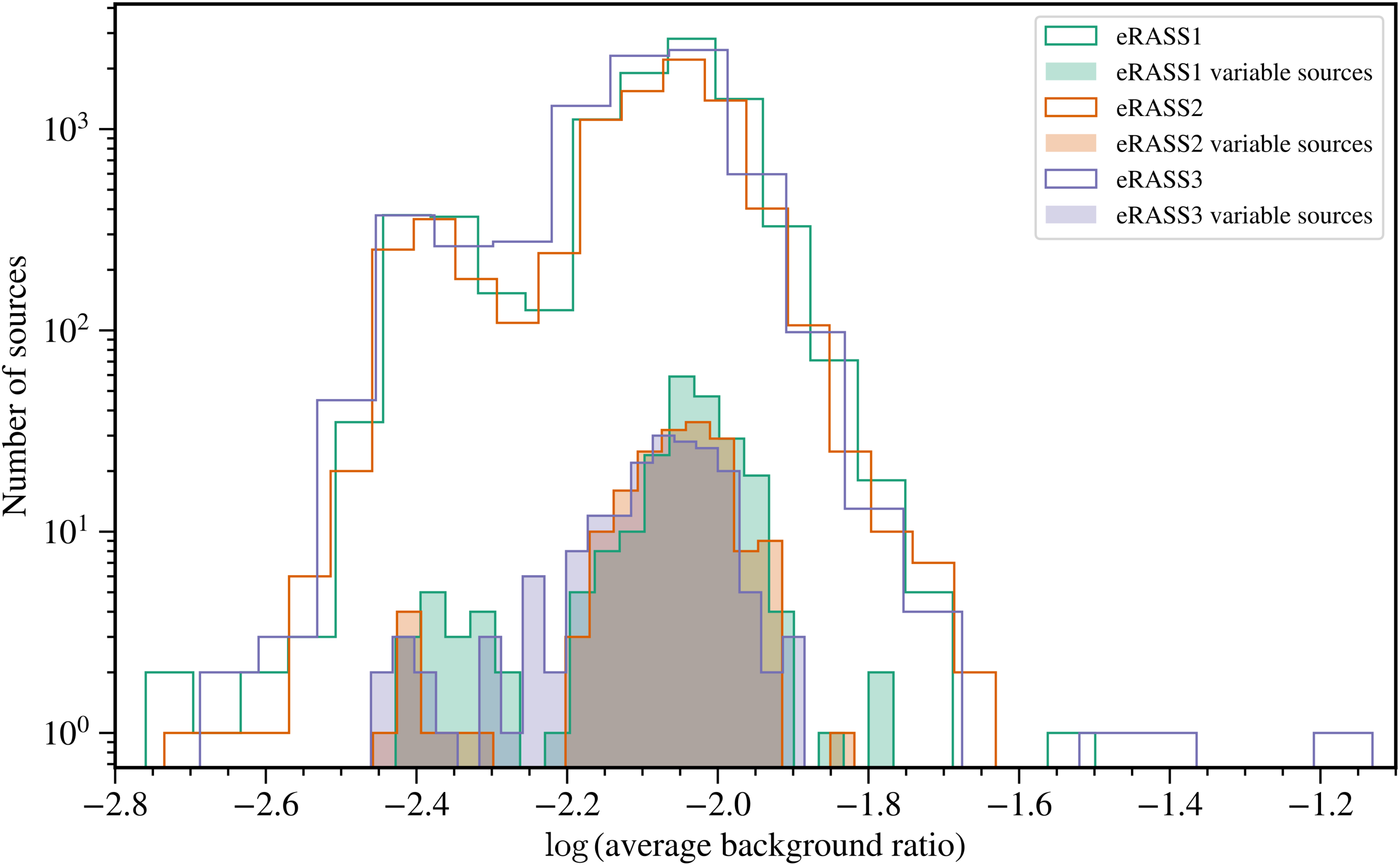}}
\caption{The distribution of the average ratio of the area of the source to the background extraction region, for each source detected in the SEP field, in eRASS1, 2, and 3. The background ratio exhibits slight variations throughout each eRASS, and across eRASSs, but usually remained close to the average value. 
 \label{ABAhist}}
\end{figure}

The background ratio describes the ratio of the source extraction area to the background extraction area. Fig. \ref{ABAhist} shows that the background extraction region was always significantly larger than the source extraction region. The most common value of the background ratio, per logarithmic interval, was $0.0089$. Nevertheless, the distribution of background ratios is broad, spanning the interval from $0.0017$ to $0.074$, and features a second prominent peak at a value of $0.0040$. 

\begin{figure}[h]
\resizebox{\hsize}{!}{\includegraphics{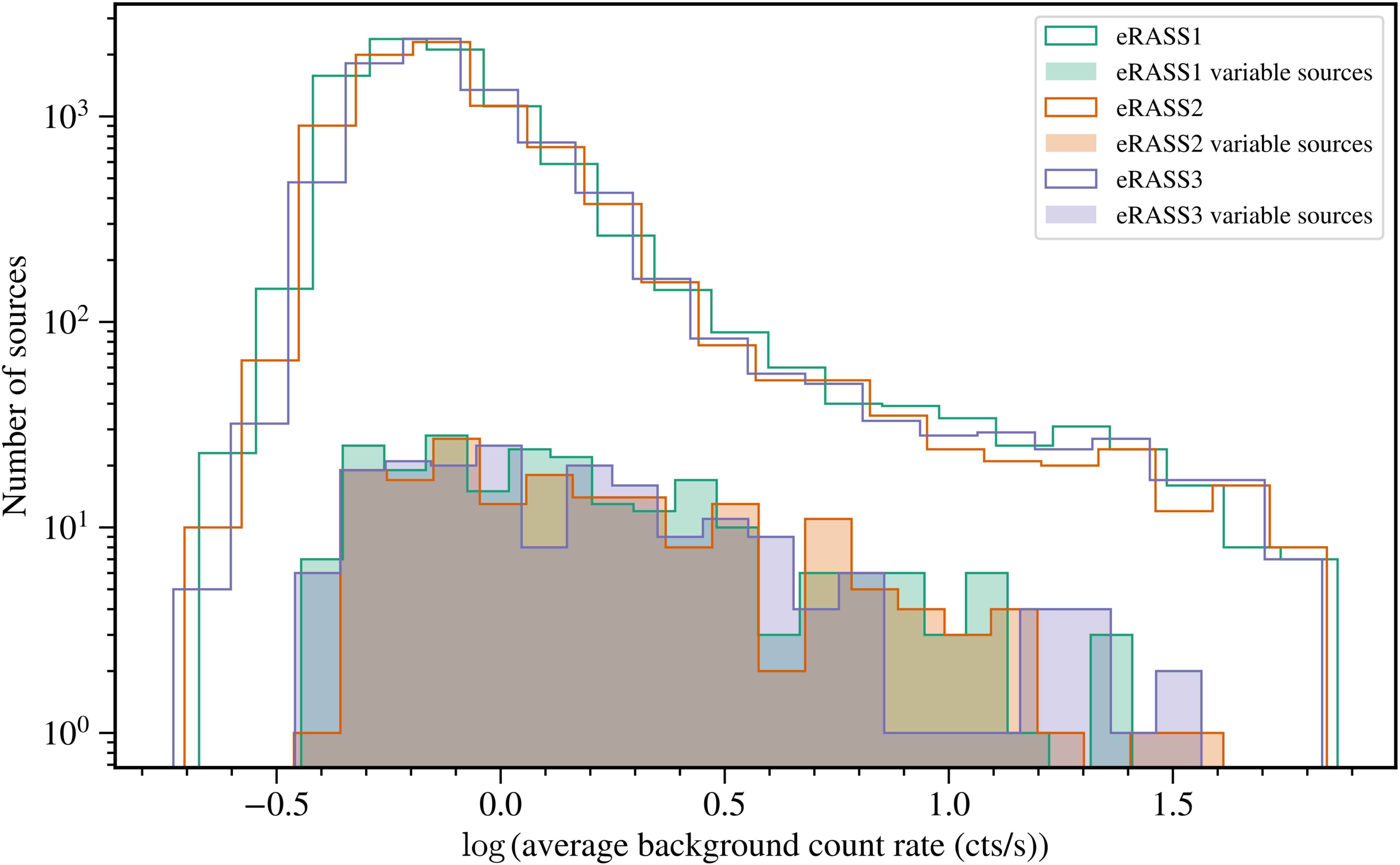}}
\caption{The distribution of the average background count rate in the background extraction regions used to determine the background subtracted source count rates. The background extraction region is typically more than $\approx100$ times larger than the source extraction region.  
 \label{ABRhist}}
\end{figure}

Fig. \ref{ABRhist} depicts the observed distribution of the background count rate for the $0.2-5.0~\mathrm{keV}$ interval, measured in the background extraction region for each detected source. Similar to the source count rate,  the average background count rate, $\overline{R_B}$, can be estimated by: 

\begin{equation}\label{backcountratedef}
\overline{R_B} \approx \frac{\sum_{i=1}^{N_{\mathrm b}}B(t_i)}{\sum_{i=1}^{N_{\mathrm b}}\epsilon(t_i) \Delta t}.
\end{equation}
\noindent

Similar to the distribution of the observed average source count rate, this follows a sharp rise from the lowest background count rate of $0.186~\mathrm{cts/s}$ up to a peak at $0.71~\mathrm{cts/s}$. However, this distribution is much more positively skewed, almost reaching a plateau at the highest background count rates of up to $73.7~\mathrm{cts/s}$. These background count rates are often much larger than the source counts rates, as they are obtained over a much larger area than the source count rate (See Fig. \ref{ABAhist}). The part of the SEP field that intersects within the LMC has a higher background count rate than the rest of the field, which accounts for some of the high background count rates observed. The width of this distribution is exacerbated by the distribution of the background ratio. This graph does not represent the distribution of the background count rate per solid angle within this field. 



\subsection{Detection likelihood}

\begin{figure}[h]
\resizebox{\hsize}{!}{\includegraphics{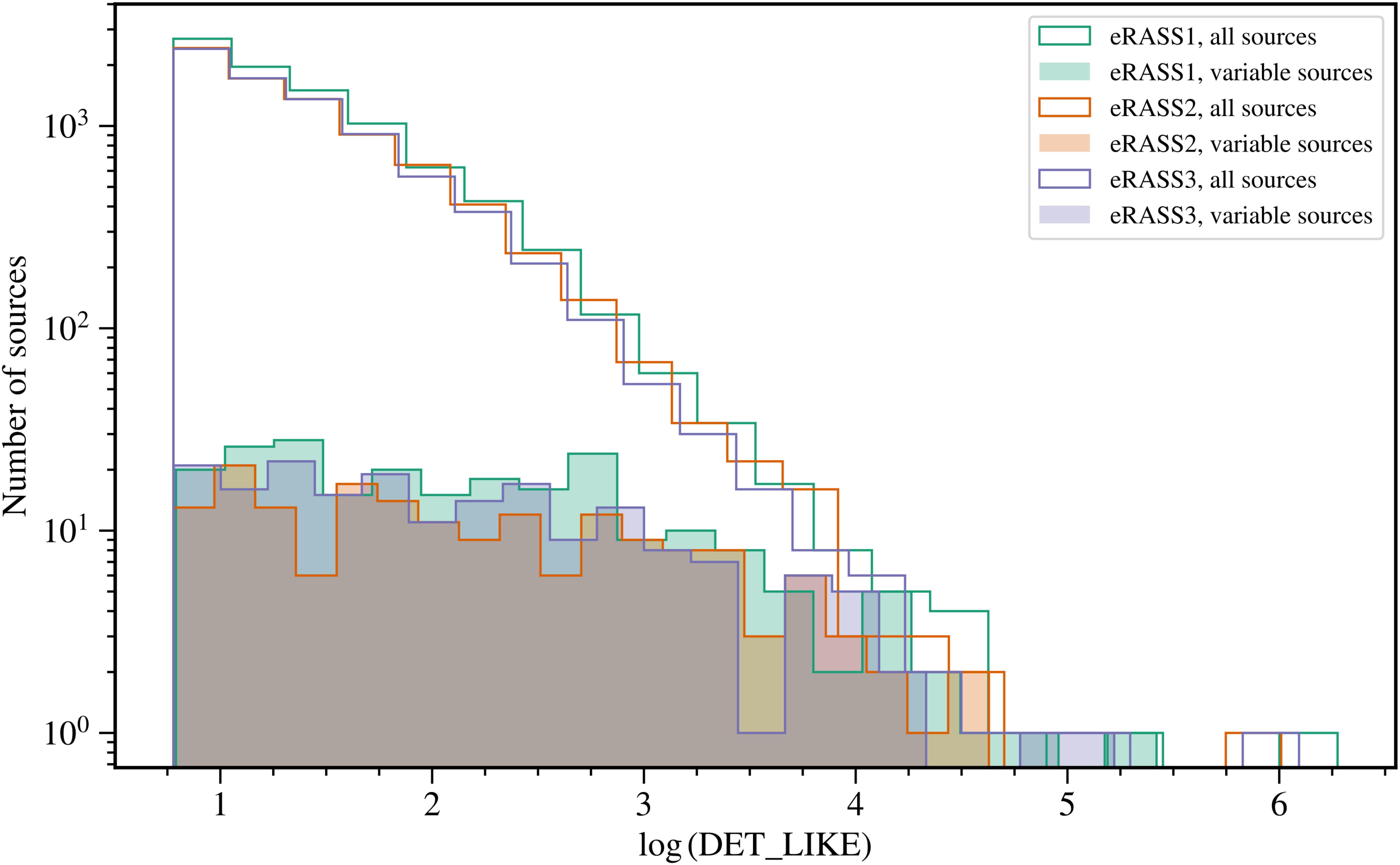}}
\caption{The distribution of the detection likelihood parameter of all sources lying within $3\degree$ of the SEP in eRASS1, 2, and 3, for the energy band $0.2-5.0~\mathrm{keV}$. 
 \label{DLhist}}
\end{figure}

The DET\_LIKE parameter quantifies the reliability of the source detection. The greater the DET\_LIKE value is, the greater the certainty that a source is genuine. In Fig. \ref{DLhist}, we plot the distribution of the detection likelihood for all detected sources in the SEP field in eRASS1, 2, and 3, for the $0.2-5.0~\mathrm{keV}$ energy band. The maximum of the distribution occurs at the lower limit of significantly detected sources, of $\mathrm{DET\_LIKE} = 6$. This indicates that a not insignificant number of sources are barely above the detection threshold, and therefore should still be considered as potentially spurious sources. The distribution drops rapidly, but extends to very high DET\_LIKE, of up to $1.89\times10^6$ in eRASS1. Low DET\_LIKE sources are observed with just a few counts above the expected background level.



Variable sources have an almost uniform DET\_LIKE distribution, as shown in Fig. \ref{DLhist}. Therefore, the higher the DET\_LIKE of a source, the more likely it is to be classified as variable.


\end{document}